\documentclass[aps,onecolumn,11pt,floatfix,altaffilletter,tightenlines,showpacs,showkeys,notitlepage,nofootinbib,preprintnumbers]{revtex4-1}
\pdfoutput=1
\usepackage[dvipdf,dvips,pdftex]{graphicx}
\usepackage[colorlinks=true,citecolor=blue,linkcolor=blue]{hyperref}
\usepackage{amsmath}
\usepackage{amssymb}
\usepackage{epsfig}
\usepackage{epstopdf}
\usepackage{url}
\usepackage{color}
\usepackage[utf8]{inputenc}
\usepackage{empheq}
\usepackage{here}
\usepackage{bm}
\usepackage{braket}

\newcommand{\be}{\begin{eqnarray}}
\newcommand{\ee}{\end{eqnarray}}
\newcommand{\al}[1]{\begin{align}#1\end{align}}
\catcode`\@=11

\def\lsim{\mathrel{\mathpalette\@versim<}}
\def\gsim{\mathrel{\mathpalette\@versim>}}
\def\@versim#1#2{\vcenter{\offinterlineskip
\ialign{$\m@th#1\hfil##\hfil$\crcr#2\crcr\sim\crcr } }}
\catcode`\@=12

\begin{document}

\title{Possibility of multi-step electroweak phase transition \\
in the two Higgs doublet models}

\author{Mayumi Aoki$^{1\,}$} \email{mayumi.aoki@staff.kanazawa-u.ac.jp}
\author{Takatoshi Komatsu$^{1\,}$} \email{t$\_$komatsu@hep.s.kanazawa-u.ac.jp}
\author{Hiroto Shibuya$^{1\,}$} \email{h$\_$shibuya@hep.s.kanazawa-u.ac.jp}

\affiliation{%
$^1$Institute for Theoretical Physics, Kanazawa University,
Kanazawa 920-1192, Japan
}

\pagestyle{plain}

\preprint{KANAZAWA-21-08}

%
\begin{abstract}
We discuss whether a multi-step electroweak phase transition (EWPT) occurs in two Higgs doublet models (2HDMs).
The EWPT is related to interesting phenomena such as baryogenesis and a gravitational wave from it.
We examine parameter regions in CP-conserving 2HDMs and find certain areas where the multi-step EWPTs occur.
The parameter search shows the multi-step EWPT prefers the scalar potential with the approximate $Z_2$ symmetry and a mass hierarchy between the neutral CP-odd and CP-even extra scalar bosons $m_A<m_H$.
By contrast, the multi-step EWPT whose first step is strongly first order favors a mass hierarchy $m_A>m_H$.
In addition, we compute the Higgs trilinear coupling in the parameter region where the multi-step EWPTs occur, which can be observed at future colliders.
We also discuss a multi-peaked gravitational wave from a multi-step EWPT.
\end{abstract}
\maketitle

%

\vspace{0.2in}
\section{Introduction}\label{sec:Introduction}
Although the standard model (SM) of particle physics is verified through various experiments, the asymmetry of baryon number in the universe is still one of the big problems. The observable of a baryon to radiation number ratio
is $\eta_B\equiv n_B/n_\gamma=(6.12\pm 0.04)\times10^{-10}$ \cite{Aghanim:2018eyx}.
To explain this asymmetry, the theory must satisfy Sakharov's three conditions \cite{Sakharov:1967dj}.
The conditions are violation of baryon numbers, violation of C and CP symmetries, and departure from thermal equilibrium.
To generate baryon asymmetry at the electroweak (EW) scale via EW baryogenesis (EWBG) \cite{Kuzmin:1985mm},
the EW phase transition (EWPT) is needed to be strongly first order. However,
the lattice simulations show that the mass of Higgs boson must be less than about $70$ GeV to make the EWPT first order in the SM \cite{Kajantie:1995kf,Csikor:1998eu}
and the observed Higgs boson with the mass of 125 GeV \cite{Aad:2012tfa,ATLAS-CONF-2012-162,Chatrchyan:2012ufa,CMS-PAS-HIG-12-045} indicates that the EWPT in the SM is cross over \cite{DOnofrio:2014rug}.
Furthermore, sufficient baryon asymmetry cannot be produced via the Cabibbo-Kobayashi-Maskawa phase,
so that the EWBG cannot be achieved successfully in the SM \cite{Gavela:1993ts, Huet:1994jb, Gavela:1994dt}.
The sufficient baryon asymmetry can be generated via the EWBG scenario by extending the scalar sector of the SM.
One of the simplest extensions is the two Higgs doublet model (2HDM), where a $SU(2)$ scalar doublet is added to the SM~\footnote{As studies for the strong 1-step PTs in the CP-conserving 2HDMs, see {\it e.g.} Refs.~\cite{Dorsch:2013wja, Basler:2016obg, Bernon:2017jgv, Wang:2018hnw, Su:2020pjw}. For the non-perturbative analyses, see Refs.~\cite{Andersen:2017ika,Kainulainen:2019kyp}.}.
Since the 2HDM has new CP-violating sources in the scalar potential, it has a possibility of achieving the EWBG.
However, the model has difficulty producing sufficient baryon number because the electric dipole moment (EDM) measurements constrain the sources strictly \cite{Haarr:2016qzq,Dorsch:2016nrg,Chen:2017com}~\footnote{In the aligned 2HDM, there are possibilities to evade the EDM constraints through cancellations among contributions to the EDMs even when CP-violating phases are unsuppressed \cite{Kanemura:2020ibp}.}.

One could come up with the idea to solve the above difficulty in the 2HDMs by considering a multi-step EWPT \cite{Blinov:2015sna}.
The sufficient baryon number is produced at the first step PT if it is the strongly first order and enough CP violation exists, while the baryon number cannot be washed out at the subsequent PT(s) if EW sphaleron processes are suppressed enough.
Consequently, sufficient baryon number is preserved at the EW vacuum even if it is a CP-conserving vacuum.
The reason why we consider baryon asymmetry produced at the first step is that the first order PT at the subsequent step occurs between $SU(2)$ broken phases.
Therefore the sphaleron processes are suppressed in both phases and the sufficient asymmetry would not be produced \cite{Hammerschmitt:1994fn, Fromme:2006cm}~\footnote{Ref.~\cite{Hammerschmitt:1994fn} shows that sufficient baryon asymmetry is difficult to be generated at the subsequent PT in the inert 2HDM. In Ref.~\cite{Fromme:2006cm}, it is also mentioned that the second step PT would not generate sufficient asymmetry in the 2HDMs.}.
Another interesting phenomenon derived from the multi-step PT is the multi-peaked gravitational wave (GW).
Since the first order PT yields a GW spectrum \cite{Witten:1984rs, Hogan:1986qda},
the superposed GW can have multiple peaks if the first order PT occurs multiple times, which could be observed by the future space-based interferometers such as the approved Laser Interferometer Space Antenna (LISA) \cite{Caprini:2015zlo, LISA:2017pwj, Caprini:2019egz}.
The previous researches concerning multi-step PTs are in the singlet extensions
\cite{
Profumo:2007wc,
Espinosa:2011ax,
Curtin:2014jma,
Jiang:2015cwa,
Huang:2015bta,
Kurup:2017dzf,
Kang:2017mkl,
Matsui:2017ggm,
Chiang:2017nmu,
Hashino:2018zsi,
Huang:2018aja,
Chiang:2019oms,
Carena:2019une,
Ghorbani:2020xqv,
Niemi:2021qvp},
inert 2HDMs \cite{
Land:1992sm,
Hammerschmitt:1994fn,
Blinov:2015sna,
Friedlander:2020tnq,
Fabian:2020hny},
2HDMs \cite{
Fromme:2006cm,
Bernon:2017jgv,
Wang:2019pet},
triplet extensions \cite{
Patel:2012pi,
Chala:2018opy,
Bell:2020gug,
Niemi:2020hto},
and the other models \cite{
Patel:2013zla,
Inoue:2015pza,
Huang:2017laj,
Chao:2017vrq,
Ramsey-Musolf:2017tgh,
Vieu:2018nfq,
Vieu:2018zze,
Bian:2018bxr,
Zhou:2018zli,
Bell:2019mbn,
Morais:2019fnm,
Zhou:2020idp,
Baum:2020vfl,
Ghosh:2020ipy,
Matsui:2021khj}.

In this paper, we study the multi-step EWPTs in the CP-conserving 2HDMs.
Because of the absence of the new CP-violating source,
the EWBG does not work and we would not discuss it.
The study for the CP-violating case remains as future work.
The main purpose of this paper is to reveal features of the multi-step PTs.
By performing parameter searches, we find certain parameter spaces where the multi-step PT occurs.
Furthermore to examine the possibility of verification of the multi-step PT at collider experiments,
we compute the deviation of the Higgs trilinear coupling from that in the SM.
It is known that the deviation can be large in the 2HDMs \cite{Kanemura:2002vm, Kanemura:2004ch} (see also Ref.~\cite{Braathen:2019zoh, Arco:2020ucn} for recent work),
and it would be observed more precisely in future colliders like High-Luminosity Large Hadron Collider (HL-LHC) \cite{Cepeda:2019klc} and International Linear Collider (ILC) \cite{Fujii:2015jha}.
We find that the deviation has a tendency to be large in certain regions when the multi-step PTs occur.
In addition,
we calculate a two-peaked GW spectrum yielded by a 2-step PT, which can be observed
by using LISA, Big Bang Observer (BBO) \cite{Corbin:2005ny}, and Ultimate Deci-Hertz Interferometer Gravitational Wave Observatory (U-DECIGO) \cite{Kudoh:2005as}.

The outline of this paper is as follows: In section \ref{sec:2HDMs} we introduce the generic characteristics of the 2HDMs.
Section \ref{sec:VatFiniteTemp} is dedicated to give the thermal effective potential.
Theoretical constraints considered in our numerical analyses are briefly introduced in Section \ref{sec:parameterconstraints}.
In section \ref{sec:numericalresults} we show the results of the parameter search for the multi-step PT.
Moreover, in section \ref{sec:PhysicalSignature}, we discuss the predictions for the Higgs trilinear couplings as the collider signatures and for the multi-peaked GW as the cosmological signature for the multi-step PT.
Our conclusions are given in section \ref{sec:Conclusion}.

\section{Two Higgs Doublet Model}\label{sec:2HDMs}
The tree-level scalar potential of the CP-conserving 2HDMs with a softly broken $Z_2$ symmetry is written as
\begin{equation}
  \begin{aligned}
V(\Phi_1,\Phi_2)&=m_1^2\Phi_1^\dagger\Phi_1+m_2^2\Phi_2^\dagger\Phi_2
-m_3^2(\Phi_1^\dagger\Phi_2+\Phi_2^\dagger\Phi_1)
+\frac{\lambda_1}{2}(\Phi_1^\dagger\Phi_1)^2+\frac{\lambda_2}{2}(\Phi_2^\dagger\Phi_2)^2
\\
&\ \ +\lambda_3(\Phi_1^\dagger\Phi_1)(\Phi_2^\dagger\Phi_2)
+\lambda_4(\Phi_1^\dagger\Phi_2)(\Phi_2^\dagger\Phi_1)
+\frac{\lambda_5}{2}\left[(\Phi_1^\dagger\Phi_2)^2+(\Phi_2^\dagger\Phi_1)^2\right],
  \end{aligned}\label{eq:potential}
\end{equation}
where $\Phi_i\ (i=1,2)$ are the $SU(2)$ scalar doublets
\begin{align}
\Phi_i&=\left(
\begin{array}{c}
w_i^+ \\
\frac{v_i+h_i+iz_i}{\sqrt{2}}
\end{array}
\right).
\end{align}
We here assume that only the neutral CP-even scalar fields have the vacuum expectation values (VEVs) $v_i$, which are real and positive, and satisfy $v\equiv\sqrt{v_1^2+v_2^2}=246$ GeV.
The third term with $m_3^2$ on the right-hand side in Eq.~(\ref{eq:potential})
breaks the $Z_2$ symmetry in the potential softly.
The coefficients are taken to be real,
although $m_3^2$ and $\lambda_5$ are complex parameters in general.
Regarding only the neutral CP-even fields $\phi_i$, the $\Phi_i$ become
\begin{align}
\Phi_i&=\left(
\begin{array}{c}
0 \\
\frac{\phi_i}{\sqrt{2}}
\end{array}
\right).
\label{eq:doublet}
\end{align}
Consequently, the tree-level scalar potential (\ref{eq:potential}) with the doublets (\ref{eq:doublet}) is
\begin{align}
V_0(\phi_1,\phi_2)&=\frac{m_1^2}{2}\phi_1^2+\frac{m_2^2}{2}\phi_2^2
-m_3^2\phi_1\phi_2
+\frac{\lambda_1}{8}\phi_1^4
+\frac{\lambda_2}{8}\phi_2^4+\frac{1}{4}(\lambda_3
+\lambda_4+\lambda_5)(\phi_1\phi_2)^2.
\label{eq:treepotential}
\end{align}
The minimum value of $V_0(\phi_1,\phi_2)$ given by $\phi_i=v_i$.
From the minimum conditions, $\left.\partial V_0/\partial \phi_i\right|_{\phi_i=v_i}=0$, we obtain
\begin{align}
m_1^2&=m_3^2\frac{v_2}{v_1}
-\frac{\lambda_1}{2}v_1^2-\frac{1}{2}(\lambda_3+\lambda_4+\lambda_5)v_2^2,\\
m_2^2&=m_3^2\frac{v_1}{v_2}-\frac{\lambda_2}{2}v_2^2-\frac{1}{2}(\lambda_3+\lambda_4+\lambda_5)v_1^2.
\label{eq:extrema2}
\end{align}

To calculate the effective potential, we introduce the field-dependent masses because we need the masses of all fields at each of the coordinates ($\phi_1, \phi_2$), which contribute to the potential at the loop-level.
The field-dependent mass matrices of the charged and the neutral CP-odd scalar fields in the gauge basis are respectively given by
\begin{equation}
\begin{aligned}
&\mathcal{M}_{w^\pm}^2=\frac{1}{2}
\left(
\begin{array}{cc}
2m_1^2+\lambda_1\phi_1^2+\lambda_3\phi_2^2 & -2m_3^2+(\lambda_4+\lambda_5)\phi_1\phi_2 \\
-2m_3^2+(\lambda_4+\lambda_5)\phi_1\phi_2 & 2m_2^2+\lambda_2\phi_2^2+\lambda_3\phi_1^2
\end{array}
\right),\\
&\mathcal{M}_z^2=\frac{1}{2}
\left(
\begin{array}{cc}
2m_1^2+\lambda_1\phi_1^2+(\lambda_3+\lambda_4-\lambda_5)\phi_2^2
& -2m_3^2+2\lambda_5\phi_1\phi_2 \\
-2m_3^2+2\lambda_5\phi_1\phi_2 &
2m_2^2+\lambda_2\phi_2^2+(\lambda_3+\lambda_4-\lambda_5)\phi_1^2
\end{array}
\right).
\end{aligned}\label{eq:bosonmassmatrices1}
\end{equation}
By taking $\phi_i=v_i$ and diagonalizing these matrices, the physical masses of the charged scalar field $H^\pm$ and the neutral CP-odd scalar field $A$ are respectively
obtained as
\begin{subequations}
\begin{align}
&m_{H^\pm}^2
=\frac{m_3^2}{\sin\beta\cos\beta}-\frac{1}{2}(\lambda_4+\lambda_5)v^2
,\\
&m_A^2
=\frac{m_3^2}{\sin\beta\cos\beta}-\lambda_5 v^2\label{eq:mA},
\end{align}\label{eq:scalarmasses1}
\end{subequations}\\
where we have introduced the angle $\beta$ as $\tan\beta \equiv v_2/v_1$.
On the other hand,
the physical squared-masses
of the neutral CP-even scalar fields $H$ and $h$ can be derived by diagonalizing the mass matrix
\begin{equation}
\begin{aligned}
&\mathcal{M}_h^2=\frac{1}{2}
\left(
\begin{array}{cc}
2m_1^2+3\lambda_1\phi_1^2+(\lambda_3+\lambda_4+\lambda_5)\phi_2^2 &
-2m_3^2+2(\lambda_3+\lambda_4+\lambda_5)\phi_1\phi_2 \\
-2m_3^2+2(\lambda_3+\lambda_4+\lambda_5)\phi_1\phi_2 &
2m_2^2+3\lambda_2\phi_2^2+(\lambda_3+\lambda_4+\lambda_5)\phi_1^2 \end{array}
\right),
\end{aligned}\label{eq:bosonmassmatrices2}
\end{equation}
with $\phi_i=v_i$ as
\begin{align}
\left(
\begin{array}{cc}
m_H^2 & 0 \\
0 & m_h^2
\end{array}
\right)
&=R(-\alpha)\mathcal{M}_h^2 R(\alpha),
~~~{\rm with}~~
R(\alpha)\equiv
\left(
\begin{array}{cc}
\cos\alpha & -\sin\alpha \\
\sin\alpha & \cos\alpha
\end{array}
\right).
\end{align}
Here, the squared masses $m_H^2$ and $m_h^2$ are obtained by
\begin{equation}
\begin{aligned}
&m_H^2=\frac{1}{2}\left[A+C+\sqrt{(A-C)^2+4B^2}\right],\\
&m_h^2=\frac{1}{2}\left[A+C-\sqrt{(A-C)^2+4B^2}\right],\\
\end{aligned}\label{eq:scalarmasses2}
\end{equation}
with
\begin{align}
&A=m_3^2\tan\beta+\lambda_1v^2\cos^2\beta, \\
&B=-m_3^2+(\lambda_3+\lambda_4+\lambda_5)v^2\sin\beta\cos\beta, \\
&C=\frac{m_3^2}{\tan\beta}+\lambda_2v^2\sin^2\beta.
\end{align}
Throughout this paper we take $h$ as the SM-like Higgs boson with $m_h=125$ GeV.

The field-dependent masses of the W boson, the Z boson, and the photon can be written as
\begin{equation}
\begin{aligned}
&m_{W}=\frac{1}{2}g\sqrt{\phi_1^2+\phi_2^2},\\
&m_{Z}=\frac{1}{2}\sqrt{g^2+g'^2}\sqrt{\phi_1^2+\phi_2^2},\\
&m_{\gamma}=0,
\end{aligned}
\end{equation}
where $g$ and $g'$ are the gauge couplings of $SU(2)_L$ and $U(1)_Y$ gauge symmetry, respectively.
The physical masses of the gauge bosons are derived by taking $\phi_i=v_i$.

The most general Yukawa term is
\begin{align}
\mathcal{L}_{\rm Yukawa}=-\bar Q_LY_u\tilde\Phi_uu_R-\bar Q_LY_d\Phi_dd_R-\bar L_LY_l\Phi_ll_R+{\rm h.c.},
\label{eq:Yukawa}
\end{align}
where $Q_L$ and $L_L$ are $SU(2)_L$ doublets of quarks and leptons, respectively, $Y_f\ \ (f=u,d,l)$ are the Yukawa matrices of the fermions, and each of $\Phi_f$ is either $\Phi_1\ {\rm or}\ \Phi_2$.
Since the 2HDMs have the two $SU(2)$ scalar doublets, we assume one of the doublets couples each of the fermions to avoid the tree-level flavor changing neutral current.
One of ways to accomplish this is assuming 2HDMs have a $Z_2$ symmetry.
In this case, there are 4 types in the 2HDMs distinguished by the $Z_2$ charges for each of the fermions as in the Tab.~\ref{table:Yukawa} \cite{Barger:1989fj,Grossman:1994jb,Aoki:2009ha}.
\begin{table}[t]
\centering
\begin{tabular}{|c||c|c|c|c|c|c|}
\hline
& $\Phi_1$ & $\Phi_2$ & $u_R$ & $d_R$ & $l_R$ & $Q_L,L_L$ \\ \hline\hline
Type-I & + &$-$&$-$&$-$&$-$&+ \\ \hline
Type-I\hspace{-.1em}I & + &$-$&$-$&+&+&+ \\ \hline
Type-X&+&$-$&$-$&$-$&+&+ \\ \hline
Type-Y&+&$-$&$-$&+&$-$&+ \\ \hline
\end{tabular}
\caption{Four types in the 2HDMs distinguished by $Z_2$ charges for each of the fermions.}
\label{table:Yukawa}

\end{table}
In the Type-I 2HDM, all quarks and charged leptons obtain their masses from the VEV of $\Phi_2$.
In the Type-II 2HDM, the VEV of $\Phi_2$ gives the masses of the up-type quarks, while that of $\Phi_1$ provides those of the down-type quarks and the charged leptons.
In the Type-X 2HDM, the charged leptons and quarks obtain their masses from the VEV of $\Phi_1$ and $\Phi_2$, respectively.
In the Type-Y 2HDM, the masses of the down-type quarks are generated by the VEV of $\Phi_1$, while those of the up-type quarks and the charged leptons are obtained by that of $\Phi_2$.
The field-dependent masses of fermions can be described as
\begin{align}
m_f=\frac{1}{\sqrt{2}}y_f \phi_i,
\end{align}
where which value assigned to $i$ depends on the types of Yukawa interactions.
The physical masses of the fermions are obtained by taking $\phi_i=v_i$.

\section{The effective potential at finite temperature}\label{sec:VatFiniteTemp}
\subsection{The one-loop corrected effective potential}
An EWPT is caused by the temperature change of the effective scalar potential.
To study the PT, we consider the thermal effective potential.
The one-loop corrected effective potential at the finite temperature $V^\beta $ is
\begin{align}
V^\beta = V_0 + V_{\rm CW} + V_{\rm CT} + \overline{V}_1^\beta,
\label{eq:totalpotential}
\end{align}
where $V_0,\ V_{\rm CW}$, $V_{\rm CT}$, and $\overline{V}_1^\beta$ are the tree-level potential (\ref{eq:treepotential}), the one-loop level potential at zero temperature (the Coleman-Weinberg potential), the counterterm potential, and the one-loop level potential at the finite temperature, respectively.

The Coleman-Weinberg potential in the $\overline{{\rm MS}}$ scheme is written by \cite{Quiros:1999jp}
\begin{align}
V_{\rm CW}(\phi_1, \phi_2)=\pm \frac{1}{64\pi^2}\sum_k
n_km_k^4(\phi_1, \phi_2)\left[\log\frac{m_k^2(\phi_1, \phi_2)}{\mu^2}-c_k
\right],
\label{eq:cwpotential}
\end{align}
where $k$ indicates scalar and gauge bosons and fermions, and $n_k$, $m_k$, and $\mu$ are the degrees of freedom of each fields, the field-dependent masses of each fields, and the renormalization scale which we set $\mu=246$ GeV, respectively.
The upper (lower) sign corresponds to the bosonic (fermionic) contribution.
The corresponding degrees of freedom are $n_k=2, 1, 1, 1, 6, 3, 2, 12, 12$, and 4 for $k=H^\pm, H, h, A, W, Z, \gamma, t, b$, and $\tau$, respectively.
We only consider the fermions which have the non-negligible contributions.
The constant $c_k$ are equal to $1/2$ for transverse gauge bosons and $3/2$ for the other particles in the $\overline{{\rm MS}}$ scheme.

The $V_{\rm CW}$ changes the coordinate of the global minimum of the potential from that of $V_0$.
We introduce the counterterm potential $V_{\rm CT}$ for fixing the coordinate, the masses and the mixing angles of the scalar fields to be equal to the tree-level ones.
Thus, we impose the following five conditions to determine $V_{\rm CT}$:
\begin{subequations}
  \begin{align}
&\left.\frac{\partial V_{\rm CT}(\phi_1,\ \phi_2)}{\partial\phi_i}\right|_{(\phi_1,\phi_2)=(v_1,v_2)}
=-\left.\frac{\partial V_{\rm CW}(\phi_1,\ \phi_2)}{\partial\phi_i}\right|_{(\phi_1,\phi_2)=(v_1,v_2)},\\
&\left.\frac{\partial^2 V_{\rm CT}(\phi_1,\ \phi_2)}{\partial\phi_i\partial\phi_j}\right|_{(\phi_1,\phi_2)=(v_1,v_2)}
=-\left.\frac{\partial^2 V_{\rm CW}(\phi_1,\ \phi_2)}{\partial\phi_i\partial\phi_j}\right|_{(\phi_1,\phi_2)=(v_1,v_2)}\ \ (i, j=1, 2).
  \end{align}\label{eq:CTconditions}
\end{subequations}\\
Following Ref. \cite{Bernon:2017jgv},
we set $V_{\rm CT}$ with five parameters $\delta m_1^2$, $\delta m_2^2$, $\delta \lambda_1$, $\delta \lambda_2$, and $\delta \lambda_{345}$,
\begin{align}
V_{\rm CT}=\delta m_1^2\phi_1^2+\delta m_2^2\phi_2^2+\delta\lambda_1\phi_1^4+\delta\lambda_2\phi_2^4+\delta\lambda_{345}\phi_1^2\phi_2^2.
\label{eq:CT}
\end{align}
Hence, the conditions (\ref{eq:CTconditions}) give
\begin{equation}
  \begin{aligned}
&\delta m_1^2=-\frac{3}{4v_1}V_1+\frac{1}{4}V_{11}+\frac{1}{4}\frac{v_2}{v_1}V_{12},\\
&\delta m_2^2=-\frac{3}{4v_2}V_2+\frac{1}{4}V_{22}+\frac{1}{4}\frac{v_1}{v_2}V_{12},\\
&\delta \lambda_1=\frac{1}{8v_1^3}(V_1-v_1V_{11}),\\
&\delta \lambda_2=\frac{1}{8v_2^3}(V_2-v_2V_{22}),\\
&\delta \lambda_{345}=-\frac{V_{12}}{4v_1v_2},
  \end{aligned}
\end{equation}
where $V_i\equiv\left.\partial V_{\rm CW}/\partial \phi_i\right|_{(v_1,v_2)}$ and $\left.V_{ij}\equiv\partial^2 V_{\rm CW}/(\partial \phi_i\partial \phi_j)\right|_{(v_1,v_2)}$.
We calculate $\delta m_1^2$, $\delta m_2^2$, $\delta \lambda_1$, $\delta \lambda_2$, and $\delta \lambda_{345}$ numerically and substitute them for $V_{\rm CT}$.
However, there are infrared divergences in the second derivatives of $V_{\rm CW}$, which are proportional to $\log m_{\rm NG}^2$ where $m_{\rm NG}$ indicate the masses of Nambu-Goldstone (NG) bosons.
To avoid these divergences, we use the approximation which is shown in Ref.~\cite{Cline:2011mm}.
In this approximation, $m_{\rm NG}$ are approximated as the mass of the SM-like Higgs boson, i.e. $m_{\rm NG}\rightarrow m_h$.
This approximation is justified because the divergences are only logarithmic, hence the changes of the masses of the NG bosons do not make large differences.

The one-loop thermal contributions to the potential can be written as \cite{Dolan:1973qd}
\begin{align}
\overline{V}_1^\beta(\phi_1, \phi_2)=
\pm\frac{T^4}{2\pi^2}\sum_k
\int dx\ x^2\ln\left[1\mp \exp\left(-\sqrt{x^2+\frac{m_k^2(\phi_1, \phi_2)}{T^2}}\right)\right],
\label{eq:thermalpotential}
\end{align}
where $T$ represents the temperature and the upper (lower) sign indicates the bosonic (fermionic) contribution.
We calculate the integral in Eq.~(\ref{eq:thermalpotential}) numerically.
The squared-masses of the scalar bosons in Eq.~(\ref{eq:thermalpotential}) can become negative for certain sets of the coordinate $(\phi_1, \phi_2)$ and $T$~\footnote{
We comment on the region for the negative scalar squared-masses in Appendix~\ref{sec:ComplexPotential}.
The negative quadratic parameters sometimes yield the negative squared-masses at finite temperature, so we would discuss the region involved with Fig.~\ref{fig:m2AllPTImAI}.
}. 
In that case, we adopt a method that is discarding the imaginary part of the thermal potential, which is related to the instability of the field configuration \cite{PhysRevD.36.2474}, and taking only the real part (e.g.~Ref.~\cite{Basler:2016obg}).

\subsection{Resummation}\label{sec:resummation}
Although $V^\beta$ contains the corrections to the one-loop level, the contributions of higher loop diagrams get larger as the temperature rises.
The dominant diagrams at the high temperature are called daisy diagrams \cite{Dolan:1973qd}.
We perform resummation which is the method for taking into account the corrections from the diagrams \cite{Parwani:1991gq, Arnold:1992rz}.
Although there are two methods for the resummation, we apply the Parwani method \cite{Parwani:1991gq}~\footnote{
There is another method of resummation, called the Arnold-Epinosa (AE) method \cite{Arnold:1992rz}.
The method takes into account only bosonic Matsubara zero-modes which are involved in infrared divergences and adds cubic terms to the potential.
The procedure uses the high-temperature expansion when dividing the thermal contributions into those of the zero and non-zero modes.
Hence, it would be unsuccessful in regions where the high-temperature expansion is not valid.
In our calculation of the multi-step EWPT, we need to consider the PT near the EW vacuum in some cases (cf.~Fig.~\ref{fig:minimumofmulti-stepmAI}), where the high-temperature expansion is broken because the condition of the expansion, $m_B/T<1$, would not be satisfied. Therefore, the AE method is not suitable for the computation of the multi-step EWPT.
In contrast to the AE method, the Parwani method can take the non-relativistic limit smoothly even if theories include heavy particles \cite{Cline:2011mm, Laine:2017hdk} since the method does not contain the high-temperature expansion.
}.
The resummation is achieved by appending the corrections from the scalar and gauge boson polarization tensors in the infrared limit $\Pi_B(T)$ to the masses of bosons $m_B^2$
\begin{align}
m_B^2(\phi_1, \phi_2)\rightarrow m_B^2(\phi_1, \phi_2)+\Pi_B(T),
\end{align}
and inserting these corrected masses to $\overline{V}_1^\beta$ in Eq.~(\ref{eq:thermalpotential}) \cite{Carrington:1991hz}.
The index $B$ represents the species of bosons.

In the 2HDMs, we carry out the resummation concretely as the following.
The resummation for scalar fields are performed
by adding the contributions of the two-point functions to the mass parameters
$m_1$ and $m_2$ in the mass matrices Eqs.~(\ref{eq:bosonmassmatrices1}) and (\ref{eq:bosonmassmatrices2}) \cite{Blinov:2015vma}
\begin{align}
m_i^2\ \rightarrow m_i^2+c_iT^2,
\end{align}
where $c_i$ are the coefficients of correction terms and determined by
$\Pi_B(T)$ which depends on the types of Yukawa interactions.
In the Type-I 2HDM, they can be written by \cite{Bernon:2017jgv}
\newpage
\begin{subequations}
  \begin{align}
&c_1=\frac{1}{8}g^2+\frac{1}{16}(g^2+g'^2)+\frac{1}{4}\lambda_1+\frac{1}{6}\lambda_3+\frac{1}{12}\lambda_4,\\
&c_2=\frac{1}{8}g^2+\frac{1}{16}(g^2+g'^2)+\frac{1}{4}\lambda_2+\frac{1}{6}\lambda_3+\frac{1}{12}\lambda_4 +\frac{1}{4}y_t^2+\frac{1}{4}y_b^2+\frac{1}{12}y_\tau^2.
  \end{align}\label{eq:thermalcorrection}
\end{subequations}\\
For the other Yukawa types, one can obtain the coefficients by apportioning the Yukawa coupling terms in Eq.~(\ref{eq:thermalcorrection}) to $c_1$ and $c_2$ according to Tab.~\ref{table:Yukawa}.
And then, we append the correction terms to the non-diagonalized scalar matrices $\mathcal{M}_{w^\pm}^2$, $\mathcal{M}_z^2$, and $\mathcal{M}_h^2$ in Eqs.~(\ref{eq:bosonmassmatrices1}) and (\ref{eq:bosonmassmatrices2}) as
\begin{align}
\mathcal{M}_{w^\pm}^2+\left(
\begin{array}{cc}
c_1&0\\
0&c_2\\
\end{array}
\right)T^2
,\ \ \
\mathcal{M}_z^2+\left(
\begin{array}{cc}
c_1&0\\
0&c_2\\
\end{array}
\right)T^2
,\ \ \
\mathcal{M}_h^2+\left(
\begin{array}{cc}
c_1&0\\
0&c_2\\
\end{array}
\right)T^2,
\label{eq:thermaleffect}
\end{align}
and obtain the corrected scalar masses by diagonalizing them.

For the gauge fields, one can carry out the resummation by appending the contributions to only the longitudinal component of the mass matrices.
Following Ref. \cite{Blinov:2015vma}, the corrected masses of the longitudinal $W$ boson can be written by
\begin{equation}
M_{W_L}^2=\frac{g^2}{4}(\phi_1^2+\phi_2^2) +2g^2T^2.
\end{equation}
The corrected mass matrix of the longitudinally polarized $Z$ boson and photon in the gauge basis is
\begin{align}
\frac{1}{4}(\phi_1^2+\phi_2^2)
\left(
\begin{array}{cc}
g^2&-gg'\\
-gg'&g'^2\\
\end{array}
\right)
+
\left(
\begin{array}{cc}
2g^2T^2&0\\
0&2g'^2T^2\\
\end{array}
\right)\,.
\end{align}
By diagonalizing it,
the corrected masses of the $Z$ boson and the photon are obtained as
\begin{equation}
  \begin{aligned}
&M_{Z_L}^2=\frac{1}{8}(g^2+g'^2)(\phi_1^2+\phi_2^2+8T^2)
+\Delta,\\
&M_{\gamma_L}^2=\frac{1}{8}(g^2+g'^2)(\phi_1^2+\phi_2^2+8T^2)
-\Delta,
  \end{aligned}\label{eq:gaugethermaleffect}
\end{equation}
with
\begin{align}
\Delta=\sqrt{\left[\frac{1}{8}(g^2+g'^2)(\phi_1^2+\phi_2^2+8T^2)\right]^2-g^2g'^2T^2(\phi_1^2+\phi_2^2+4T^2)}.
\end{align}


\section{Theoretical constraints and EW-vacuum stability}\label{sec:parameterconstraints}

For the theoretical constraints on the model,
we consider constraints from the boundedness from below (BFB) of $V_0$, which is described as \cite{Deshpande:1977rw, Sher:1988mj, Nie:1998yn, Kanemura:1999xf}
\begin{align}
\lambda_1>0,~~ \lambda_2>0,~~ -\sqrt{\lambda_1\lambda_2}<\lambda_3,~~ -\sqrt{\lambda_1+\lambda_2}<\lambda_3+\lambda_4-\lambda_5,
\end{align}
the perturbativity,
\begin{align}
|\lambda_n|<4\pi\ (n=1,2,\cdots 5),
\end{align}
and the tree-level unitarity \cite{Kanemura:1993hm, Akeroyd:2000wc}.

Furthermore the absolute tree-level stability of the EW vacuum is required~\cite{Barroso:2013awa, Ivanov:2015nea}, where the negative $m_3^2$ is disfavored.
In following analyses, we confirm numerically that the EW vacuum is the global minimum in the region for $|\phi_i|\leq10$ TeV,
and remove the cases where $\sqrt{\phi_1^2+\phi_2^2}= 246$ GeV is not satisfied at the global minimum.

\section{Numerical results} \label{sec:numericalresults}

\begin{table}[t]
\centering
\begin{tabular}{|c||c|c|c|c|c|}
\hline
 & $m_A$ [GeV] & $m_H$ [GeV] & $\tan\beta$ & $\cos(\beta-\alpha)$ & $m_3$ [GeV$$] \\ \hline\hline
Type-I ($m_A=m_{H^\pm}$) & 180--1000(/10) & 130--1000(/10) & 2--10(/0.5) & $-$0.25--0.25(/0.05) & 0--100 \\ \hline
Type-I ($m_H=m_{H^\pm}$) & 130--1000(/10) & 180--1000(/10) & 2--10(/0.5) & $-$0.25--0.25(/0.05) & 0--100 \\ \hline
Type-X ($m_A=m_{H^\pm}$) & 180--1000(/10) & 130--1000(/10) & 2--10(/0.5) & 0 & 0--100 \\ \hline
Type-X ($m_H=m_{H^\pm}$) & 130--1000(/10) & 180--1000(/10) & 2--10(/0.5) & 0 & 0--100 \\ \hline
\end{tabular}
\caption{Parameter regions studied in the Type-I and Type-X 2HDMs with $m_A$ or $m_H=m_{H^\pm}$.
We perform analyses in every 10 GeV in the masses of the neutral scalar bosons, $0.5$ in $\tan\beta$, $0.05$ in $\cos(\beta-\alpha)$ (though we set $\cos(\beta-\alpha)=0$ in the Type-X 2HDMs), and 5 GeV in $m_3$.}
\label{tb:allparameterregion}

\end{table}


In this section, we discuss the parameter space where the multi-step PT occurs.
To study the PT, CosmoTransitions \cite{Wainwright:2011kj} is used in the analyses.
We also study the region
where the strongly first order PT occurs in the multi-step PT.
The strength of the PT $\xi$ is defined by
\begin{align}
\xi\equiv\frac{v_{c}}{T_{c}},
\label{eq:xi1}
\end{align}
where $T_c$ is the critical temperature, at which minima degenerate between two phases, and
$v_c$ is the critical value of $\sqrt{\phi_1^2+\phi_2^2}$ at $T_c$.
As the criterion for a strong PT, we consider $\xi \geq 1$, where the sphaleron processes are suppressed enough in the $SU(2)$ broken phase.
We especially focus on the cases that
the first step PTs of the 2-step PTs are strongly first order~\footnote{We take into account the first order, the second order, and the cross-over PTs as the first step PTs.}, which we name ``the strong 2-step PTs"~\footnote{
The strong 3-step PTs (i.e. the 3-step PTs where the first step PT is strongly first order) are found in the Type-I 2HDMs, where they occur in the region that the strong 2-step PTs happen in.
However the number of points for such PTs is much smaller than that for the strong 2-step PTs, hence we do not discuss the results.
Note that we cannot find such PTs in the Type-X 2HDMs.}.
In this case, the sphaleron rate is suppressed in the broken phase, and it is expected that $v_{c}$ and $T_{c}$ at the subsequent step PTs are respectively larger and smaller than the previous PT.
Therefore, the inequality $\xi \geq 1$ is kept and the sphaleron rate is also suppressed at the subsequent step PTs~\footnote{If baryon number is generated at the first step PT (although our model cannot generate baryon number since the CP is conserved), it remains unwashed-out.}.

In the following analysis,
instead of the eight parameters
($m_1^2,\ m_2^2,\ m_3^2,\ \lambda_{1-5}$)
in $V_0$,
we take the following set as the input parameters:
\begin{align}
m_{H^\pm},\ m_A,\ m_H,\ \tan\beta,\ \cos(\beta-\alpha),\ m_3,\ m_h,\ v.
\end{align}
Here $m_h=125$ GeV and $v=246$ GeV.
In order to put restrictions on the range of the other input parameters,
we consider the experimental constraints from the EW precision data, the $B\rightarrow X_s\gamma$ decays,
 the $H^\pm\rightarrow \tau\nu$ decays,
and the coupling measurements of the Higgs boson.
The EW precision data can be satisfied by assuming the mass degeneracy between
the charged scalar boson and at least one of the extra neutral scalar bosons, $m_{H^\pm} \simeq m_A\ {\rm or}\ m_H$,
which makes the custodial symmetry recovered and hence the $\rho$ parameter $\rho\simeq1$ like in the SM~\cite{Haber:2010bw}.
For $m_{H^\pm}$,
the range $m_{H^\pm}< 590$ GeV is excluded from $B\rightarrow X_s\gamma$ decays in the Type-II and -Y 2HDMs \cite{Haller:2018nnx}, while $m_{H^\pm}\leq 170$ GeV is excluded from $H^\pm\rightarrow \tau\nu$ decays
in the Type-X 2HDM \cite{Arhrib:2018ewj}.
The constraints from the coupling measurements of the Higgs boson~\cite{Aad:2019mbh} show that {\it e.g.},
$|\cos(\beta-\alpha)|> 0.25$ (0.3) is excluded at $\tan\beta=2$ (10) in the Type-I 2HDM, and
$|\cos(\beta-\alpha)|> 0.15$ (0.05) is excluded at $\tan\beta=2$ (10) in the Type-X 2HDM.
In the Type-II and -Y 2HDMs, the constraints are stricter than those in the Type-I and -X 2HDMs.

In our analyses, based on the above constraints, we take the range for the input parameters in the Type-I and -X 2HDMs
as shown in Tab.~\ref{tb:allparameterregion}.
Imposing the mass degeneracy $m_\Phi=m_{H^\pm}$ $(\Phi=H$ or $A)$,
the ranges for $m_A$ and $m_H$ are taken as 180 GeV--1 TeV for $m_\Phi$ and 130 GeV--1 TeV for the other extra neutral scalar boson.
For the mixing angles, we take $\tan\beta =2-10$, and $|\cos(\beta-\alpha)|\leq 0.25$ in the Type-I 2HDM, while the \textit{alignment limit},
$\cos(\beta-\alpha)=0$, in the Type-X 2HDM.
In the analyses we focus on the relatively small value for $m_3$ as $0 \leq m_3 \leq 100$ GeV, since the strong 2-step PTs
which we are interested in prefer to occur for the smaller $m_3$ and
do not occur for $m_3\simeq100$ GeV (cf.~Fig.~\ref{fig:m3xi1AllPTImAI}).
As depicted in Tab.~\ref{tb:allparameterregion},
we take every 10 GeV in $m_A$ and $m_H$, $0.5$ in $\tan\beta$, $0.05$ in $\cos(\beta-\alpha)$ (though we set $\cos(\beta-\alpha)=0$ in the Type-X 2HDMs), and 5 GeV in $m_3$.

In the Type-II and -Y 2HDMs,
we take the same ranges for the input parameters with those in the Type-X 2HDM, but $m_\Phi=m_{H^\pm}\geq590$ GeV by the constraint from $B\rightarrow X_s\gamma$.
In these cases, we have found that
the stability of the EW vacuum is not realized
because the contributions of the heavy extra scalar fields lift up the potential significantly at the EW vacuum (see Eq.~(\ref{eq:cwpotential}))
and the origin $(\phi_1, \phi_2)=(0,0)$ becomes the global minimum.
Hence we discuss only Type-I and -X 2HDMs hereafter.

\subsection{Type-I}
\subsubsection{Type-I $(m_A=m_{H^\pm})$}
\label{sec:AnalysismAI}

\begin{figure}[t]
 \centering
 \begin{tabular}{c}

 \begin{minipage}{0.5\hsize}
 \centering
 \includegraphics[clip, width=7.cm]{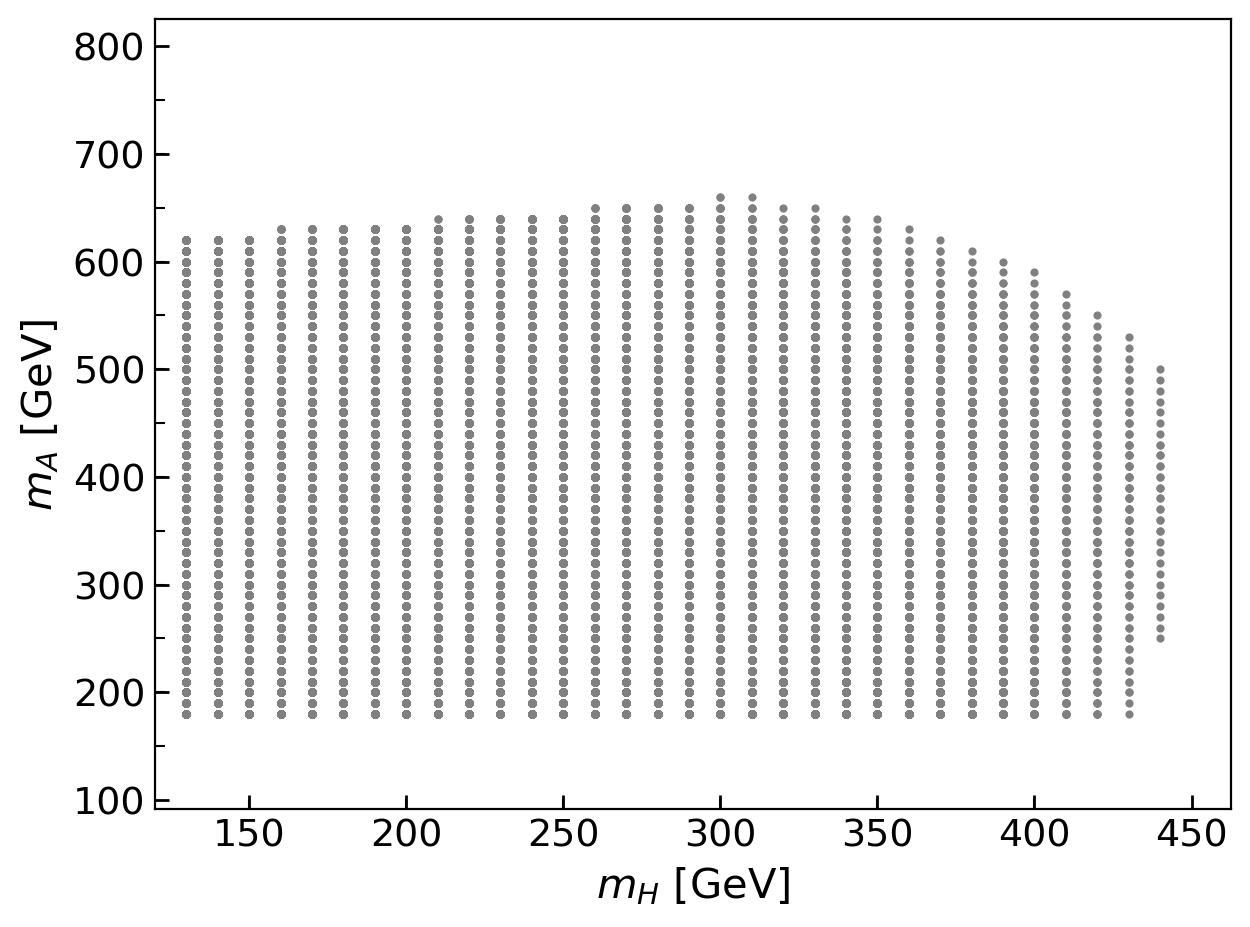}
 \end{minipage}

 \begin{minipage}{0.5\hsize}
 \centering
 \includegraphics[clip, width=7.cm]{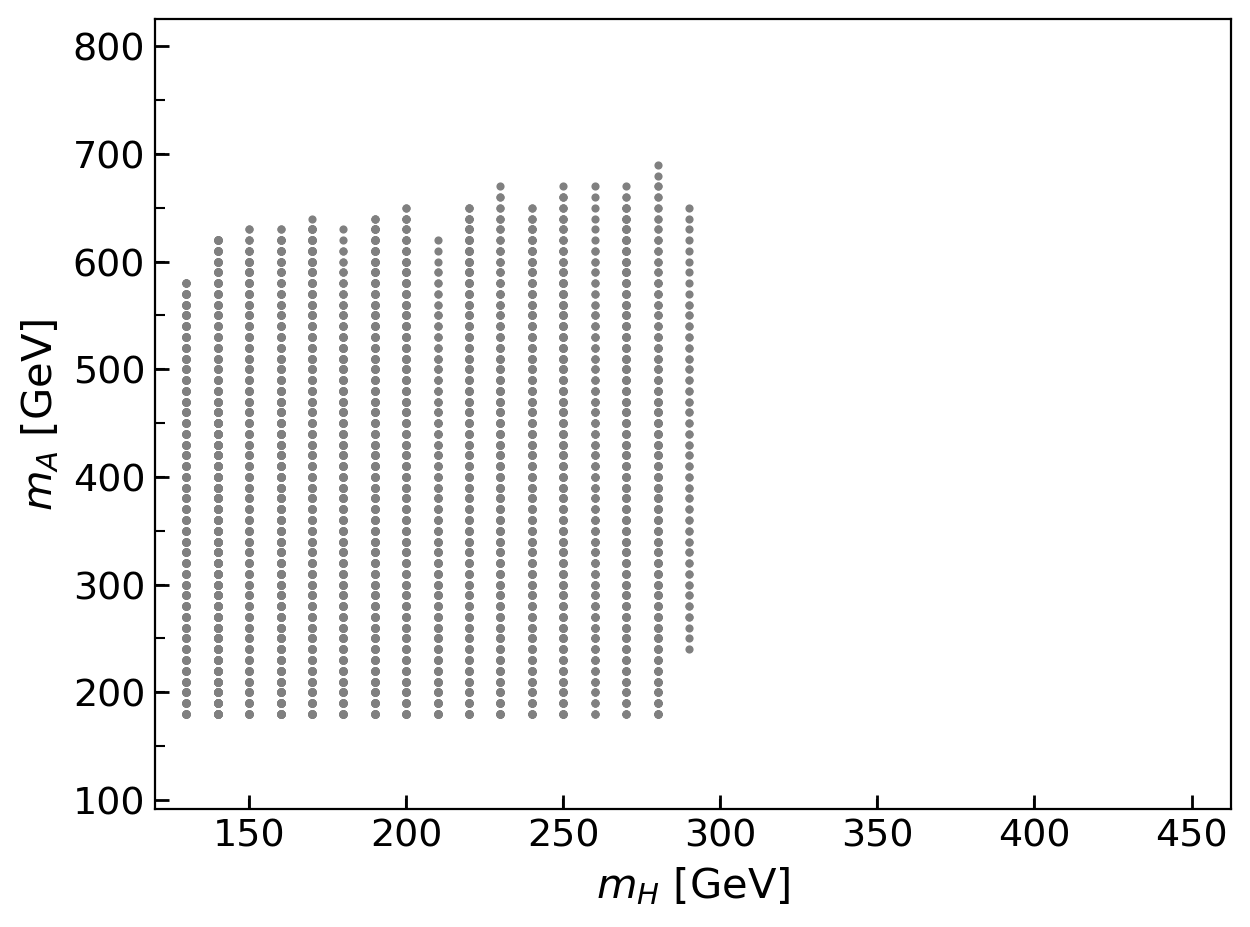}
 \end{minipage}

 \end{tabular}
 \caption{Parameter regions in the $m_A$ vs.~$m_H$ plane allowed by the theoretical constraints (the BFB, the perturbativity, and the tree-level unitarity) in the Type-I 2HDM with $m_A=m_{H^\pm}$. The left and right panel show the regions in the cases of $\tan\beta=2$ and 7, respectively. The other input parameters follow Tab.~\ref{tb:allparameterregion}.}
 \label{fig:theoricalconstraints}
\end{figure}

\begin{figure}[t]
\centering
\begin{tabular}{c}

\begin{minipage}{0.5\hsize}
\centering
\includegraphics[clip, width=7.5cm]{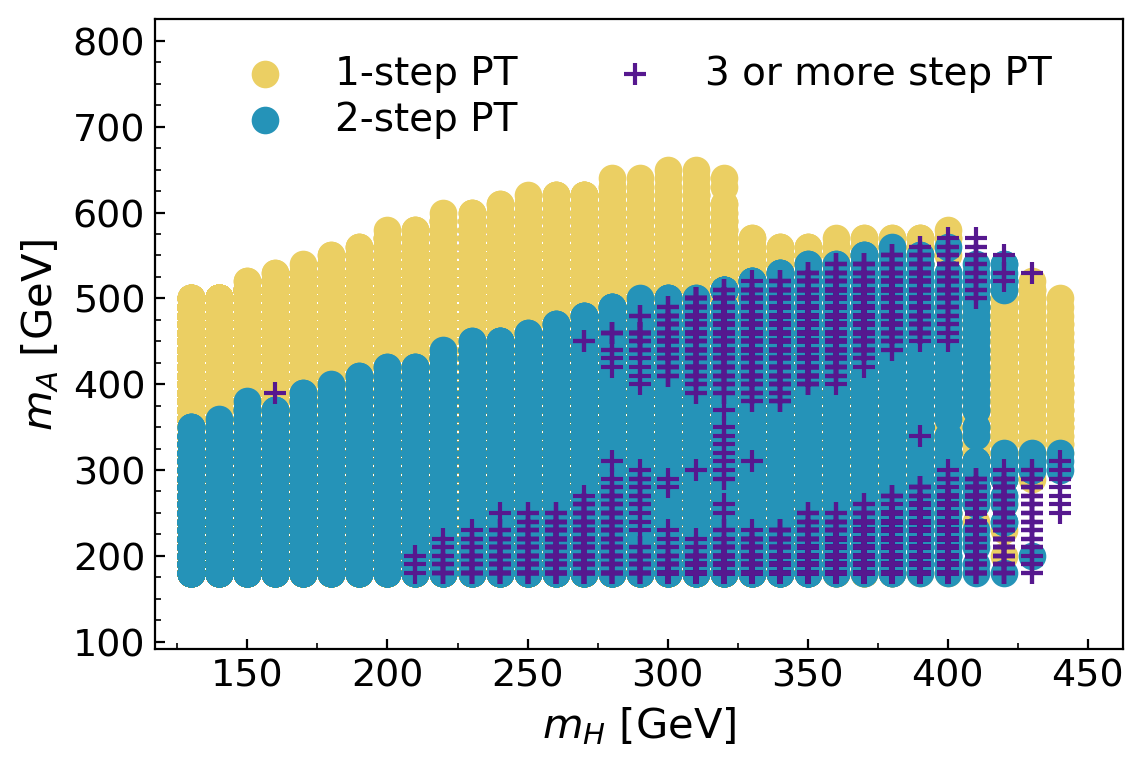}
\end{minipage}

\begin{minipage}{0.5\hsize}
\centering
\includegraphics[clip, width=7.5cm]{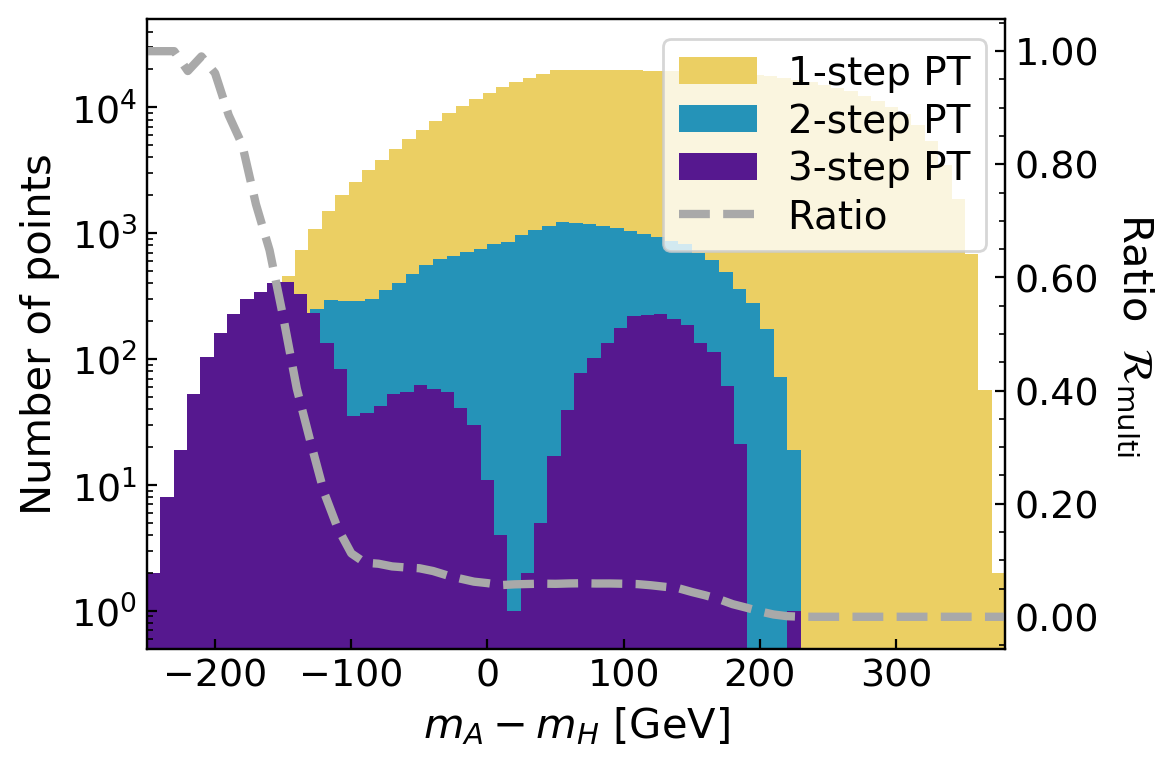}
\end{minipage}\\

\begin{minipage}{0.5\hsize}
\centering
\includegraphics[clip, width=7.5cm]{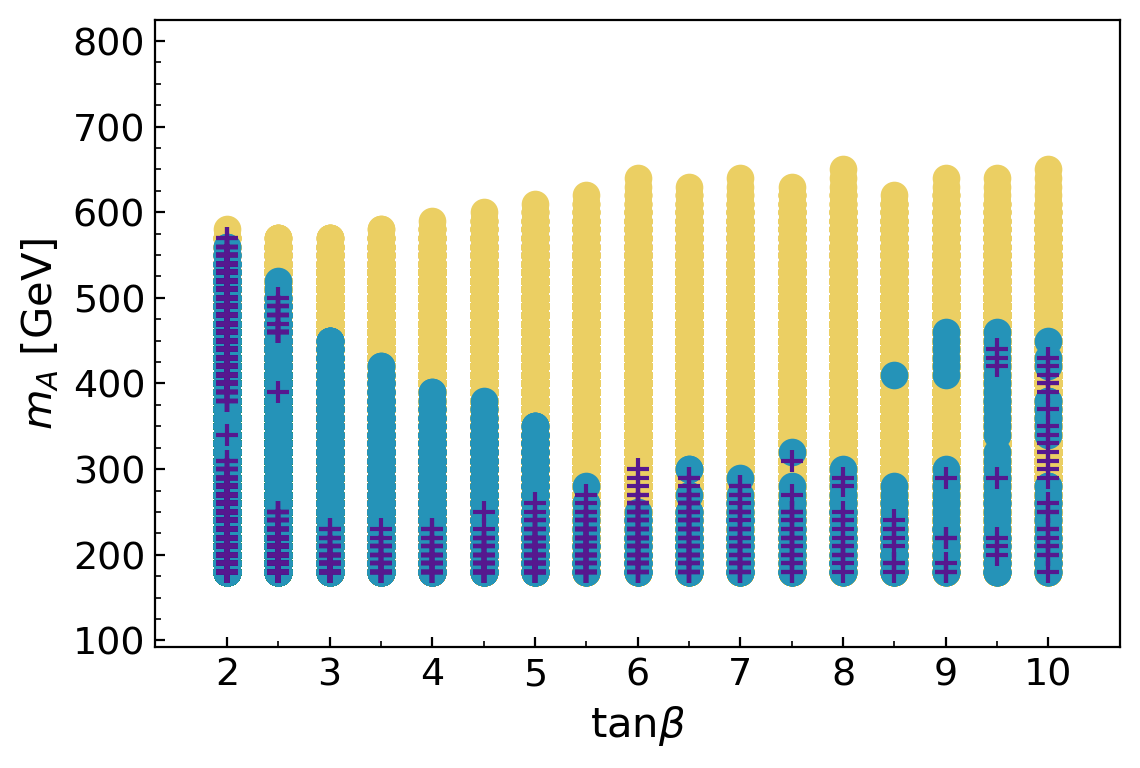}
\end{minipage}

\begin{minipage}{0.5\hsize}
\centering
\includegraphics[clip, width=7.5cm]{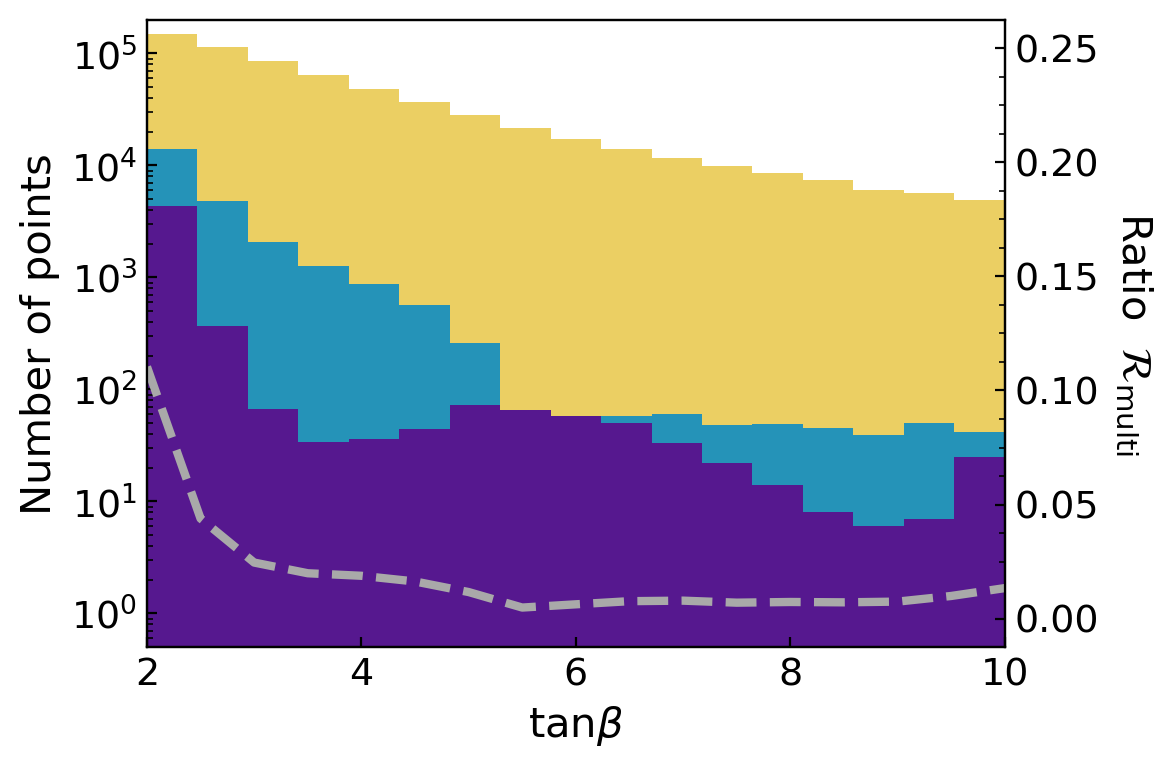}
\end{minipage}\\

\begin{minipage}{0.5\hsize}
\centering
\includegraphics[clip, width=7.5cm]{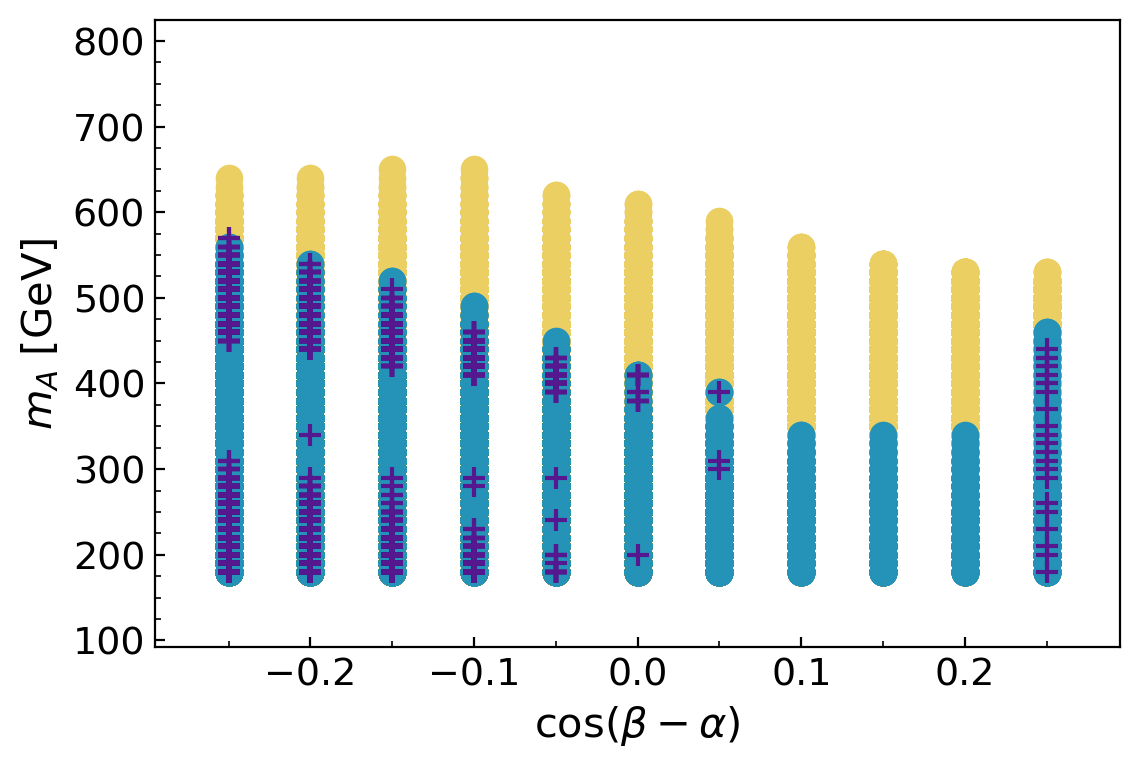}
\end{minipage}

\begin{minipage}{0.5\hsize}
\centering
\includegraphics[clip, width=7.5cm]{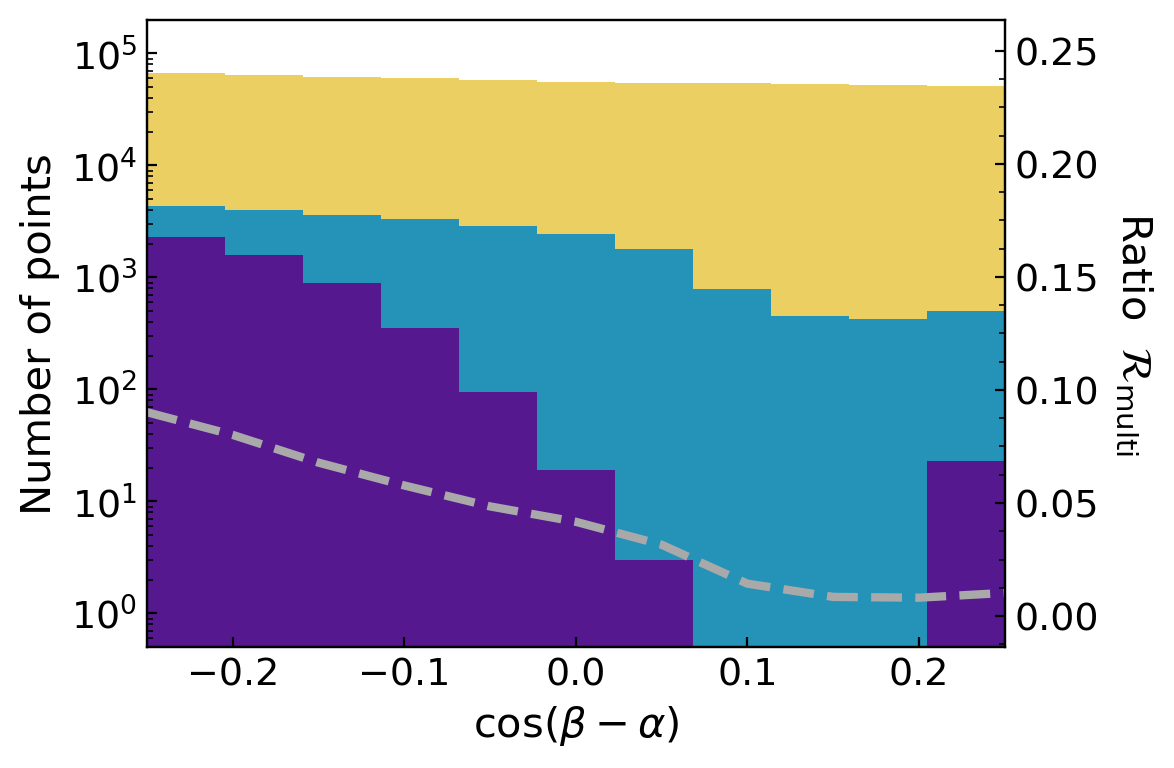}
\end{minipage}\\

\end{tabular}
\caption{
Left: Parameter points where the 1-step and multi-step PTs occur in the $m_A\ {\rm vs.}~ m_H$ (top),
$m_A\ {\rm vs.}~ \tan\beta$ (middle), and $m_A\ {\rm vs.}~ \cos(\beta-\alpha)$ (bottom) planes in the Type-I 2HDM with $m_A=m_{H^\pm}$.
The yellow, blue, and purple points show the results for the 1-step, 2-step, and 3 or more step PTs, respectively.
Right: Number of points where the 1-step and multi-step PTs occur as a function of $m_A-m_H$ (top), $\tan\beta$ (middle) and $\cos(\beta-\alpha)$ (bottom).
The 1-step, 2-step, and 3 or more step PTs are colored by yellow, blue, and purple, respectively.
The grey dashed lines in the panels represent ${\cal R}_{\rm multi}$, which are the ratios of the number of points for the multi-step PTs to that for all PTs.
}
\label{fig:AllPTImAI}
\end{figure}


In this subsection, we show the results in the Type-I 2HDM with $m_A=m_{H^\pm}$.
Fig.~\ref{fig:theoricalconstraints} represents the allowed parameter region by the theoretical constraints (the BFB, the perturbativity, and the tree-level unitarity) in the $m_A$ vs.~$m_H$ plane at $\tan\beta=2$ (left) and 7 (right).
It shows in the case of $\tan\beta=7$ the upper limit on $m_H$ is lower than that in the case of $\tan\beta=2$, as $m_H \lesssim 290$ (440) GeV for $\tan\beta=7$ (2).

The left panels of Fig.~\ref{fig:AllPTImAI} exhibit the parameter points where the 1-step and multi-step PTs (left) occur in the $m_A\ {\rm vs.}~ m_H$ (top), $m_A\ {\rm vs.}~ \tan\beta$ (middle), and $m_A\ {\rm vs.}~ \cos(\beta-\alpha)$ (bottom) planes.
The yellow, blue, and purple points show the results for the 1-step, 2-step, and 3 or more step PTs, respectively
\footnote{
The ``3 or more step PTs" includes the 4-step PTs.
The numbers of points for 3-step and 4-step PTs account for about $1\%$ and $0.1\%$ of the number of all points in this case, respectively.
}.
Here in addition to the theoretical constraints considered in Fig.~\ref{fig:theoricalconstraints},
the constraint for the stability of the EW vacuum is further imposed.
Compared the top left panel in Fig.~\ref{fig:AllPTImAI} with
Fig.~\ref{fig:theoricalconstraints}, we can see that the region with the larger $m_A$ (and $m_H$)
is excluded by the constraint.
In the left panels of Fig.~\ref{fig:AllPTImAI},
the range of $m_A$ where the multi-step PTs occur gets larger,
as $m_H$ increases, and $\tan\beta$ and $\cos(\beta-\alpha)$
(except for $\tan\beta \simeq 10$ and $\cos(\beta-\alpha) \simeq 0.25)$
respectively decreases.
In the top left panel of Fig.~\ref{fig:AllPTImAI},
the parameter region of the multi-step PTs overlaps with that of the 1-step PTs,
but for the region where $m_H \simeq 420$ GeV with $m_A \simeq 550$ GeV or 200--300 GeV, the 3 or more step PTs occur mostly.
The right panels in Fig.~\ref{fig:AllPTImAI} represent the number of points for the 1-step (yellow), 2-step (blue), 3 or more step PTs (purple), respectively, as a function of $m_A-m_H$ (top), $\tan\beta$ (middle) and $\cos(\beta-\alpha)$ (bottom).
The ratios ${\cal R}_{\rm multi}$ in the panels, plotted as grey dashed lines, are the ratios of the number of points for the multi-step PTs to that for all PTs,
\begin{align}
{\cal R}_{\rm multi}=\frac{\# {\rm ~of ~points ~for~ the~ multi\mathchar`-step~ PTs}}{\# {\rm ~of ~points ~for~ all~ PTs}}.
\label{eq:Rmulti}
\end{align}
We see that,
in the top right panel of Fig.~\ref{fig:AllPTImAI},
the multi-step PTs favor $m_A-m_H <0$, and
the ratio ${\cal R}_{\rm multi}\simeq 1$ is obtained at $m_A-m_H\simeq -210$ GeV.
Hence, when ${\cal R}_{\rm multi}$ becomes around 1, $m_H\gtrsim390$ GeV and $m_A$ is around 200 GeV.
Moreover, the middle and bottom right panels of Fig.~\ref{fig:AllPTImAI} show that
${\cal R}_{\rm multi}$ becomes larger for the smaller $\tan\beta$ and $\cos(\beta-\alpha)$,
and reaches about 10\% at $\tan\beta=2$ and $\cos(\beta-\alpha)=-0.25$, respectively.
For $\tan\beta \simeq 10$ and $\cos(\beta-\alpha)\simeq 0.25$, we can see that ${\cal R}_{\rm multi}$ are only a few \%, respectively, although the allowed ranges of $m_A$ for the multi-step PTs are wide in the middle and bottom left panels in
Fig.~\ref{fig:AllPTImAI}.

As expected from Fig.~\ref{fig:theoricalconstraints}, the regions with $m_H \simeq 420$ GeV are realized for $\tan\beta \simeq 2$ in this analysis.
In such a low $\tan\beta$ case, the $B\rightarrow\mu^+\mu^-$ process gives the constraint on $m_{H^\pm}$ as {\it e.g.},
$m_{H^\pm}> 340$ (125) GeV at $\tan\beta\simeq2$ (3) at 95$\%$CL in the Type-I and -X 2HDMs \cite{Haller:2018nnx}.
Therefore the region where the 3 or more step PTs occur mostly
with $m_H\simeq 420$ GeV and $m_A(=m_{H^\pm})\simeq 200$--300 GeV is excluded by the constraint from $B\rightarrow\mu^+\mu^-$.
We have found that even when the constraint is taken into account in this analysis,
the multi-step PTs favor the mass hierarchy $m_A < m_H$ and {\it e.g.},
${\cal R}_{\rm multi}\simeq$ 100\% (10\%) is obtained at $m_A-m_H \simeq -150$ ($-$80) GeV.
Moreover, we should note that if $m_A-m_H < -m_Z$,
the region would be constrained by the extra Higgs boson search $H\to AZ$
at the LHC \cite{Khachatryan:2016are,Aaboud:2018eoy,Sirunyan:2019wrn}.
It is generally more severe for the low $\tan\beta$ and the $\cos(\beta -\alpha)$ closer to zero \cite{Kling:2020hmi,Benbrik:2020nys}.
We leave the detailed analyses including the constraints from such extra Higgs boson searches for future work.


\begin{figure}[t]
\centering
\begin{tabular}{c}

\begin{minipage}{0.5\hsize}
\centering
\includegraphics[clip, width=7.5cm]{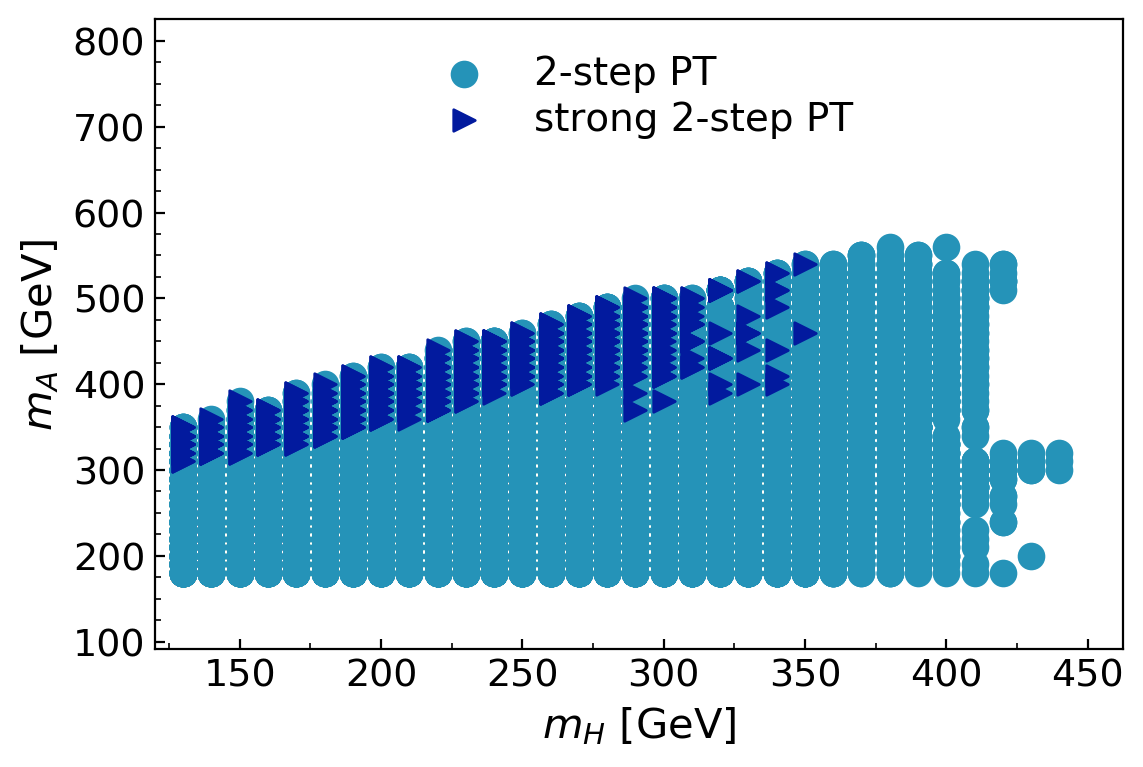}
\end{minipage}

\begin{minipage}{0.5\hsize}
\centering
\includegraphics[clip, width=7.5cm]{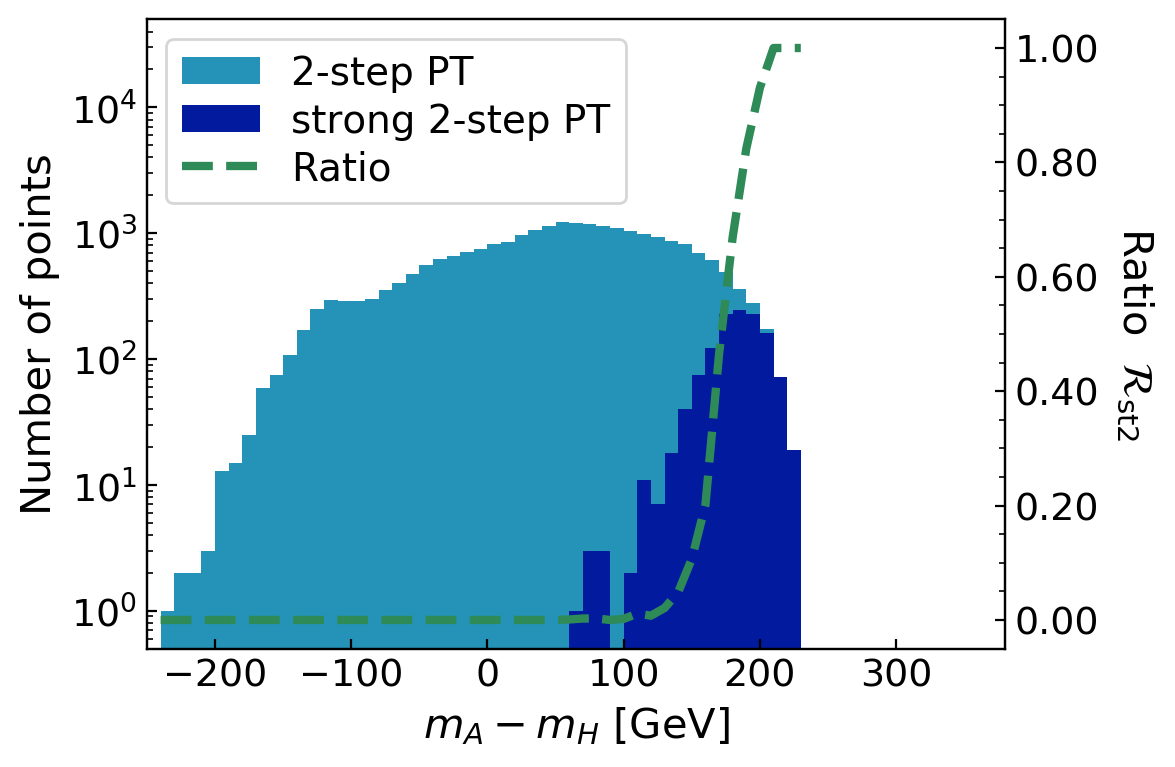}
\end{minipage}\\

\begin{minipage}{0.5\hsize}
\centering
\includegraphics[clip, width=7.5cm]{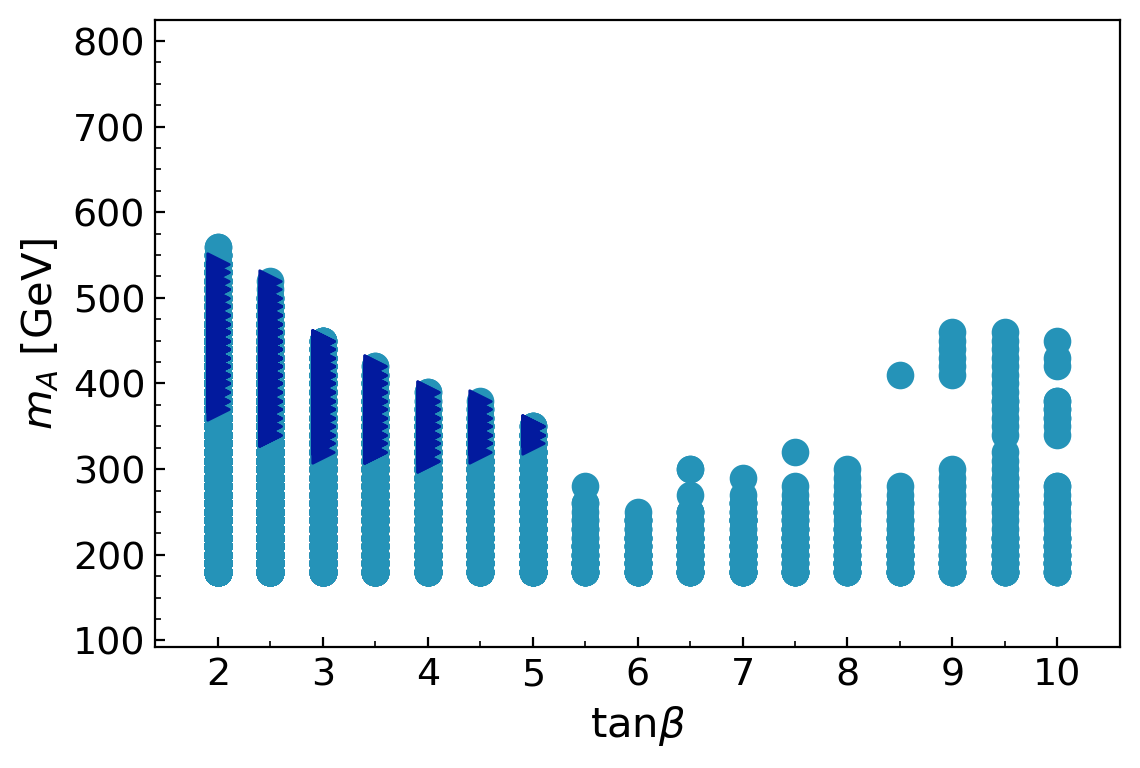}
\end{minipage}

\begin{minipage}{0.5\hsize}
\centering
\includegraphics[clip, width=7.5cm]{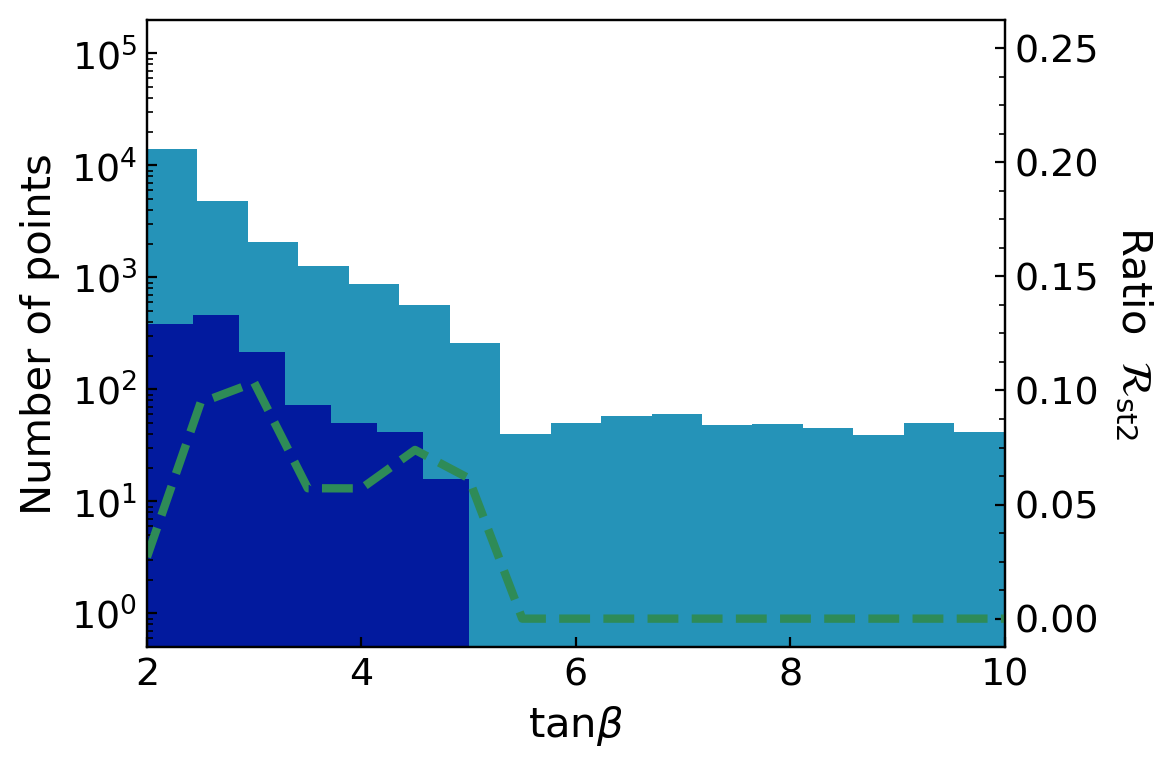}
\end{minipage}\\

\begin{minipage}{0.5\hsize}
\centering
\includegraphics[clip, width=7.5cm]{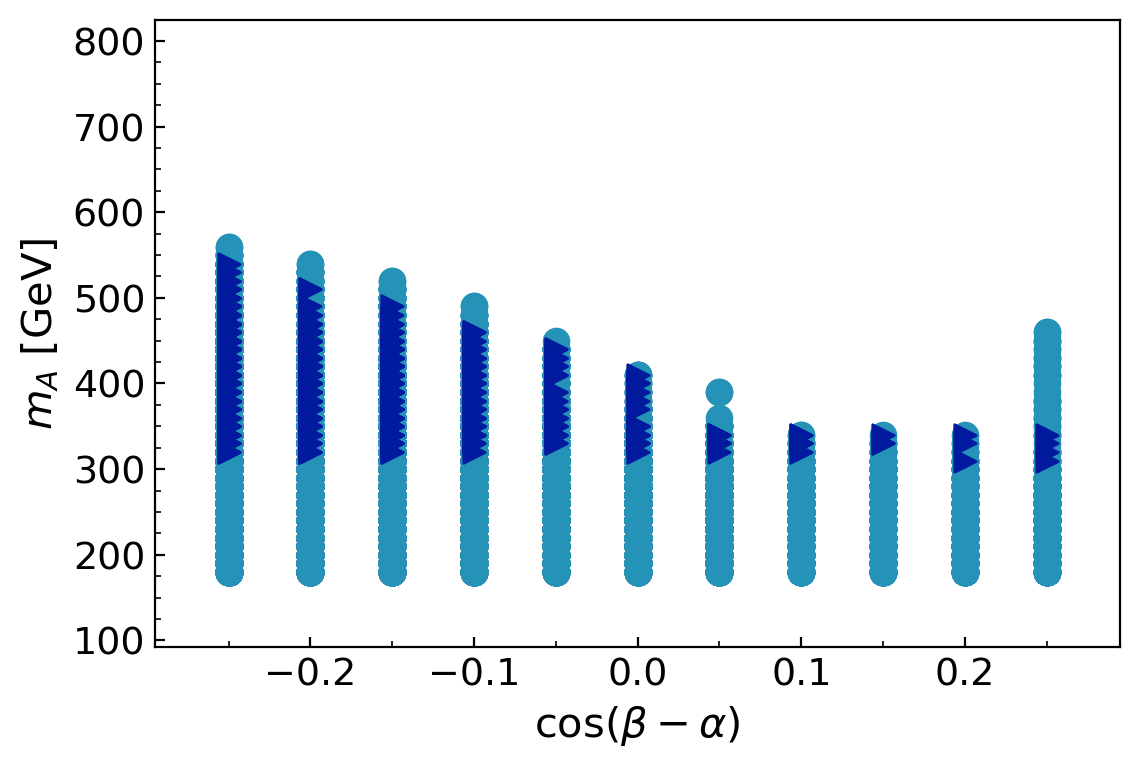}
\end{minipage}

\begin{minipage}{0.5\hsize}
\centering
\includegraphics[clip, width=7.5cm]{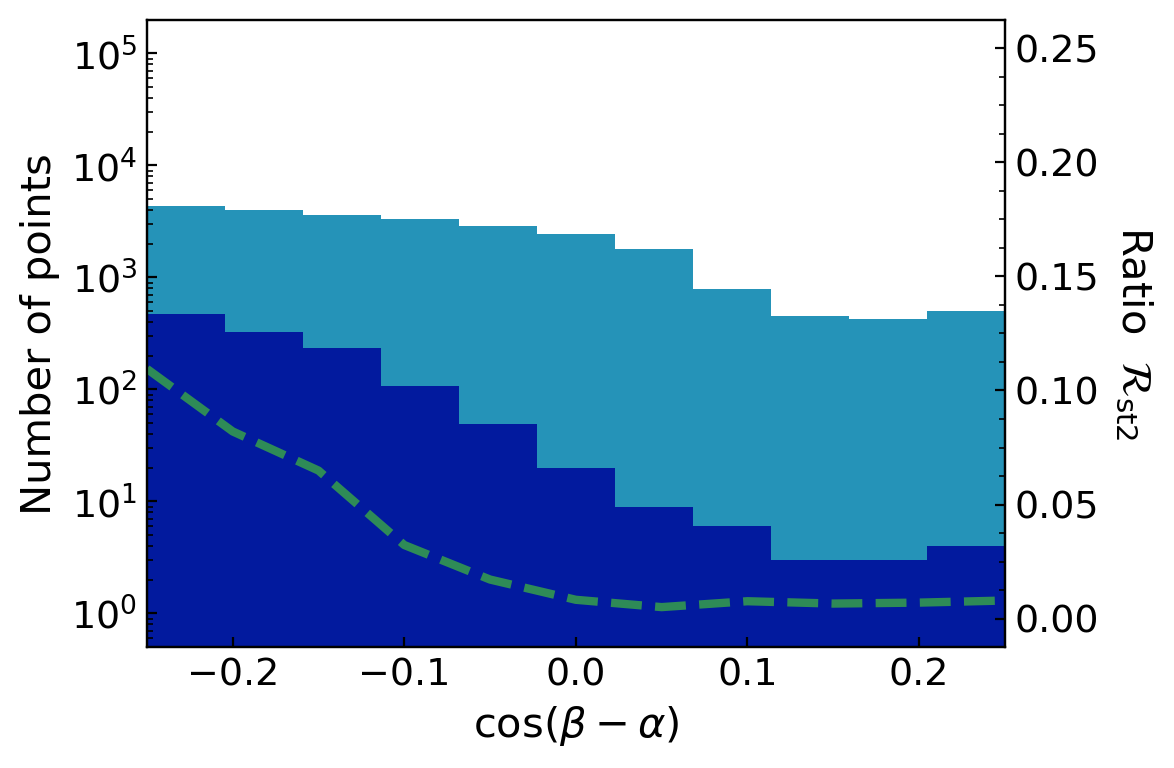}
\end{minipage}

\end{tabular}
\caption{
Left: Parameter points where the 2-step and strong 2-step PTs occur in the $m_A\ {\rm vs.}~ m_H$ (top),
$m_A\ {\rm vs.}~ \tan\beta$ (middle), and $m_A\ {\rm vs.}~ \cos(\beta-\alpha)$ (bottom) planes in the Type-I 2HDM with $m_A=m_{H^\pm}$.
The blue and dark-blue points present the parameter points where the 2-step and strong 2-step PTs occur, respectively.
Right: Number of points where the 2-step and strong 2-step PTs occur as a function of $m_A-m_H$ (top), $\tan\beta$ (middle) and $\cos(\beta-\alpha)$ (bottom).
The 2-step and strong 2-step PTs are colored by blue and dark-blue, respectively.
The green dashed lines in the panels represent ${\cal R}_{\rm st2}$, which are the ratios of the number of points for the strong 2-step PTs to that for 2-step PTs.
}
\label{fig:strongPTImAI}
\end{figure}


The left panels of Fig.~\ref{fig:strongPTImAI} show the parameter points where the 2-step and strong 2-step PTs occur in the $m_A\ {\rm vs.}~ m_H$ (top), $m_A\ {\rm vs.}~ \tan\beta$ (middle), and $m_A\ {\rm vs.}~ \cos(\beta-\alpha)$ (bottom) planes as well as Fig.~\ref{fig:AllPTImAI}.
The blue and dark-blue points present the parameter points where the 2-step and strong 2-step PTs occur, respectively.
From the top left panel of Fig.~\ref{fig:strongPTImAI}, we can see that the strong 2-step PTs happen in the region where $m_A\gtrsim300$ GeV and $m_H\lesssim350$ GeV, and the mass hierarchy $m_A>m_H$ exists.
Additionally, in the middle and bottom left panels of Fig.~\ref{fig:strongPTImAI}, the range of $m_A$ where the strong 2-step PTs happen increases
as $\tan\beta$ and $\cos(\beta-\alpha)$ becomes smaller, respectively. Here the strong 2-step PTs occur only in $\tan\beta\lesssim5$.
The right panels in Fig.~\ref{fig:strongPTImAI} exhibit the number of points for the 2-step (blue) and strong 2-step PTs (dark-blue), respectively, as a function of $m_A-m_H$ (top), $\tan\beta$ (middle) and $\cos(\beta-\alpha)$ (bottom).
The ratios ${\cal R}_{\rm st2}$ in the right panels are the ratios of the number of points for the strong 2-step PTs to that for the 2-step PTs,
\begin{align}
{\cal R}_{\rm st2}=\frac{\# {\rm ~of ~points ~for~ the ~strong~2\mathchar`-step~ PTs}}{\# {\rm ~of ~points ~for~ the~2\mathchar`-step~ PTs}},
\label{eq:Rst2}
\end{align}
which are shown by green dashed lines.
It is notable that in the top right panel of Fig.~\ref{fig:strongPTImAI}
the strong 2-step PTs favor $m_A-m_H >0$ and
${\cal R}_{\rm st2}$ becomes ${\cal R}_{\rm st2}=100$\% at $m_A-m_H \simeq 210$ GeV.
Moreover, the strong 2-step PTs are likely to occur at the small $\tan\beta$ and $\cos(\beta-\alpha)$, respectively,
as shown in the middle and bottom right panels of Fig.~\ref{fig:strongPTImAI}.
The parameter regions of the strong 2-step PTs do not receive the constraint from $B\rightarrow\mu^+\mu^-$, while
some of them with $m_A-m_H > m_Z$ could be constrained from the $A\to HZ$ decay at the LHC \cite{Kling:2020hmi,Benbrik:2020nys}.


\begin{figure}[t]
\centering
\begin{tabular}{c}

\begin{minipage}{0.5\hsize}
\centering
\includegraphics[clip, width=7.5cm]{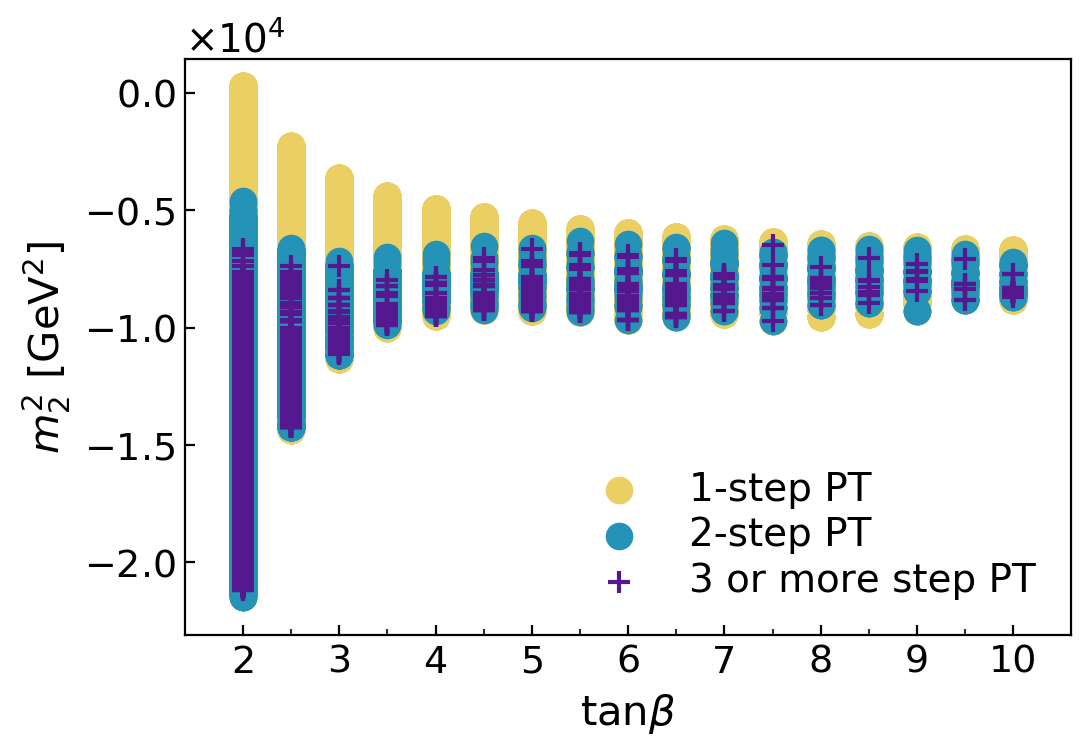}
\end{minipage}

\begin{minipage}{0.5\hsize}
\centering
\includegraphics[clip, width=7.5cm]{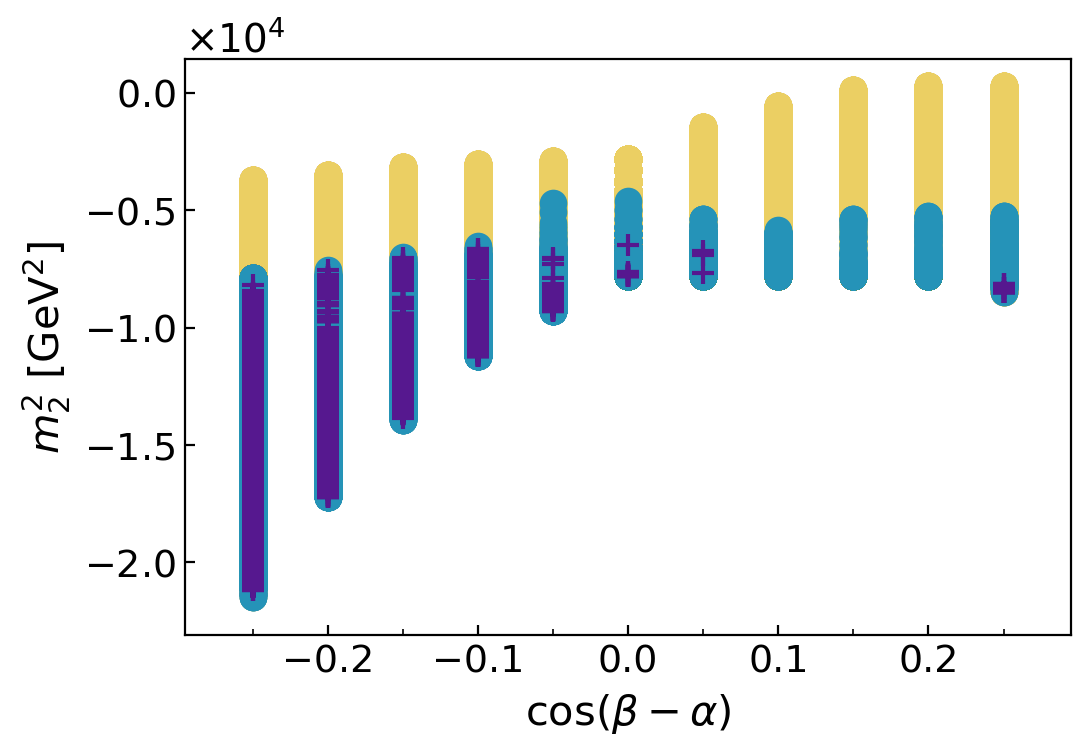}
\end{minipage}

\end{tabular}
\caption{
Parameter points where the 1-step, 2-step, and 3 or more step PTs occur in the $m_2^2$ vs.~$\tan\beta$ (left) and $m_2^2$ vs.~$\cos(\beta-\alpha)$ (right) planes in the Type-I 2HDM with $m_A=m_{H^\pm}$.
}
\label{fig:m2AllPTImAI}
\end{figure}

\begin{figure}[t]
\centering
\begin{tabular}{c}

\begin{minipage}{0.5\hsize}
\centering
\includegraphics[clip, width=7.5cm]{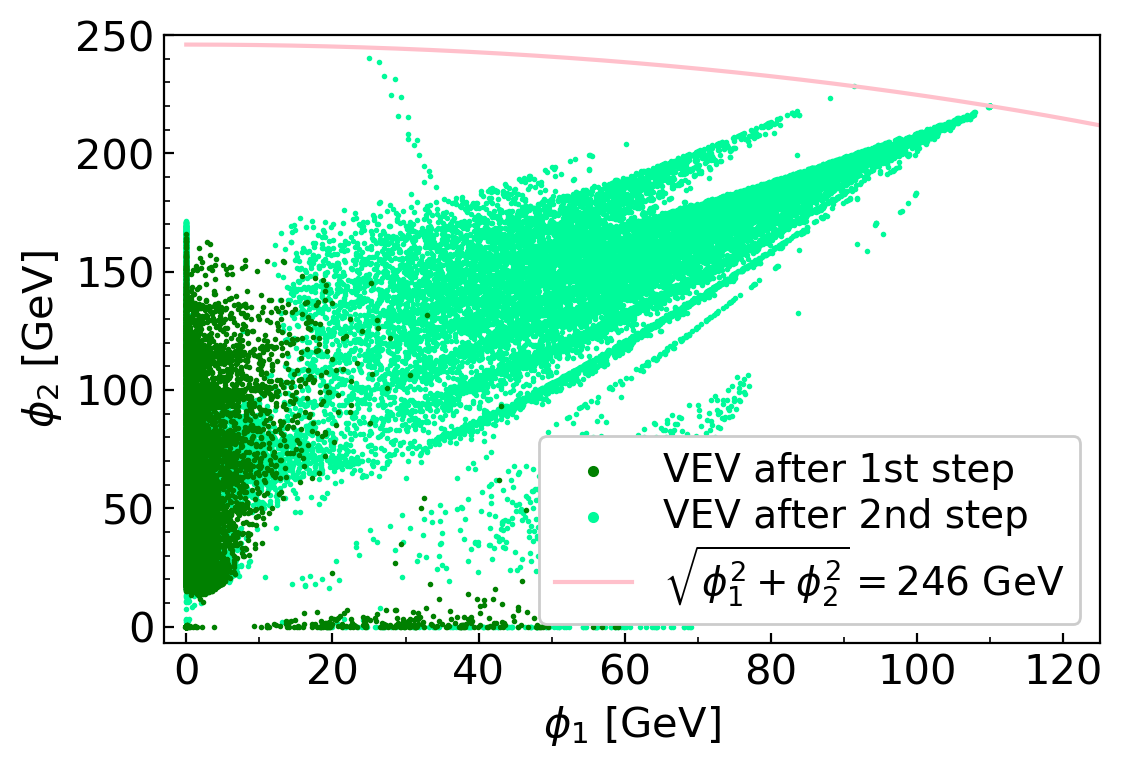}
\end{minipage}

\begin{minipage}{0.5\hsize}
\centering
\includegraphics[clip, width=7.5cm]{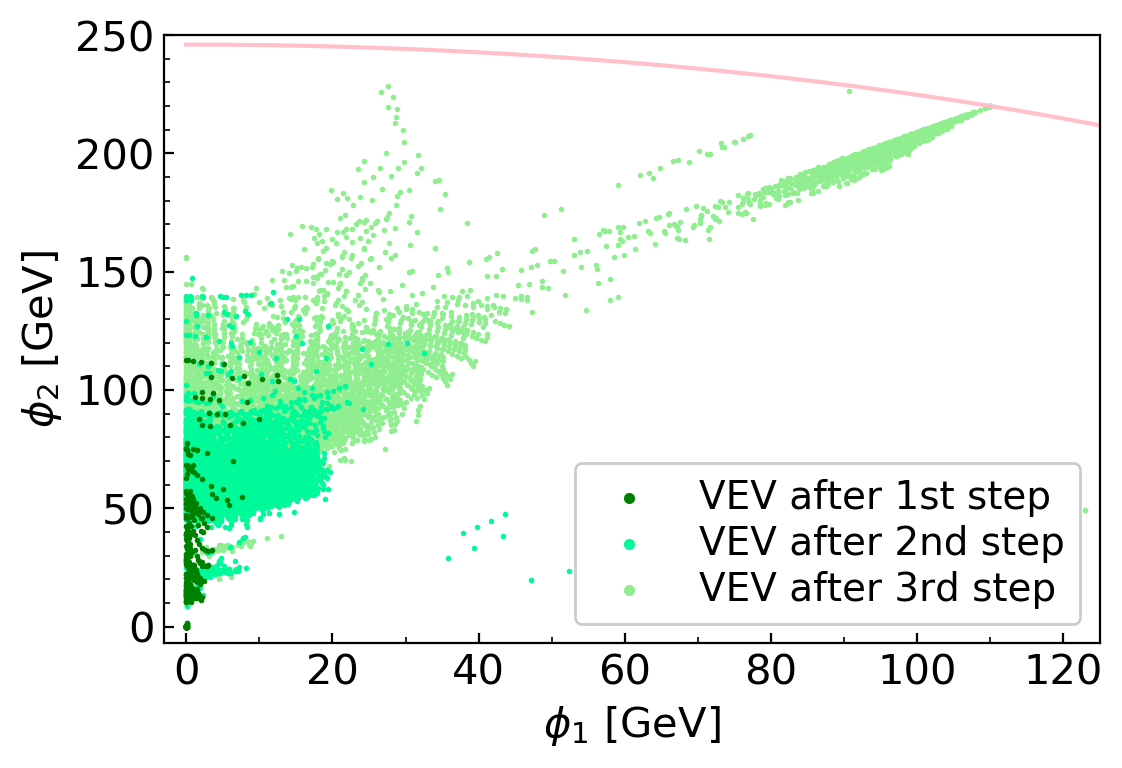}
\end{minipage}

\end{tabular}
\caption{
VEVs after each step of the 2-step (left) and 3-step (right) PTs in the Type-I 2HDM with $m_A=m_{H^\pm}$.
The dark-green, green, and light-green points represent the VEVs after the first, the second, and the third step PTs, respectively.
The pink line represents the place where $\sqrt{\phi_1^2+\phi_2^2}=246$ GeV, that the EW vacuum lies on.
}
\label{fig:minimumofmulti-stepmAI}
\end{figure}

The multi-step PTs tend to occur for the smaller $\tan\beta$ and $\cos(\beta-\alpha)$.
Next, we investigate the correlations between these parameters and $m_2^2$
that is an important parameter to determine the path of PT.
As we will see later in Fig.~\ref{fig:minimumofmulti-stepmAI},
the VEVs after the first step in the multi-step PTs have a tendency to be located mainly along the $\phi_2$ axis.
Therefore we expect that $m_2^2$ should be negative and have large enough magnitude to make the multi-step PTs occur because such $m_2^2$ makes the potential decrease in the direction of the $\phi_2$ axis.
From Eq.~\eqref{eq:extrema2}, $m_2^2$ can be written as
\begin{align}
m_2^2=\frac{1}{\tan\beta}\left[m_3^2-\frac{1}{2}(m_H^2-m_h^2)\cos\alpha\sin\alpha \right]
-\frac{1}{2}(m_h^2\cos^2\alpha+m_H^2\sin^2\alpha).
\label{eq:m2^2}
\end{align}
In Fig.~\ref{fig:m2AllPTImAI}, the regions where the 1-step, 2-step, and 3 or more step PTs occur are shown in the $m_2^2$ vs.~$\tan\beta$ (left) and $m_2^2$ vs.~$\cos(\beta-\alpha)$ (right) plane, respectively.
As can be expected, the multi-step PTs occur for the smaller $m_2^2$
as $m_2^2\lesssim -0.5\times10^4$ GeV$^2$.
The minimum value of $m_2^2$ decreases as $\tan\beta$ and $\cos(\beta-\alpha)$ are smaller, respectively.
These features can be understood from Eq.~\eqref{eq:m2^2}.
The leading term on the right-hand side of Eq.~\eqref{eq:m2^2} is the last one, $-\frac{1}{2}m_H^2\sin^2\alpha$, in our explored parameter region.
Hence, as $m_H|\sin\alpha|$ increases, the negative $m_2^2$ with the large magnitude can be obtained.
As shown in Fig.~\ref{fig:theoricalconstraints}, the larger $m_H$ is allowed for the smaller $\tan\beta$.
On the other hand, $|\sin\alpha|$ increases as $\cos(\beta-\alpha)$ gets smaller in our parameter space.
Thus,
the minimum value of $m_2^2$ decreases as $\tan\beta$ and $\cos(\beta-\alpha)$ get smaller, respectively.
Meanwhile,
at $\cos(\beta-\alpha)\simeq0.2$, $\sin\alpha$ is possible to be zero.
In this case,
the second and last terms in Eq.~\eqref{eq:m2^2} vanish and
the value of $m_H$ does not affect $m_2^2$.
However,
in the region where $\cos(\beta-\alpha)\simeq0.25$ and $\tan\beta\simeq10$,
$|\sin\alpha|$ can be large to some extent and the contribution of the last term in Eq.~\eqref{eq:m2^2} recovers,
 which leads to the negative $m_2^2$ with the slightly large magnitude.
This case is presented in Fig.~\ref{fig:m2AllPTImAI}.


\begin{figure}[t]
\centering
\begin{tabular}{c}

\begin{minipage}{0.5\hsize}
\centering
\includegraphics[clip, width=7.5cm]{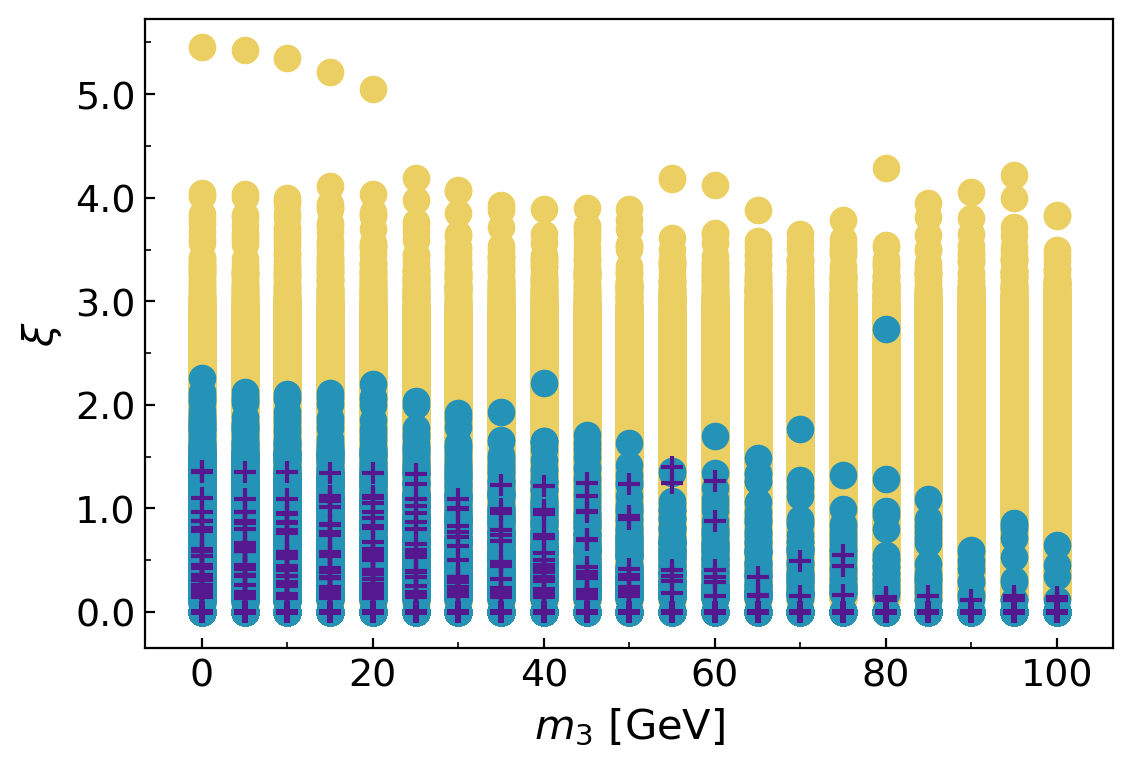}
\end{minipage}

\begin{minipage}{0.5\hsize}
  \centering
   \includegraphics[clip, width=7.5cm]{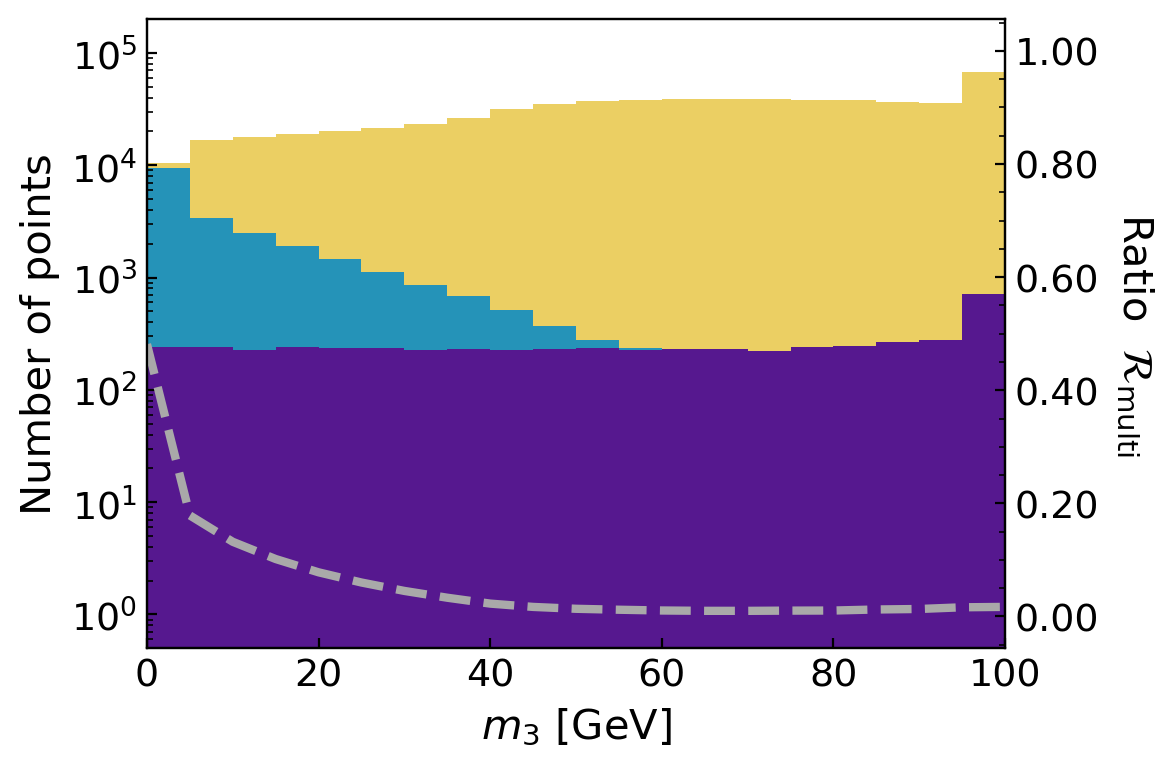}
 \end{minipage}\\

\begin{minipage}{0.5\hsize}
\centering
\includegraphics[clip, width=7.5cm]{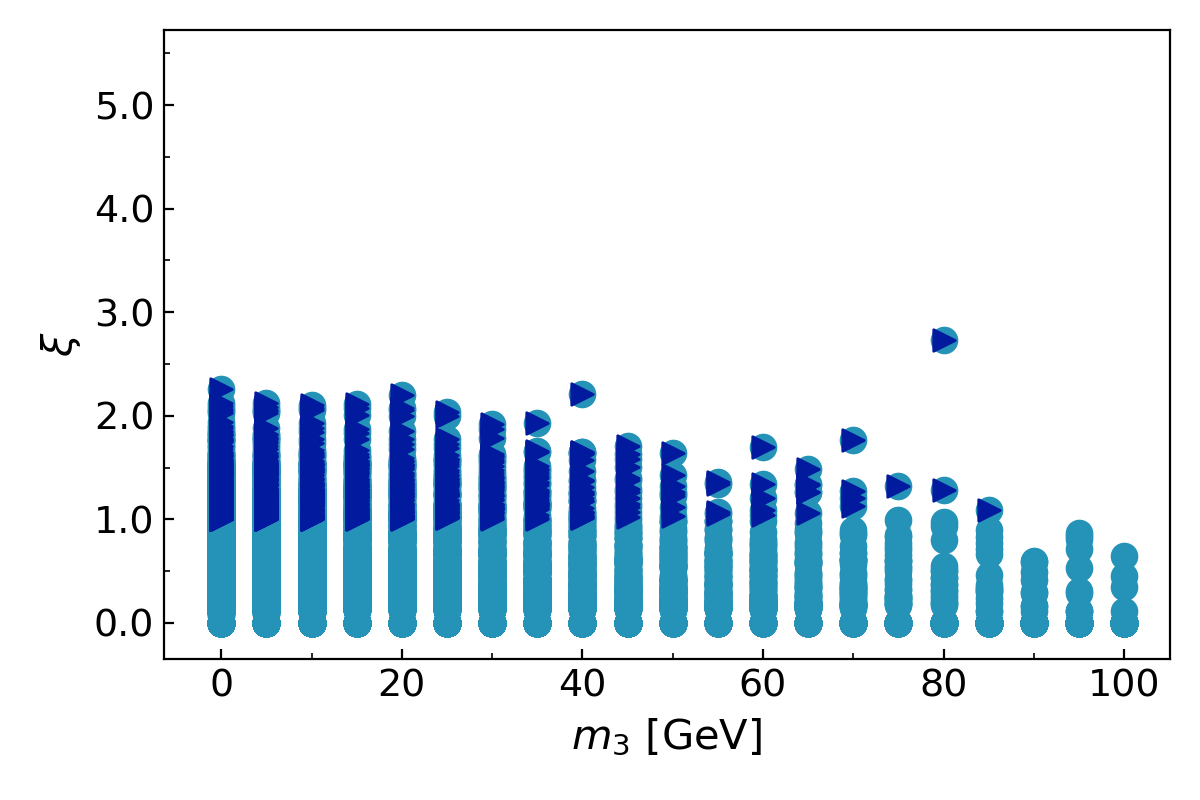}
\end{minipage}

\begin{minipage}{0.5\hsize}
 \centering
  \includegraphics[clip, width=7.5cm]{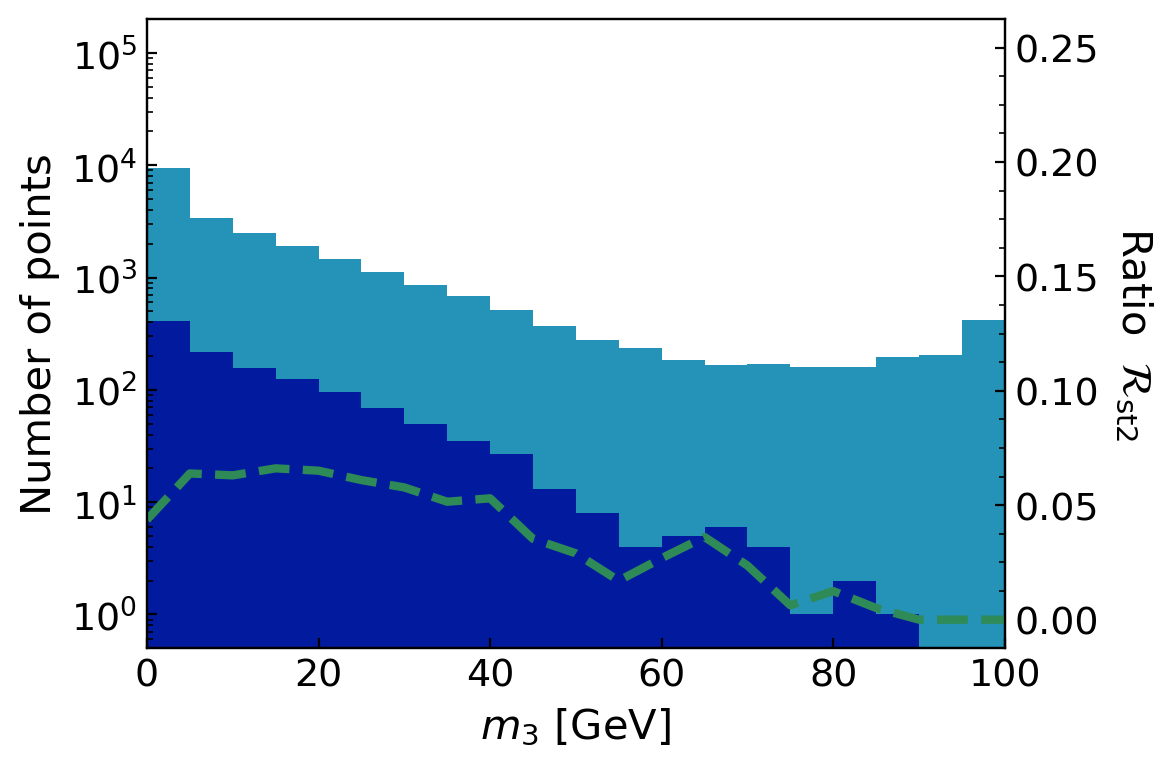}
\end{minipage}

\end{tabular}
\caption{
Left: Parameter points where the 1-step and multi-step PTs (upper), and the strong 2-step PTs (lower) occur in the $\xi$ vs.~$m_3$ plane in the Type-I 2HDM with $m_A=m_{H^\pm}$.
Right: Number of points where the 1-step and multi-step PTs (upper), and the strong 2-step PTs (lower) occur as a function of $m_3$.
The grey (upper) and green (lower) dashed lines represent ${\cal R}_{\rm multi}$ and ${\cal R}_{\rm st2}$, respectively.
The way to color is the same as in Fig.~\ref{fig:AllPTImAI}.
}
\label{fig:m3xi1AllPTImAI}
\end{figure}


As described above, when $m_2^2$ is negative with the large magnitude, the first step in the multi-step PT tends to occur along the $\phi_2$ axis.
In order to see more clearly, we show in Fig.~\ref{fig:minimumofmulti-stepmAI}
the VEVs after each step of the 2-step (left) and 3-step (right) PTs in the $\phi_2$ vs.~$\phi_1$ plane.
The dark-green points indicate the VEVs after the first step PTs and the green points show the ones
after the second step PTs. In the right panel of Fig.~\ref{fig:minimumofmulti-stepmAI},
the light-green points indicate the VEVs after the third step PTs.
The EW vacuum, where the relation $\sqrt{\phi_1^2+\phi_2^2}=246$ GeV is satisfied,
is located on the pink line.
One can see that the dark-green points have a tendency to be along the axes, and the most of VEVs after the last step PTs are located in the direction of the EW vacuum.
In the left panel of Fig.~\ref{fig:minimumofmulti-stepmAI},
the VEVs after the first step PTs extend to in the directions of the $\phi_1$ or $\phi_2$ axis, even though many of them extend to the $\phi_2$ axis.
We find that those cases have the negative $m_1^2$ or $m_2^2$ with the large magnitude as expected.
Additionally, it is found that the VEVs after the first step PTs are located on or near the $\phi_1$ axis
when $\tan\beta \simeq 2$, $\cos(\beta-\alpha)$ is large and $m_3$ is near zero.
On the other hand, for the 3-step PTs in the right panel of Fig.~\ref{fig:minimumofmulti-stepmAI},
the VEVs after the first step PTs favor going along the $\phi_2$ axis rather than $\phi_1$.
The magnitudes of those VEVs after the first step PTs are not so large, therefore
it would be difficult for the first step PT of the 3-step PTs to become the strongly first order.
Similar results to above are also seen in the Type-I 2HDM with $m_H=m_{H^\pm}$ and the Type-X 2HDM with $m_A$ or $m_H=m_{H^\pm}$.


Finally, the left panels of Fig.~\ref{fig:m3xi1AllPTImAI} show the strength of PT $\xi$ of the first step PT as a function of the $m_3$~\footnote{
In Ref.~\cite{Basler:2016obg}, $\xi$ of the 1-step PT is computed with the Parwani method.
Compared with our result ($\xi\lesssim6$ as shown in the upper left panel of Fig.~\ref{fig:m3xi1AllPTImAI}), the larger values of $\xi$ such as $\xi \gsim 20$ have been obtained,
however the explored parameter ranges in Ref.~\cite{Basler:2016obg} is larger
({\it e.g.}, $0\leq m_3^2\leq5\times 10^5$ GeV$^2$ and $1\leq\tan\beta\leq35$ in the Type-I 2HDM).
}.
Considering the tree-level potential $V_0$ in Eq.~(\ref{eq:treepotential}), the large $m_3^2$ makes the potential decrease in the region far from the axes and the magnitude of the negative $m_2^2$ small (cf.~Eq.\eqref{eq:m2^2}).
These make the directions of the first step PT toward the region far from axes,
so that it is difficult for the multi-step PTs whose first step PTs occur along the axes to happen.
From the analysis, we have found that the maximum magnitude of the VEVs after the first step of the multi-step PTs are gradually larger as $m_3$ gets smaller.
Therefore, $\xi$ of the first step PT has a tendency to be large for the smaller $m_3$.
We have also found that the larger ${\cal R}_{\rm multi}$ and ${\cal R}_{\rm st2}$ are obtained for the smaller $m_3$ as in the right panels of Fig.~\ref{fig:m3xi1AllPTImAI}.
Note that no parameter points where the strong 2-step PTs occur were found in $m_3 \gsim 90$ GeV.
We summarize in Tabs.~\ref{tb:Numbertable1} and~\ref{tb:Numbertable2} in Appendix.~\ref{sec:NumberAnalysis},
the values or ranges of input parameters where the ratios ${\cal R}_{\rm multi}$ and ${\cal R}_{\rm st2}$ have the maximum values, respectively, for Type-I and -X 2HDMs.

\subsubsection{Type-I $(m_H=m_{H^\pm})$}
\label{sec:AnalysismHI}
\begin{figure}[t]
  \centering
  \begin{tabular}{c}

  \begin{minipage}{0.5\hsize}
  \centering
  \includegraphics[clip, width=7.cm]{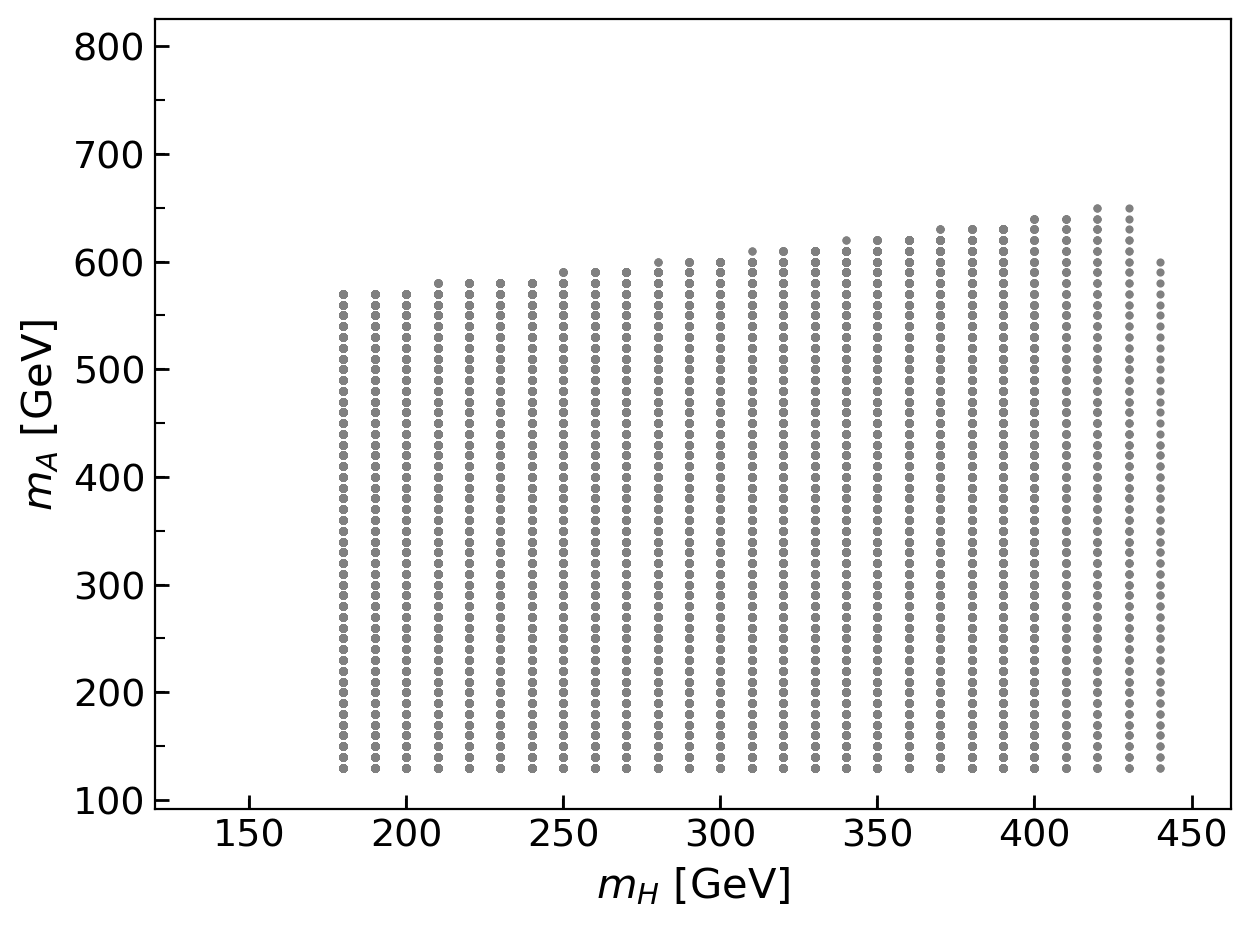}
  \end{minipage}

  \begin{minipage}{0.5\hsize}
  \centering
  \includegraphics[clip, width=7.cm]{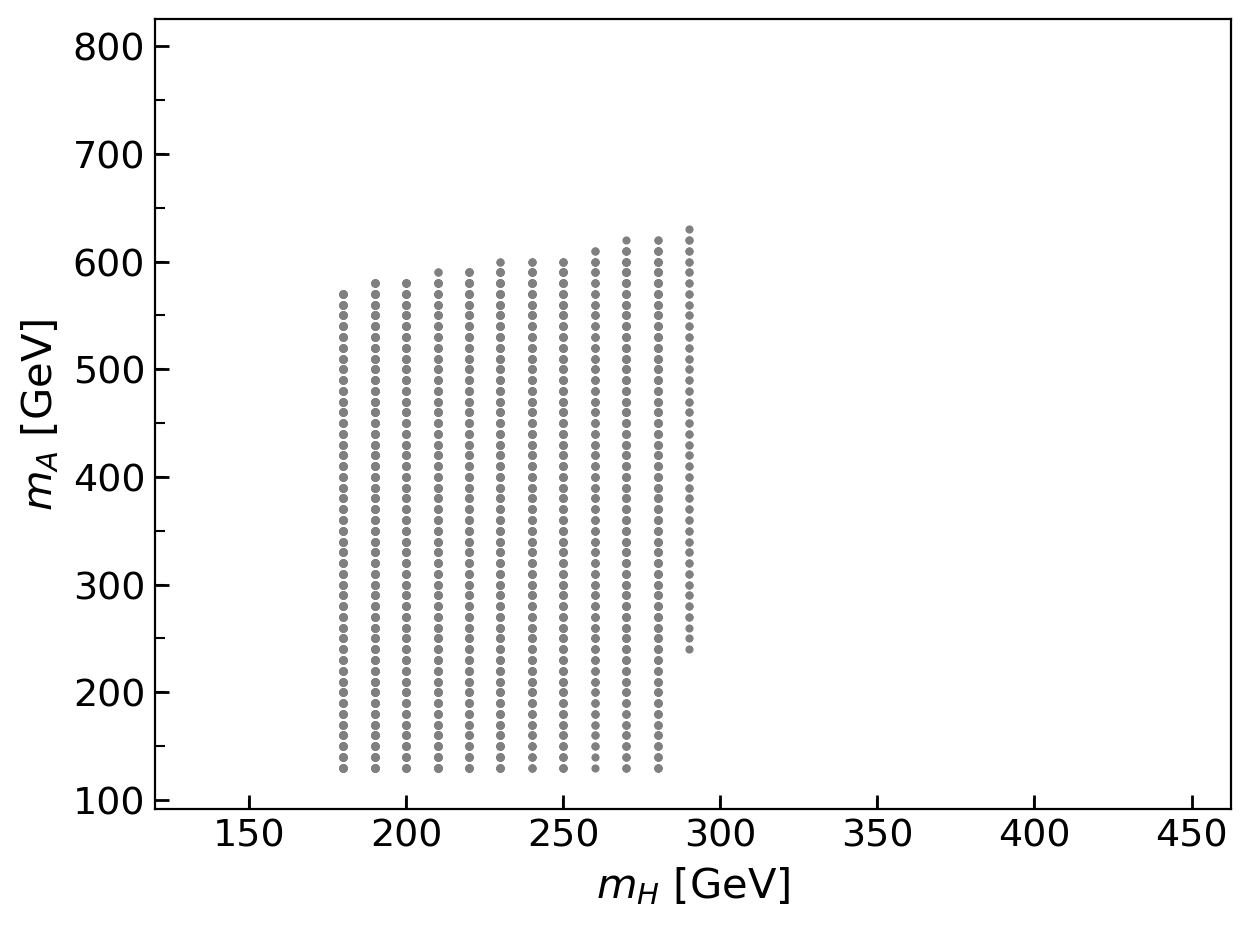}
  \end{minipage}

  \end{tabular}
  \caption{Parameter regions in the $m_A$ vs.~$m_H$ plane allowed by the theoretical constraints (the BFB, the perturbativity, and the tree-level unitarity) in the Type-I 2HDM with $m_H=m_{H^\pm}$. The left and right panel show the regions in the cases of $\tan\beta=2$ and 7, respectively. The other input parameters follow Tab.~\ref{tb:allparameterregion}.}
  \label{fig:theoricalconstraintsmHI}
 \end{figure}

Fig.~\ref{fig:theoricalconstraintsmHI} represents the allowed parameter region by the theoretical constraints (the BFB, the perturbativity, and the tree-level unitarity) in the $m_A$ vs.~$m_H$ plane at $\tan\beta=2$ (left) and 7 (right) in the Type-I 2HDM with $m_H=m_{H^\pm}$.
We can see the theoretical constraints on $m_H$ at $\tan\beta=7$ is more severe than the ones at $\tan\beta=2$, as $m_H \lesssim 290$ (440) GeV for $\tan\beta=7$ (2).


 \begin{figure}[t]
 \centering
 \begin{tabular}{c}

 \begin{minipage}{0.5\hsize}
 \centering
 \includegraphics[clip, width=7.5cm]{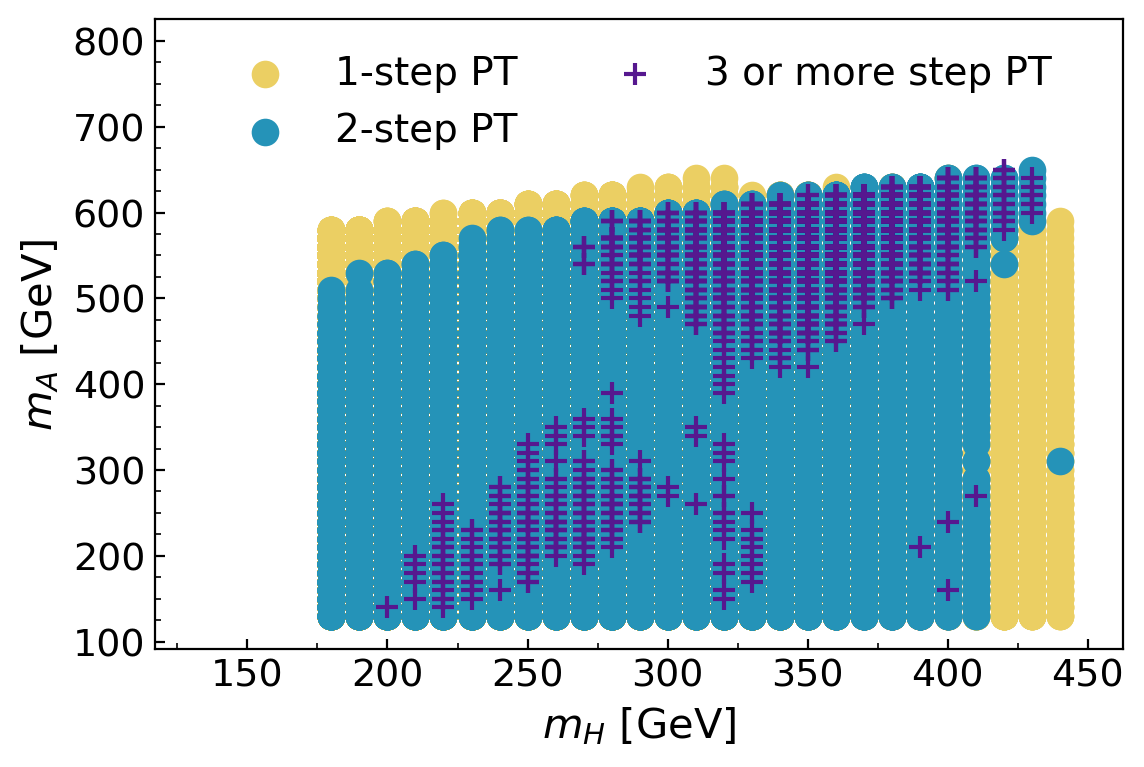}
 \end{minipage}

 \begin{minipage}{0.5\hsize}
 \centering
 \includegraphics[clip, width=7.5cm]{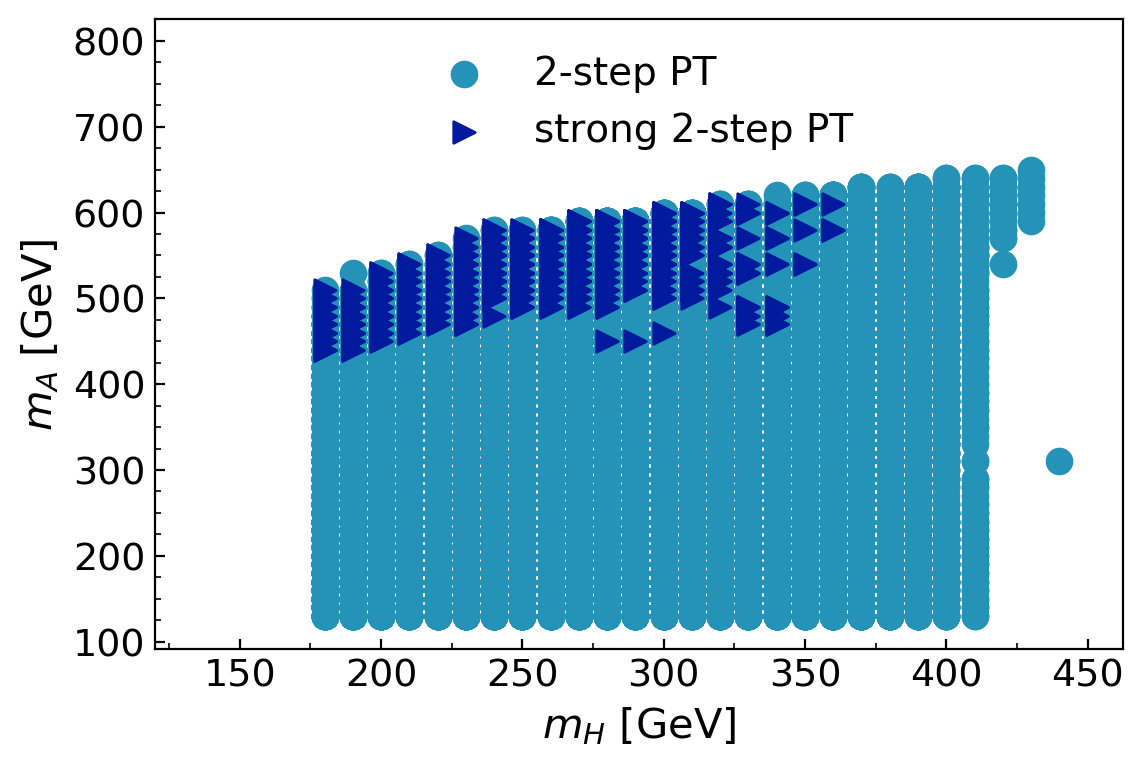}
 \end{minipage}\\

 \begin{minipage}{0.5\hsize}
  \centering
  \includegraphics[clip, width=7.5cm]{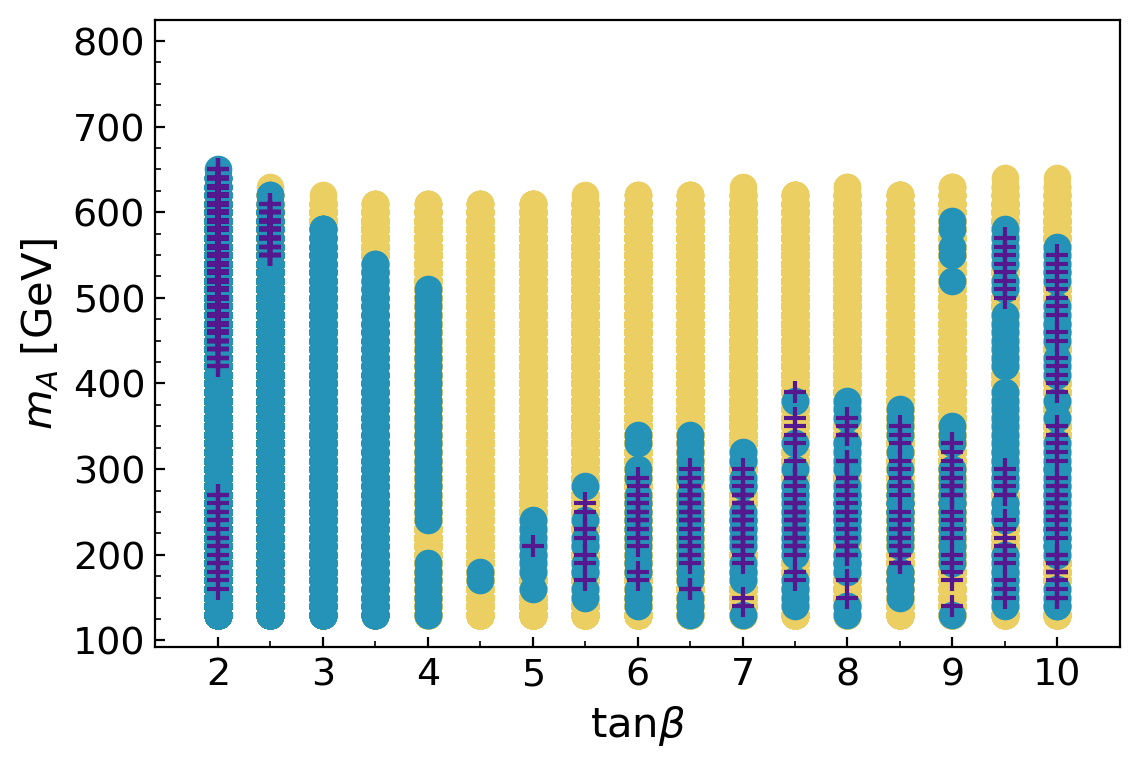}
  \end{minipage}

  \begin{minipage}{0.5\hsize}
  \centering
  \includegraphics[clip, width=7.5cm]{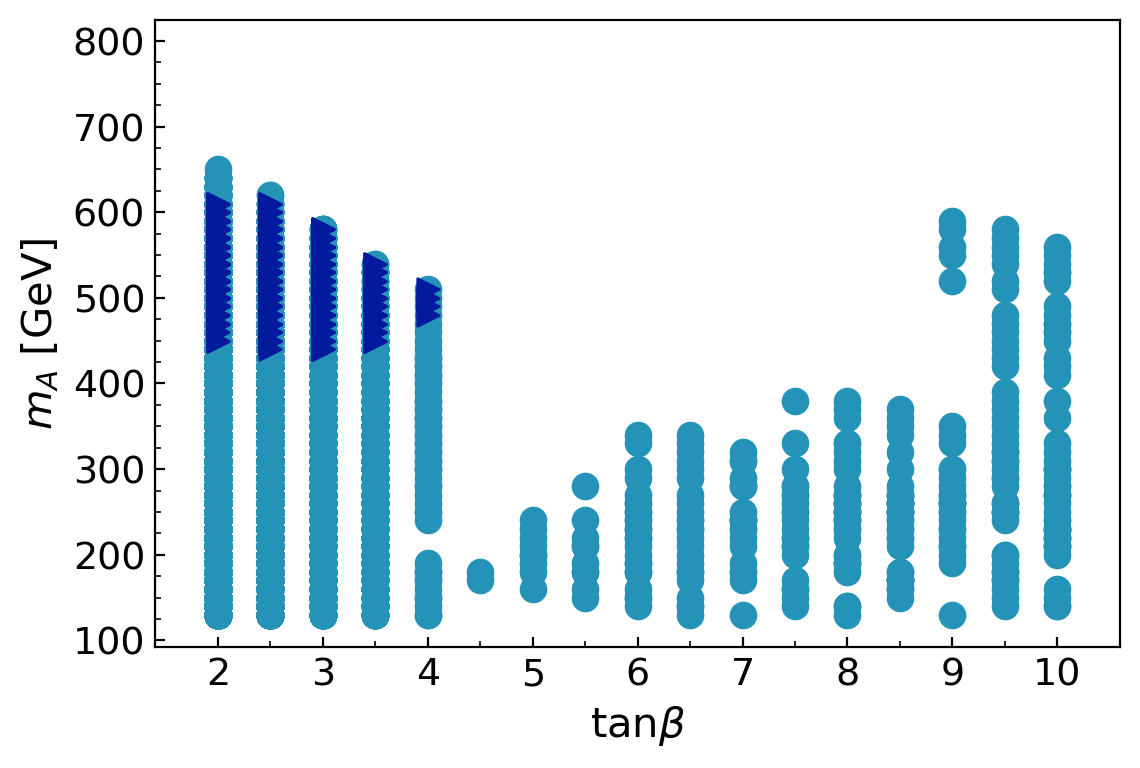}
  \end{minipage}\\

 \begin{minipage}{0.5\hsize}
 \centering
 \includegraphics[clip, width=7.5cm]{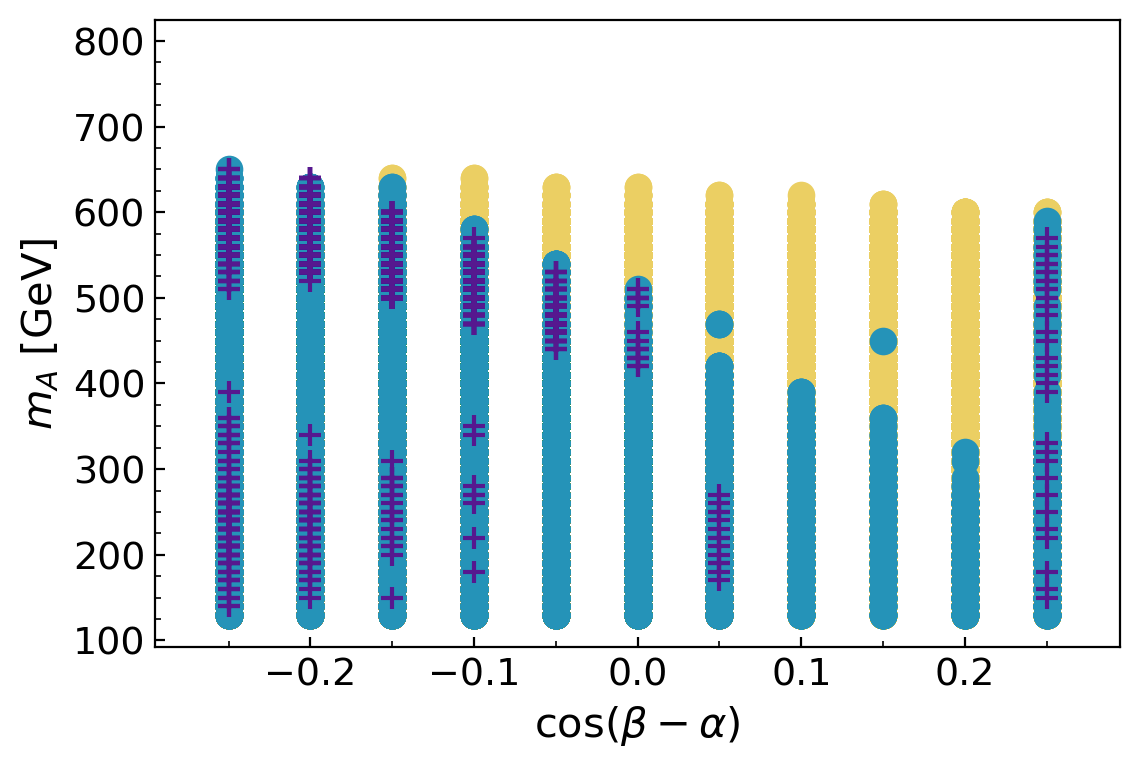}
 \end{minipage}

 \begin{minipage}{0.5\hsize}
 \centering
 \includegraphics[clip, width=7.5cm]{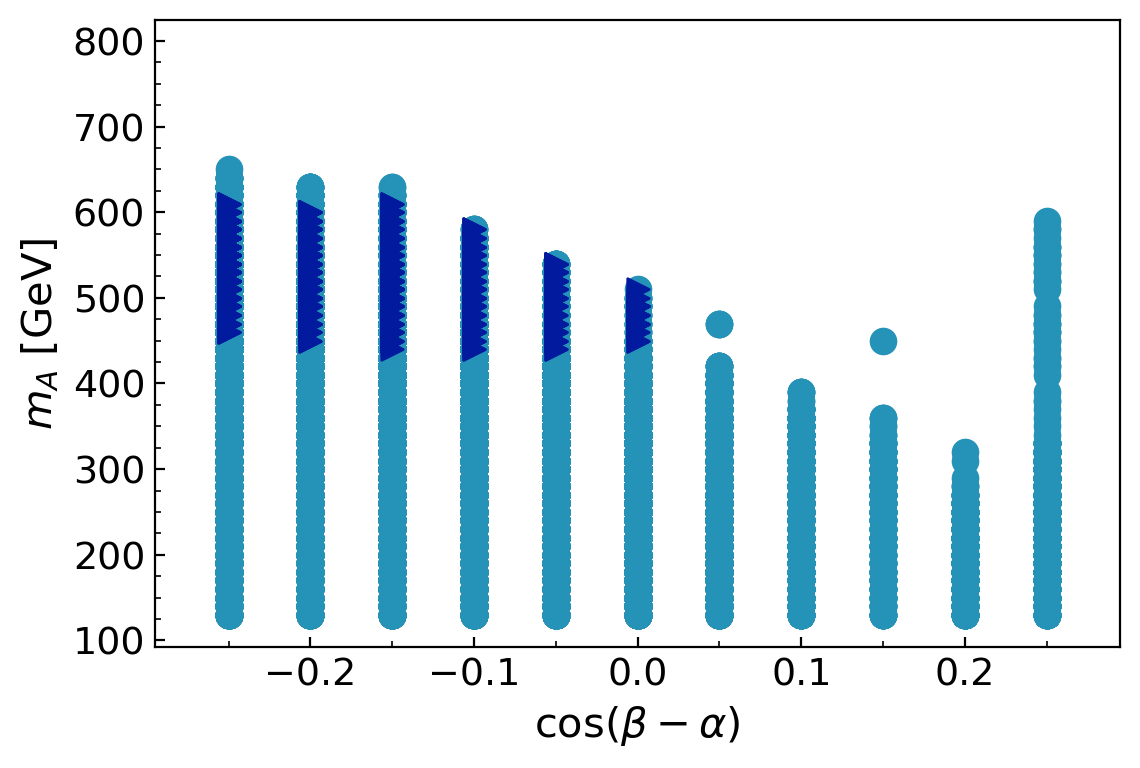}
 \end{minipage}\\

 \begin{minipage}{0.5\hsize}
 \centering
 \includegraphics[clip, width=7.5cm]{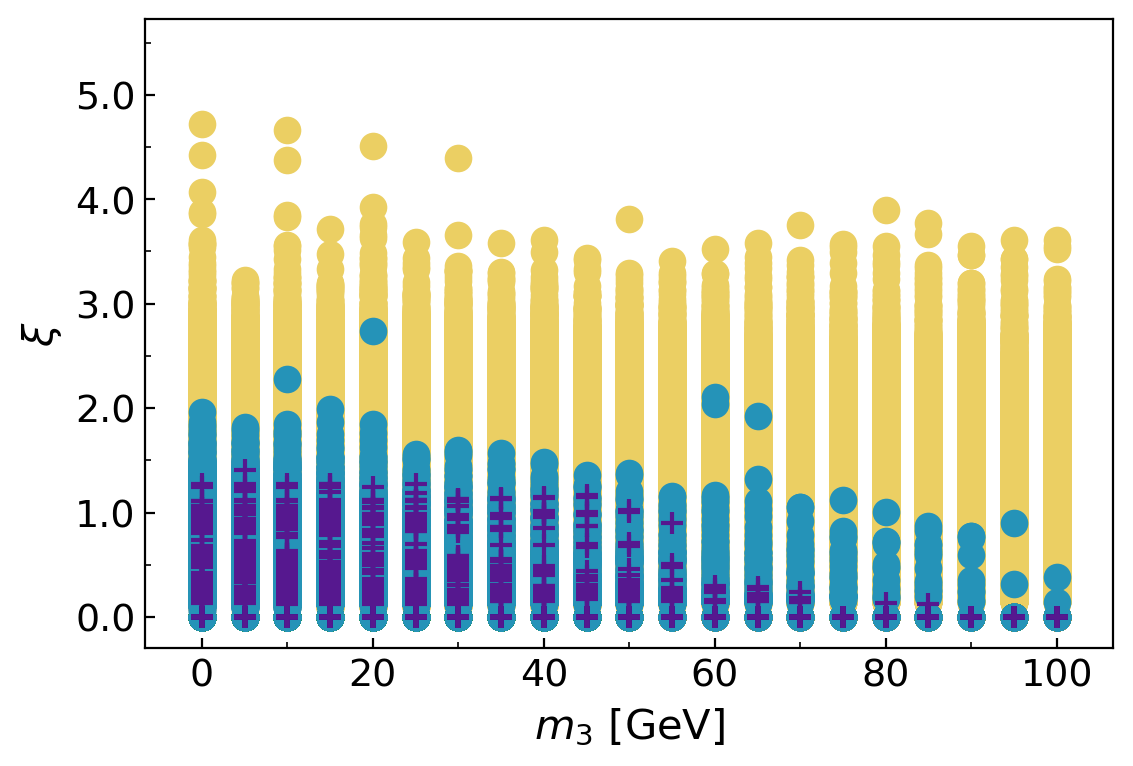}
 \end{minipage}

 \begin{minipage}{0.5\hsize}
 \centering
 \includegraphics[clip, width=7.5cm]{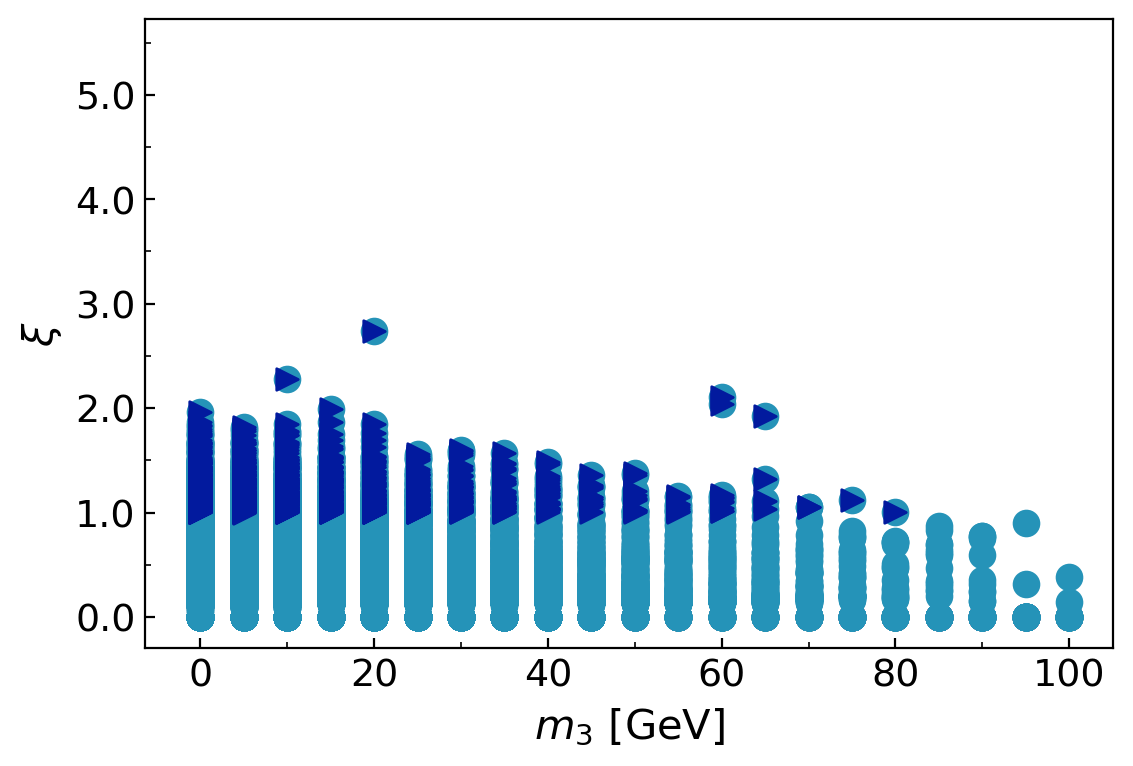}
 \end{minipage}

 \end{tabular}
 \caption{
 Same as Figs.~\ref{fig:AllPTImAI} and \ref{fig:m3xi1AllPTImAI} but for the Type-I 2HDM with $m_H=m_{H^\pm}$.
 }
 \label{fig:AllPTmHI}
\end{figure}


Fig.~\ref{fig:AllPTmHI} shows the parameter points where the 1-step, 2-step, and 3 or more step PTs (left), and the strong 2-step PTs (right)
occur in the Type-I 2HDM with $m_H=m_{H^\pm}$ for the $m_A$ vs.~$m_H$ (first line), $m_A$ vs.~$\tan\beta$ (second line), $m_A$ vs.~$\cos(\beta-\alpha)$ (third line), and $\xi$ vs.~$m_3$ (fourth line) planes, respectively.
Here the constraint from the stability of the EW vacuum is imposed.
It is weaker than that in the Type-I 2HDM with $m_A=m_{H^\pm}$,
since the maximum mass scale of the extra scalar fields is lower due to the smaller maximal value of $m_{H^\pm}(=m_H < 450$ GeV).
The ranges of $m_A$ where the multi-step PTs occur are larger as the $\tan\beta$ and $\cos(\beta-\alpha)$ decrease, respectively,
in the left panels of the second and third lines in Fig.~\ref{fig:AllPTmHI}.
Here the number of points where multi-step PTs occur in $\tan\beta\simeq10$ and $\cos(\beta-\alpha)\simeq0.25$ is respectively found to be small as in the Type-I with $m_A=m_{H^\pm}$.
On the other hand, we can see, from the upper three left panels of Fig.~\ref{fig:AllPTmHI}, there is a region that only the multi-step PTs occur
in $m_A\gtrsim 600$ GeV and $m_H\gtrsim 410$ GeV with
$\tan\beta\simeq 2$ and $\cos(\beta-\alpha)\simeq -0.25$.
Such a region could be confirmed by the extra Higgs boson search for $A\to HZ$ at the LHC.
In the bottom left panel of Fig.~\ref{fig:AllPTmHI},
the maximum value of $\xi$ for the first step PT increases as $m_3$ gets smaller and
reaches around 2.
Moreover, we have found that the ratio ${\cal R}_{\rm multi}$ has the maximum value when $m_A-m_H$ is negative with the large magnitude like $-210$ GeV as in the Type-I 2HDM with $m_A=m_{H^\pm}$ (cf.~Tab.~\ref{tb:Numbertable1} of Appendix~\ref{sec:NumberAnalysis}).
Additionally, ${\cal R}_{\rm multi}$ have maximum values at $\tan\beta\simeq2$, $\cos(\beta-\alpha)\simeq-0.25$, and $m_3\simeq0$, respectively.

The strong 2-step PTs occur in $m_A\gtrsim440$ GeV and $m_H\lesssim360$ GeV with the mass hierarchy $m_A>m_H$ in the top right panel of Fig.~\ref{fig:AllPTmHI}.
We can also see that they happen only in $\tan\beta\lesssim4$ and $\cos(\beta-\alpha)\lesssim0$ from the right panels of the second and third lines.
In addition, the small $m_3$ is favored when the strong 2-step PTs happen in the bottom right panel.
Some parameter points in the above region are excluded by the constraint from $B\rightarrow\mu^+\mu^-$, {\it e.g.} most of the points for the multi-step PTs with $m_H\simeq330$--340 GeV.
Besides, we have clarified that the ratio ${\cal R}_{\rm st2}$ is the largest in $m_A-m_H>0$, the small $\tan\beta$, $\cos(\beta-\alpha)$, and $m_3$, respectively, as shown in Tab.~\ref{tb:Numbertable2} of Appendix~\ref{sec:NumberAnalysis}.
These tendencies are not changed even if we consider the constraint from $B\rightarrow\mu^+\mu^-$.

 \subsection{Type-X}
 \label{sec:AnalysisX}
\subsubsection{Type-X $(m_A=m_{H^\pm})$}
\label{sec:AnalysismAX}
\begin{figure}[t]
 \centering
 \begin{tabular}{c}

 \begin{minipage}{0.5\hsize}
 \centering
 \includegraphics[clip, width=7.5cm]{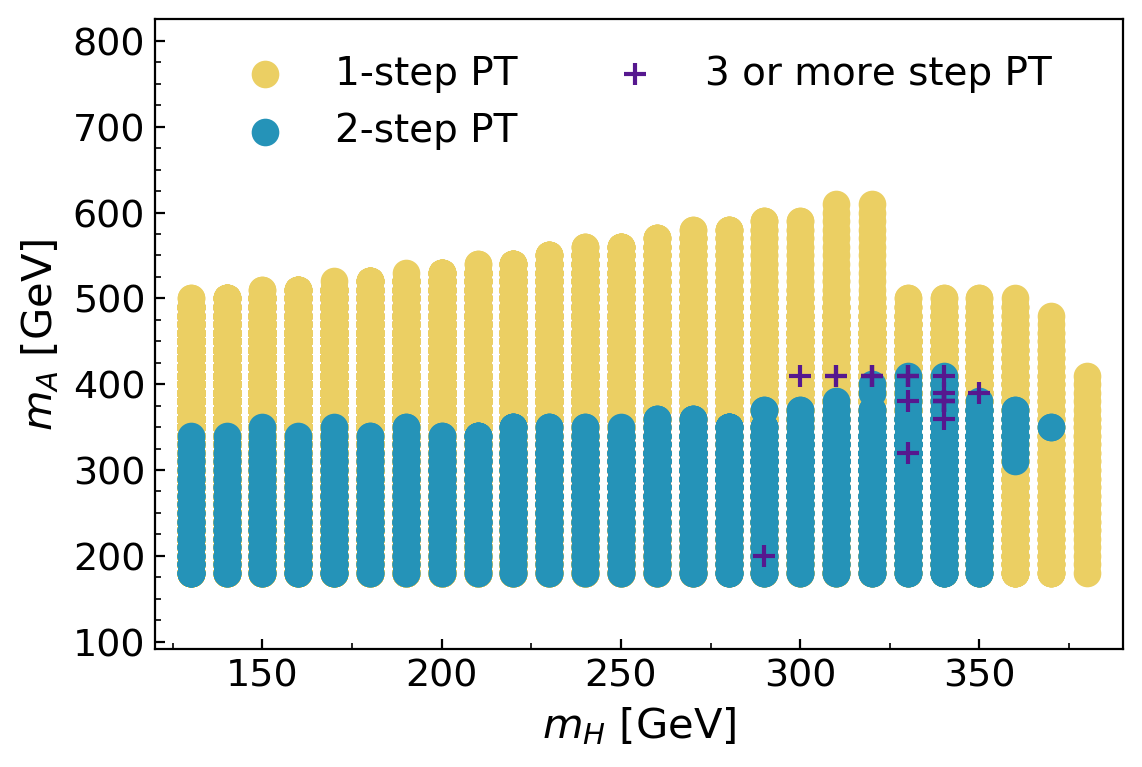}
 \end{minipage}

 \begin{minipage}{0.5\hsize}
 \centering
 \includegraphics[clip, width=7.5cm]{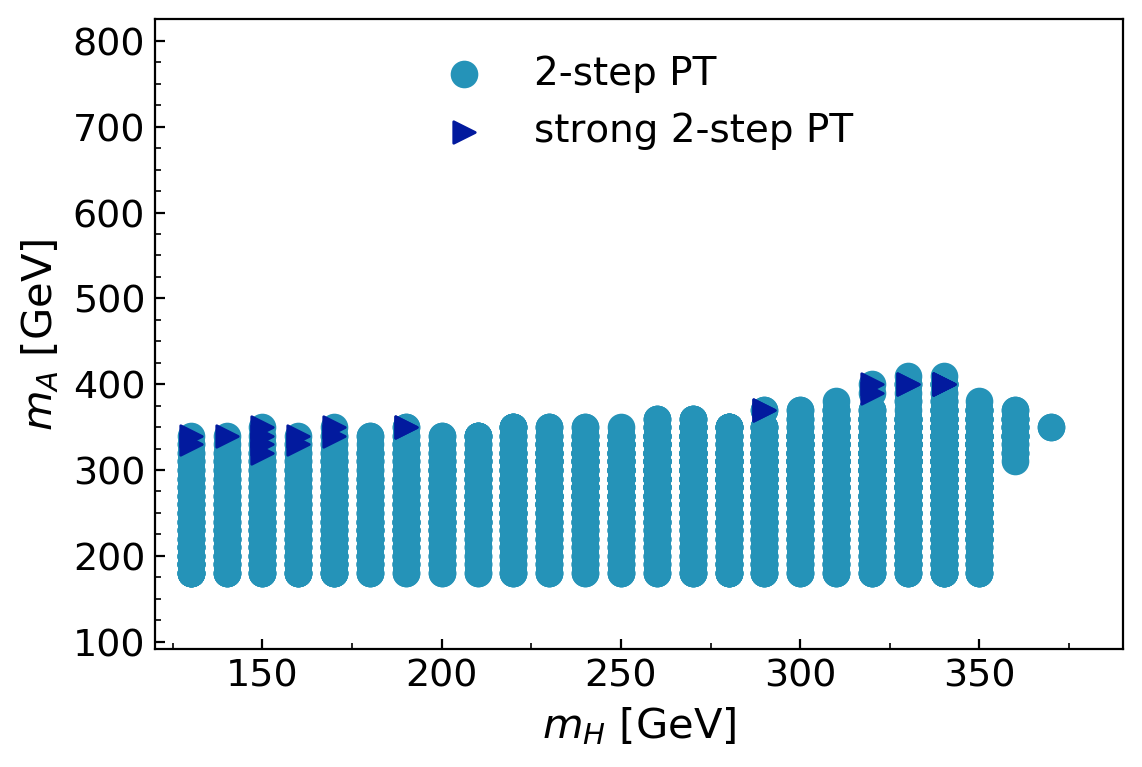}
 \end{minipage}\\

 \begin{minipage}{0.5\hsize}
 \centering
 \includegraphics[clip, width=7.5cm]{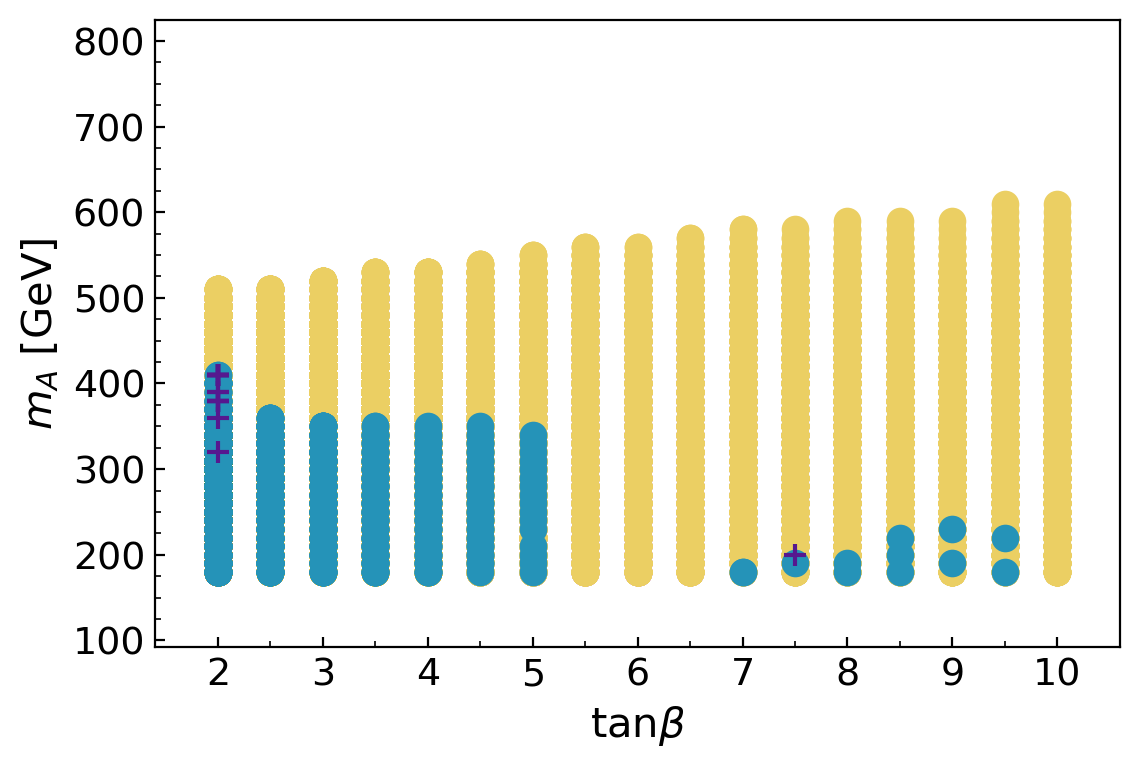}
 \end{minipage}

 \begin{minipage}{0.5\hsize}
 \centering
 \includegraphics[clip, width=7.5cm]{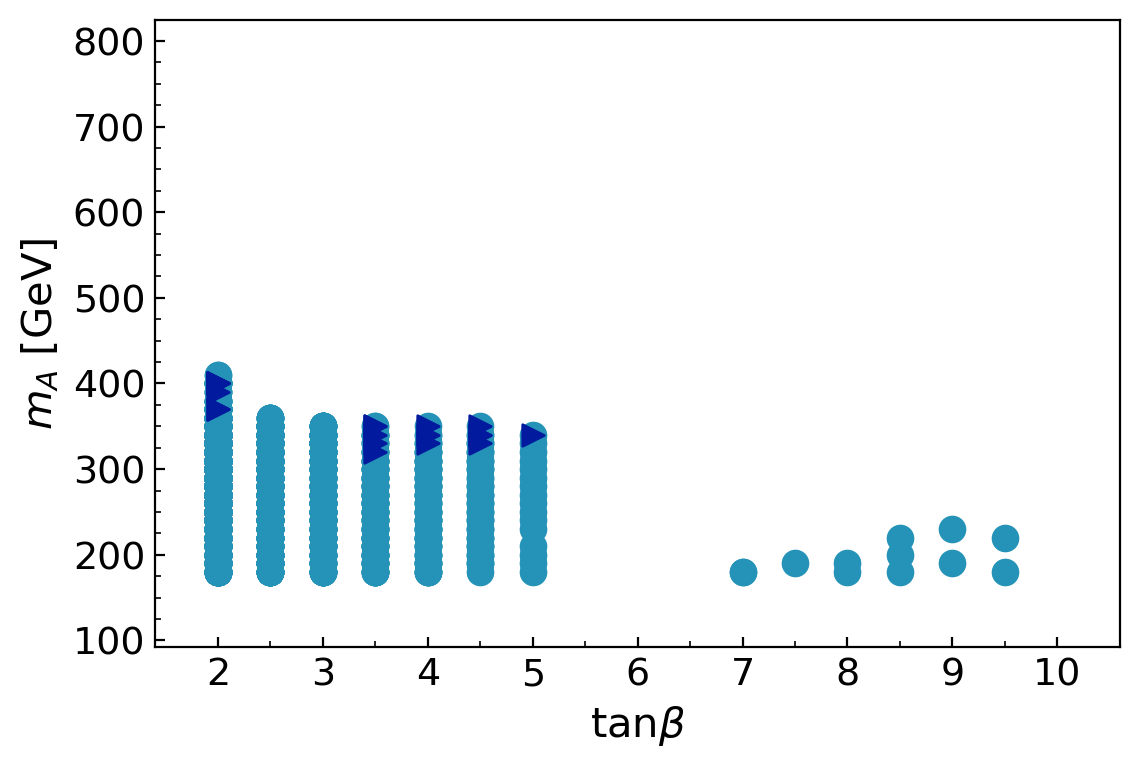}
 \end{minipage}\\

 \begin{minipage}{0.5\hsize}
 \centering
 \includegraphics[clip, width=7.5cm]{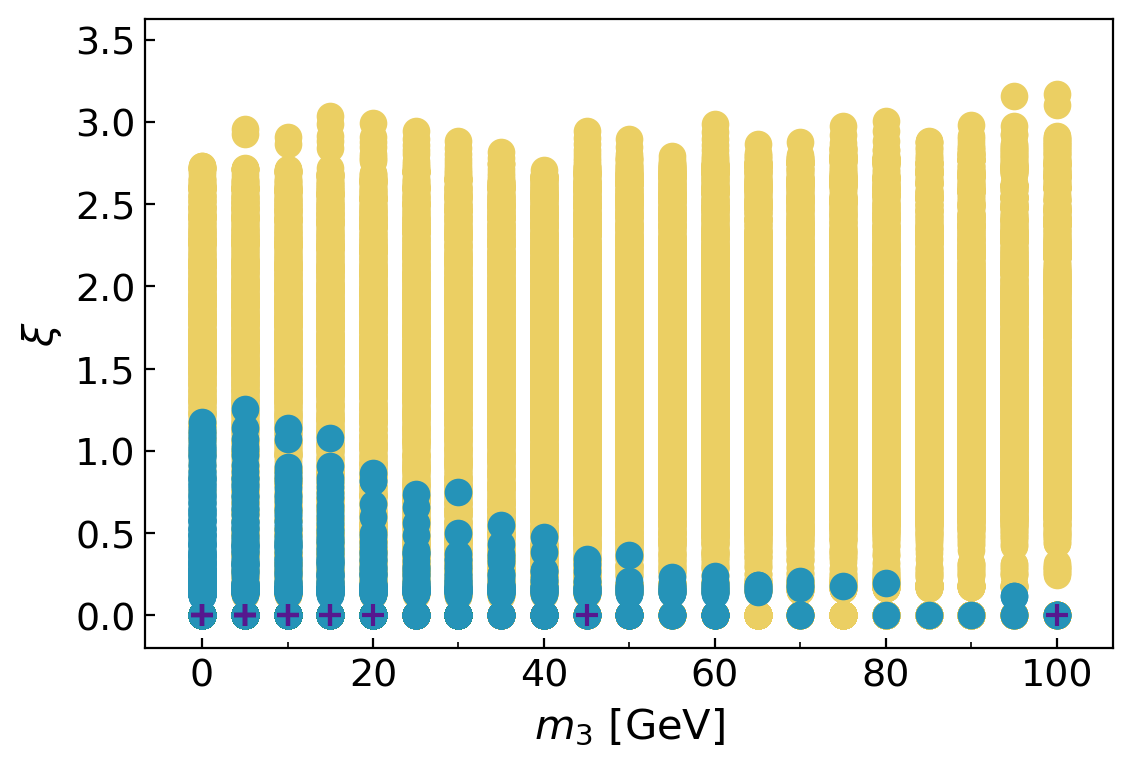}
 \end{minipage}

 \begin{minipage}{0.5\hsize}
 \centering
 \includegraphics[clip, width=7.5cm]{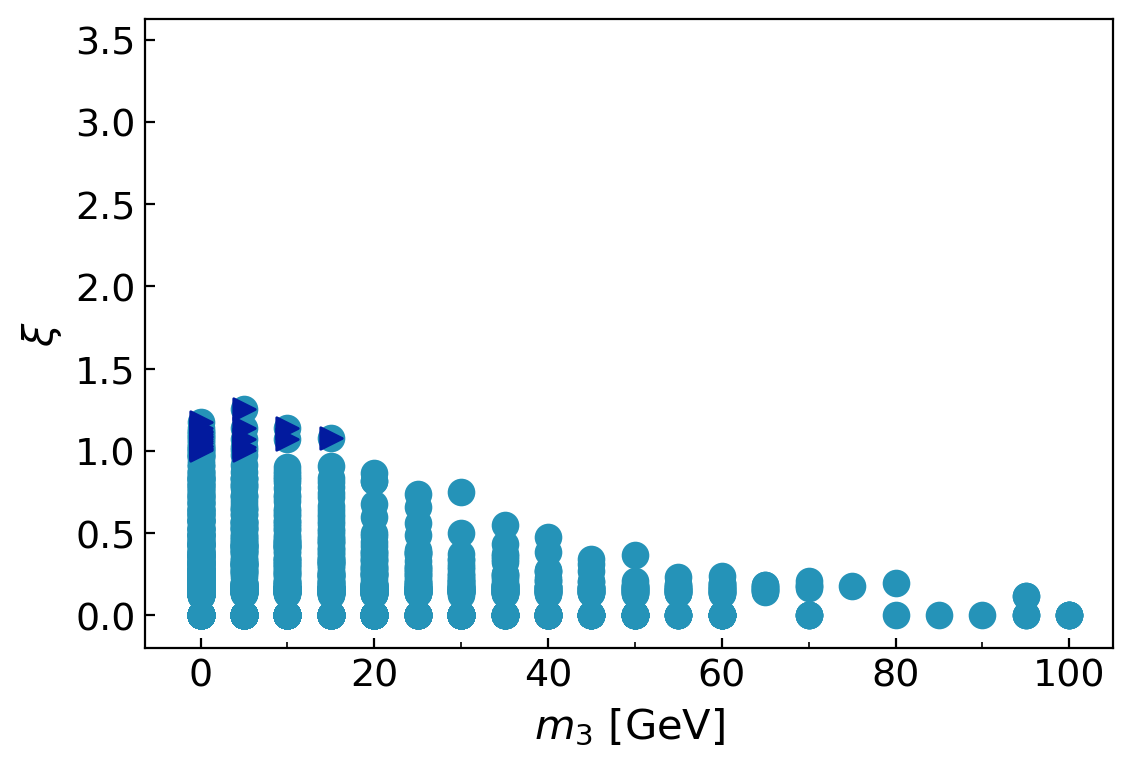}
 \end{minipage}

 \end{tabular}
 \caption{
 Parameter points where the 1-step and multi-step PTs (left), and the strong 2-step PTs (right) occur in the $m_A\ {\rm vs.}~ m_H$ (top),
 $m_A\ {\rm vs.}~ \tan\beta$ (middle), and $\xi\ {\rm vs.}~ m_3$ (bottom) planes in the Type-X 2HDM with $m_A=m_{H^\pm}$.
 The way to color points is the same as in Fig.~\ref{fig:AllPTImAI}.
 Note that we set $\cos(\beta-\alpha)=0$.
 }
 \label{fig:AllPTmAX}
\vspace{0.2in}
 \end{figure}

In the cases of the Type-X 2HDMs, we take the alignment limit $\cos(\beta-\alpha)=0$.
Fig.~\ref{fig:AllPTmAX} presents the parameter points where the 1-step and multi-step PTs (left), and the strong 2-step PTs (right) occur in the Type-X 2HDM with $m_A=m_{H^\pm}$, for the $m_A\ {\rm vs.}~ m_H$ (top), $m_A\ {\rm vs.}~ \tan\beta$ (middle), and $\xi\ {\rm vs.}~ m_3$ (bottom) planes.

 \begin{figure}[t]
 \centering
 \begin{tabular}{c}

 \begin{minipage}{0.5\hsize}
 \centering
 \includegraphics[clip, width=7.5cm]{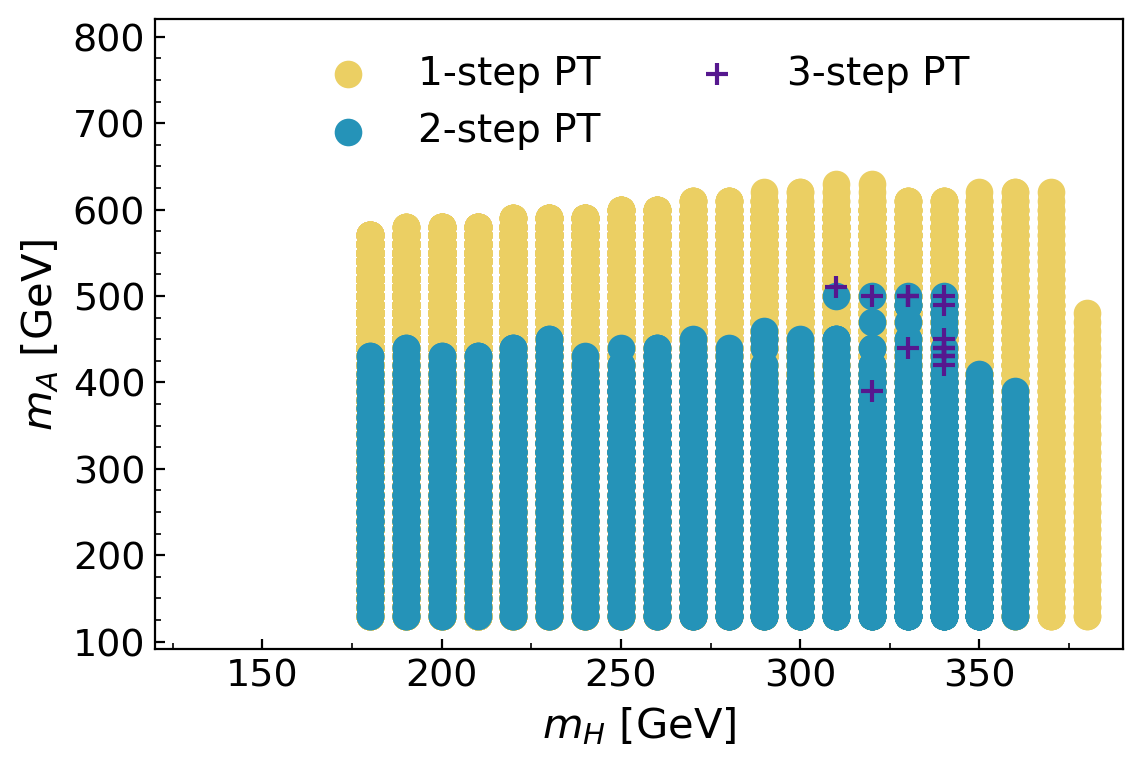}
 \end{minipage}

 \begin{minipage}{0.5\hsize}
 \centering
 \includegraphics[clip, width=7.5cm]{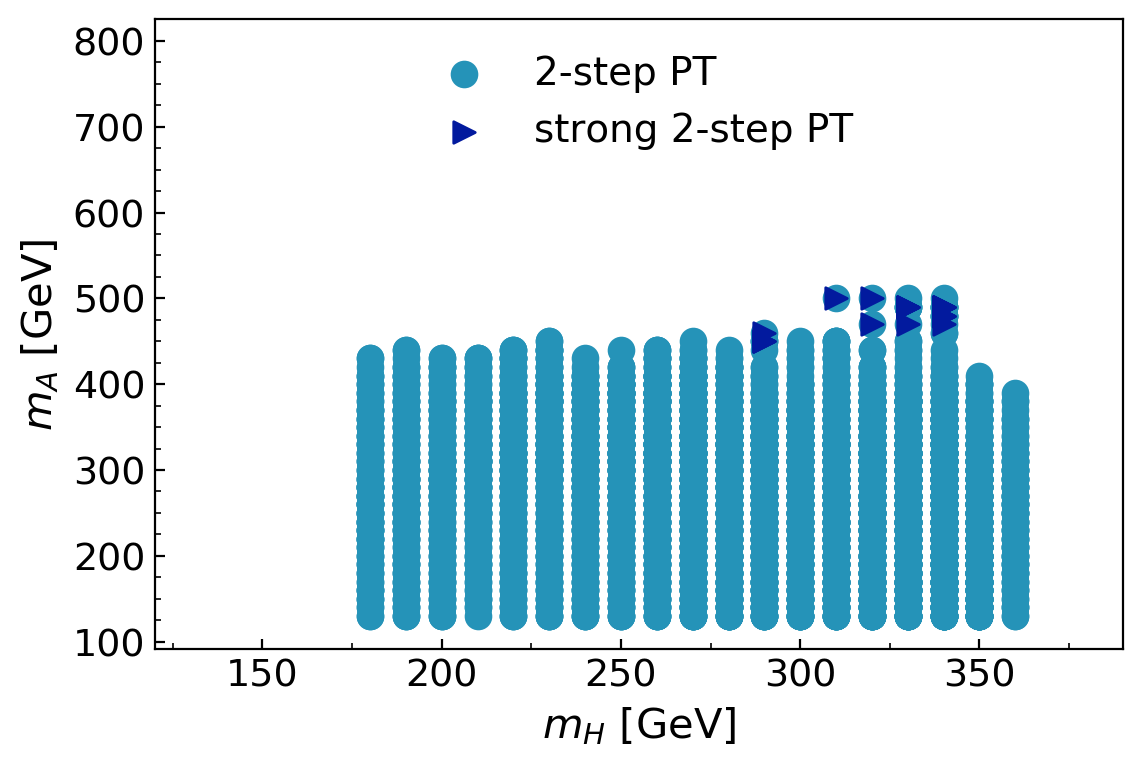}
 \end{minipage}\\

 \begin{minipage}{0.5\hsize}
 \centering
 \includegraphics[clip, width=7.5cm]{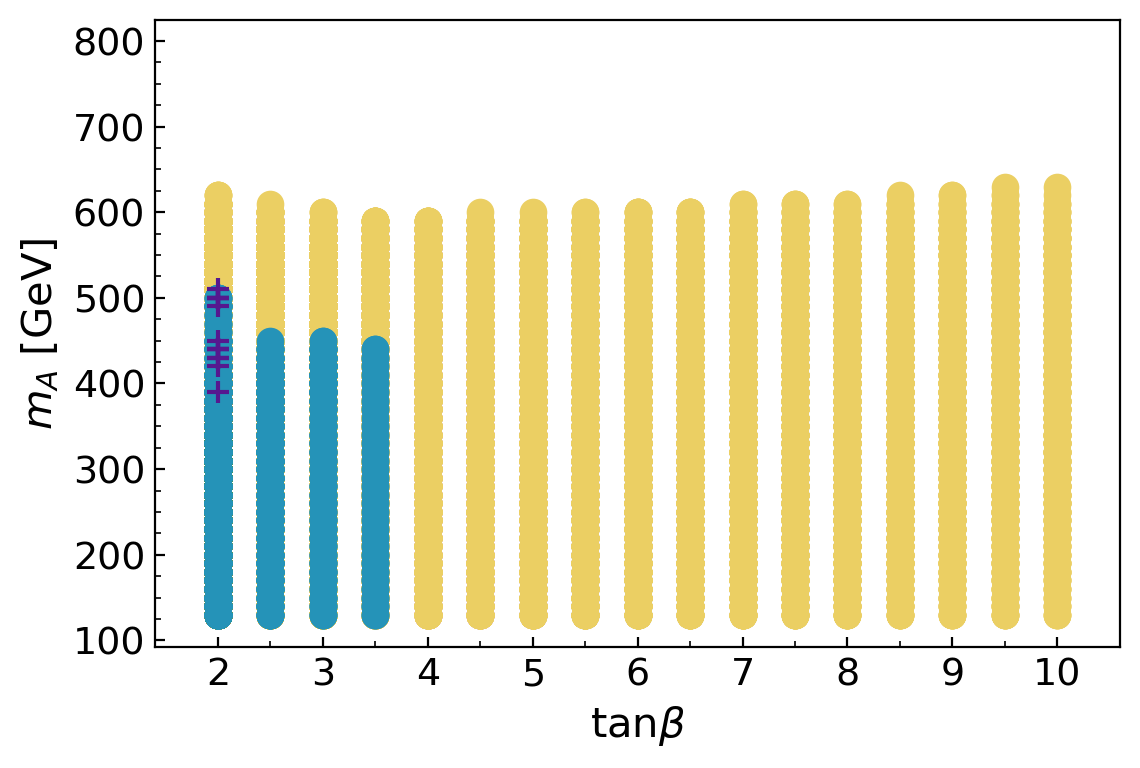}
 \end{minipage}

 \begin{minipage}{0.5\hsize}
 \centering
 \includegraphics[clip, width=7.5cm]{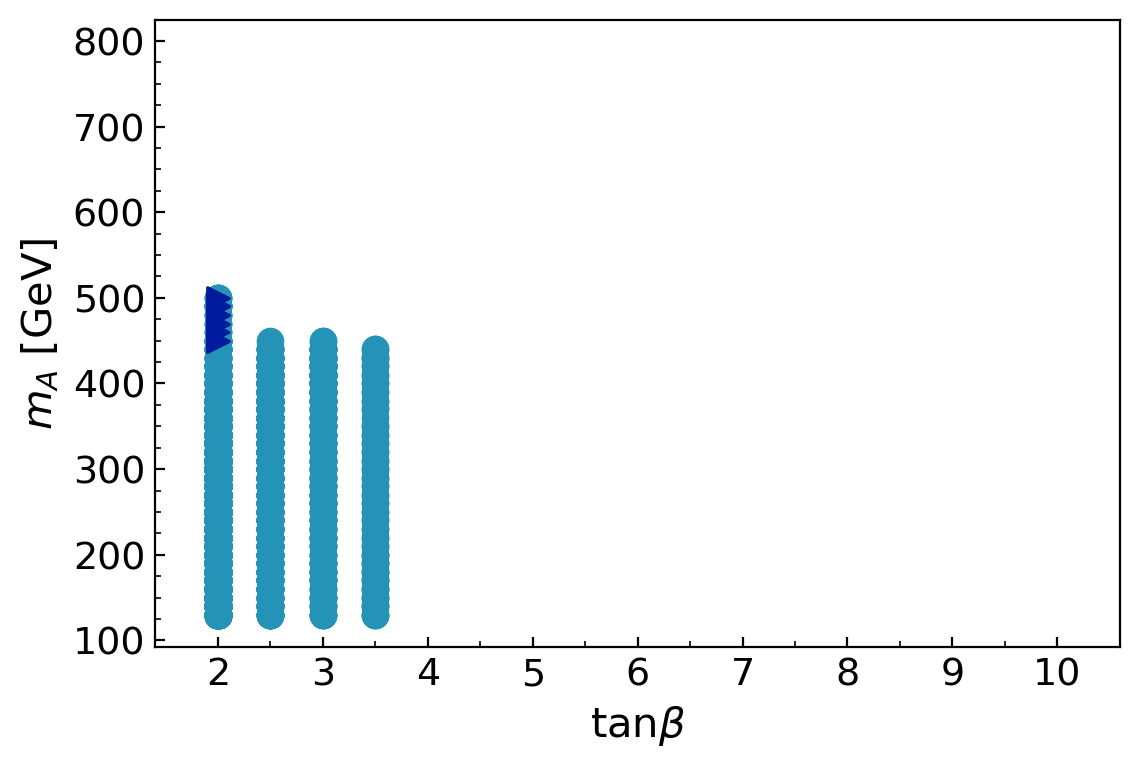}
 \end{minipage}\\

 \begin{minipage}{0.5\hsize}
 \centering
 \includegraphics[clip, width=7.5cm]{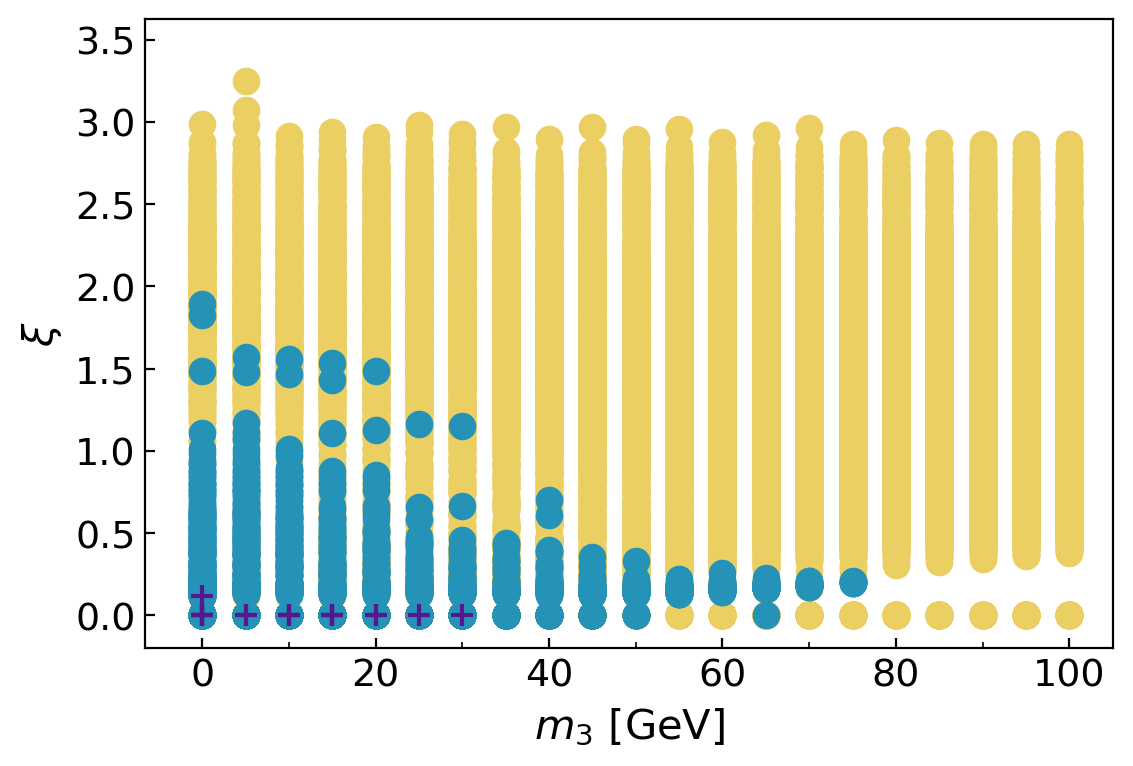}
 \end{minipage}

 \begin{minipage}{0.5\hsize}
 \centering
 \includegraphics[clip, width=7.5cm]{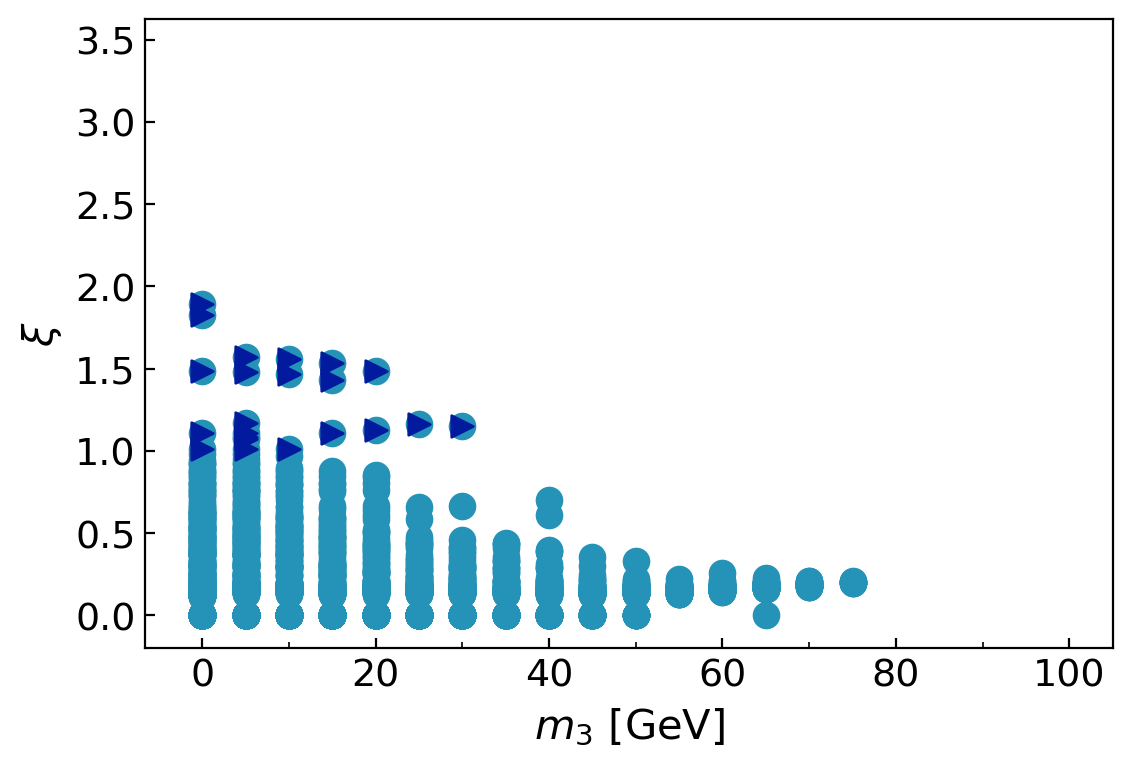}
 \end{minipage}

 \end{tabular}
 \caption{
 Same as Fig.~\ref{fig:AllPTmAX} but for the Type-X 2HDM with $m_H=m_{H^\pm}$.
 }
 \label{fig:AllPTmHX}
\vspace{0.2in}
 \end{figure}


In the top left panel of Fig.~\ref{fig:AllPTmAX}, the range of $m_A$ where the multi-step PTs occur does not change much for $m_H\lesssim350$ GeV.
The middle left panel shows that most of the multi-step PTs occur in $\tan\beta\lesssim5$.
From the bottom left panel,
we can see that the maximum value of $\xi$ for the multi-step PTs increases as $m_3$ decreases and reaches around 1.2.
From our analyses, ${\cal R}_{\rm multi}$ in
the Type-X with $m_A=m_{H^\pm}$ has the largest value as 21\% at $m_A-m_H\simeq-130$ GeV.
It also gets the maximum value at $\tan\beta\simeq2$ and $m_3\simeq0$, respectively.
We have found that the region for the multi-step PTs with $m_A-m_H\simeq-130$ GeV is realized for $\tan\beta\simeq 2$ in this case, so that
such a region is excluded by the constraints from both the $B\rightarrow\mu^+\mu^-$ and $H\to AZ$ \cite{Benbrik:2020nys}.
In the allowed region by $B\rightarrow\mu^+\mu^-$ constraint,
we have also found that the ${\cal R}_{\rm multi}$ has the larger value for the $m_A-m_H<0$, {\it e.g.} ${\cal R}_{\rm multi}= 9$\% for $m_A-m_H\simeq -90$ GeV,
and the smaller $\tan\beta$, respectively.

On the other hand, the strong 2-step PTs only occur in the narrow region with $m_A>m_H$ as shown in the top right panel of Fig.~\ref{fig:AllPTmAX}.
Moreover, the bottom right panel of Fig.~\ref{fig:AllPTmAX} shows that the strong 2-step PTs happen only in $m_3\lesssim15$ ${\rm GeV}$ and they predict $\xi\simeq$1--1.3.
Although the points for the strong 2-step PTs do not receive the constraint from $B\rightarrow\mu^+\mu^-$, those with $m_H\simeq 150$ GeV would be excluded by the $A\to HZ$ search \cite{Benbrik:2020nys}, while those with $m_H\simeq 330$ GeV remain.

As above, the parameter region where the multi-step PTs occur in this case gets narrow compared with that in the Type-I 2HDM with $m_A=m_{H^\pm}$.
The Yukawa coupling of the top quark is same among the types as in Tab.~\ref{table:Yukawa}, therefore the contribution of the top quark, which gives the fermion's largest contribution to $V_{\rm CW}$, is not dependent on the types.
According to this, the reason for the narrow region in the analysis would be the difference of the range of $\cos(\beta-\alpha)$.
As proof of that, we have confirmed that the result for $\cos(\beta-\alpha)=0$ in the Type-I 2HDM with $m_A=m_{H^\pm}$ have the same tendency as in the Type-X 2HDM with $m_A=m_{H^\pm}$.

\subsubsection{Type-X $(m_H=m_{H^\pm})$}
\label{sec:AnalysismHX}
Fig.~\ref{fig:AllPTmHX} exhibits the parameter points where the 1-step and multi-step PTs (left), and the strong 2-step PTs (right) occur in the Type-X 2HDM with $m_H=m_{H^\pm}$, for the $m_A\ {\rm vs.}~ m_H$ (top), $m_A\ {\rm vs.}~ \tan\beta$ (middle), and $\xi\ {\rm vs.}~ m_3$ (bottom) planes.
In this case we have found that there are no parameter points where the more than 3-step PTs occur, therefore the purple points show the points for only the 3-step PTs.
Although the largest value of $m_A$ for the multi-step PTs is larger than that in the Type-X with $m_A=m_{H^\pm}$, the other tendencies in Fig.~\ref{fig:AllPTmHX} are similar.
The maximum value of $\xi$ for the multi-step PTs reaches near 2 in the bottom left panel.
We have found that ${\cal R}_{\rm multi}$ is the largest at $m_A-m_H\simeq -210$ GeV, $\tan\beta\simeq2$, and $m_3\simeq0$, respectively (cf.~Tab.~\ref{tb:Numbertable1} of Appendix~\ref{sec:NumberAnalysis}).
However, the region with the large magnitude of the negative $m_A-m_H$, which is found at $\tan\beta\simeq2$, is excluded by the constraint from $H\to AZ$ decay \cite{Benbrik:2020nys}.
On the other hand, in the region where the $H\to AZ$ channel does not open,
we find that ${\cal R}_{\rm multi}$ is obtained for $m_A<m_H$
as {\it e.g.} ${\cal R}_{\rm multi}=5$\% for $m_A-m_H\simeq-80$ GeV.

Meanwhile, from the top right panel in Fig.~\ref{fig:AllPTmHX},
same as the other cases, the strong 2-step PTs only occur when the mass hierarchy $m_A>m_H$ exists with $\tan\beta \simeq 2$.
Additionally, they happen when $m_3$ is small as $m_3\lesssim 30$ GeV in the bottom right panel of Fig.~\ref{fig:AllPTmHX}.
The constraint from $B\rightarrow\mu^+\mu^-$ excludes the part of the region where the multi-step PTs happen, {\it e.g.} 290 GeV $\lesssim m_H\lesssim$ 340 GeV, hence the region for the strong 2-step PTs is excluded.

We have confirmed that the result for $\cos(\beta-\alpha)=0$ in the Type-I 2HDM with $m_H=m_{H^\pm}$ have same tendency with the ones in the Type-X 2HDM with $m_H=m_{H^\pm}$.


\begin{figure}[t]
 \centering
 \begin{tabular}{c}

 \begin{minipage}{0.5\hsize}
 \centering
 \includegraphics[clip, width=7.5cm]{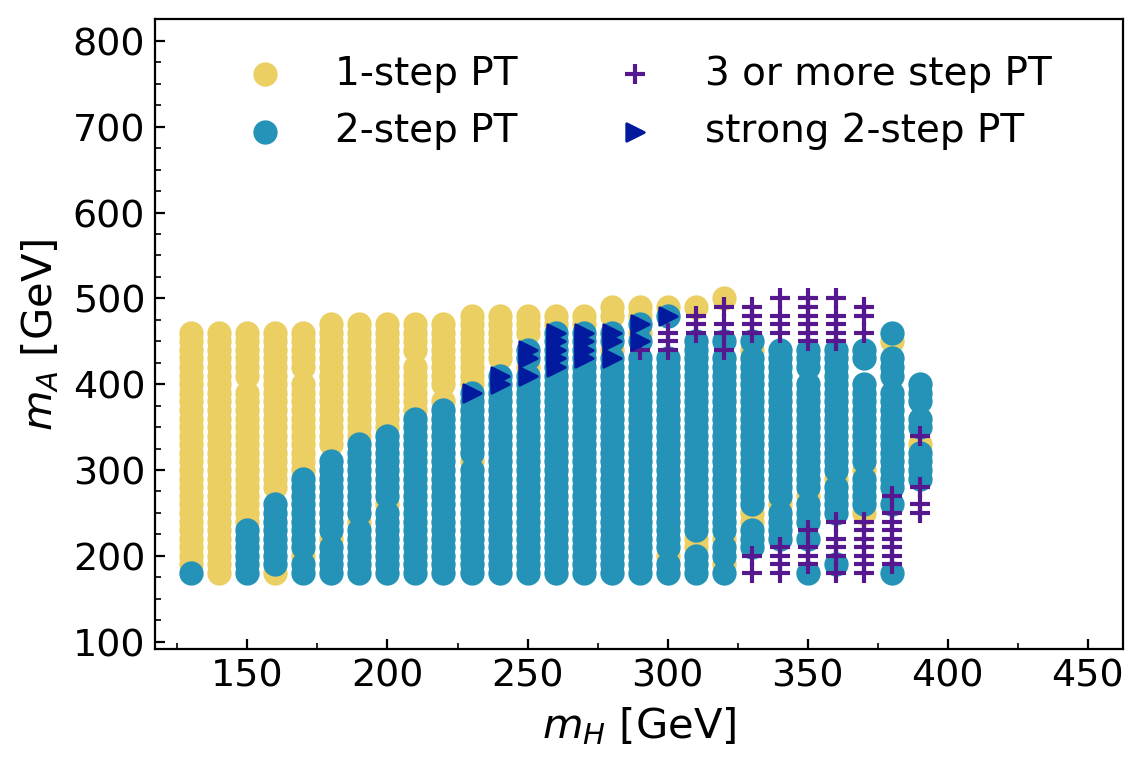}
 \end{minipage}

 \begin{minipage}{0.5\hsize}
 \centering
 \includegraphics[clip, width=7.5cm]{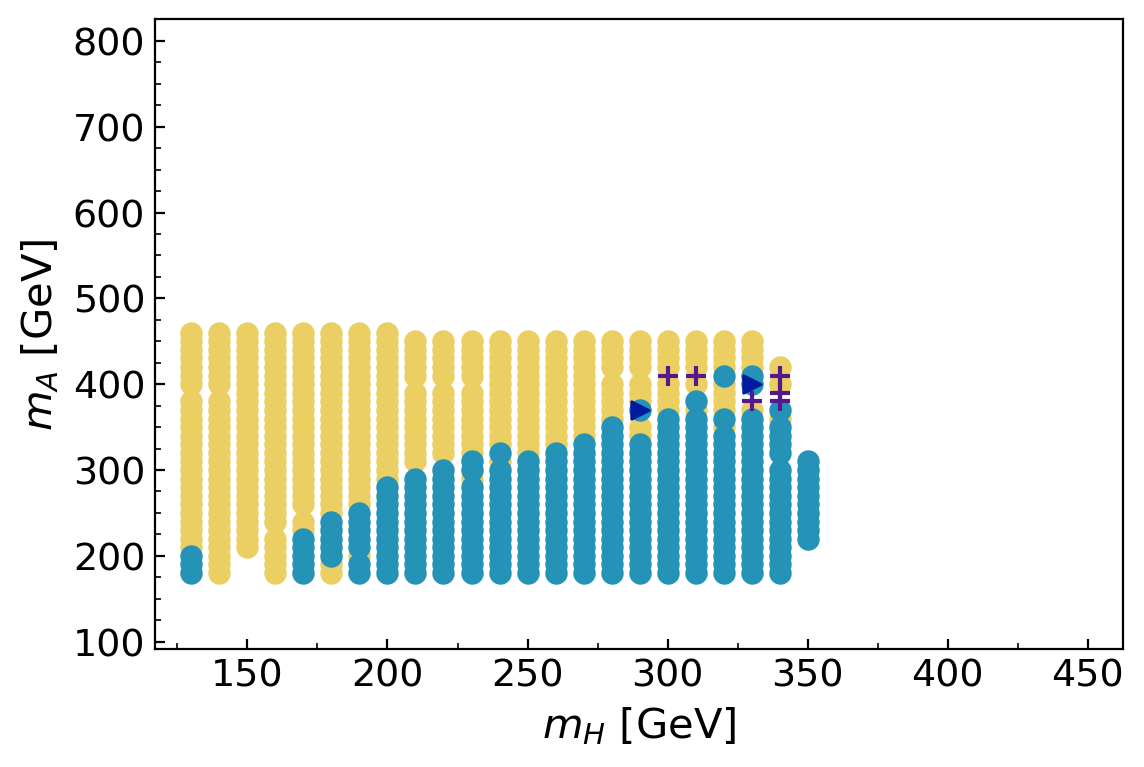}
 \end{minipage}

 \end{tabular}
 \caption{
 Parameter points where the 1-step, 2-step, strong 2-step, and 3 or more step PTs occur in the $m_A$ vs.~$m_H$ plane in the Type-I 2HDM with $m_A=m_{H^\pm}$.
 The left (right) panel shows the results at $\tan\beta=2$, $\cos(\beta-\alpha)=-0.2$ ($0$), and $m_3=0$.
 }
 \label{fig:AllPTfixedmAI}
\vspace{0.2in}
 \end{figure}


\vspace{0.5in}

To summarize briefly,
the region where the multi-step PTs
is likely to occur is where $m_A-m_H$ is negative with large magnitude,
$\tan\beta$ is small,
$\cos(\beta-\alpha)$ is negative and small in the Type-I 2HDMs (it is fixed at zero in the Type-X 2HDMs),
and $m_3$ is small, respectively.
Different from the feature of the multi-step PTs, the strong multi-step PTs occur only when the mass hierarchy $m_A>m_H$ exists,
while the tendencies for the other parameters are similar.
Finally, we show the results of two specific cases.
Fig.~\ref{fig:AllPTfixedmAI} shows the parameter points where the 1-step, 2-step, strong 2-step, and 3 or more step PTs occur in the $m_A$ vs.~$m_H$ plane in the Type-I 2HDM with $m_A=m_{H^\pm}$. The other input parameters are fixed as $\tan\beta=2$, $\cos(\beta-\alpha)=-0.2$ (left panel) or $0$ (right panel), and $m_3=0$.
We can see the regions for the 1-step and multi-step PTs are almost divided.
In addition, ${\cal R}_{\rm multi}$
in the left panel of Fig.~\ref{fig:AllPTfixedmAI} is larger than that in the right panel, which implies the multi-step PTs favor the negative values of $\cos(\beta-\alpha)$.
We can also find that the strong 2-step PTs occur only with the mass hierarchy $m_A>m_H$.
The above features are also seen in the Type-I 2HDM with $m_H=m_{H^\pm}$.
Note that the right panel of Fig.~\ref{fig:AllPTfixedmAI} has similar tendencies with the result in the Type-X 2HDM with $m_A=m_{H^\pm}$ as described before.
Taking into account the constraint
from $B\rightarrow \mu\mu$ decays,
the region of $m_A(=m_{H^\pm})\lesssim340$ GeV is excluded.
In the survival parameter space in the left panel of Fig.~\ref{fig:AllPTfixedmAI}, the multi-step PTs occur mostly for $m_H\gsim
300$ GeV.
The region might be tested by the extra Higgs boson search of $A\to HZ$ if $m_A-m_H>m_Z$.


\section{Physical signatures} \label{sec:PhysicalSignature}
\subsection{Higgs trilinear couplings}
In this section, to research the possibility of the verification of the multi-step PT by collider experiments, we discuss
the Higgs trilinear coupling $\lambda_{hhh}$.
The coupling $\lambda_{hhh}$ is derived by calculating the third derivative of the effective potential with respect to the SM-like Higgs fields at the EW vacuum as
\begin{align}
\lambda_{hhh}=\left.\frac{\partial^3 V_{\rm eff}^{T=0}(\phi_1, \phi_2)}{\partial h^3}\right|_{(\phi_1, \phi_2)=(v_1, v_2)},
\end{align}
with $V_{\rm eff}^{T=0}\equiv V_0+V_{\rm CW}+V_{\rm CT}$.
The trilinear coupling corrected by the leading 1-loop contribution of the top quarks in the SM is written by
\begin{align}
\lambda_{hhh}^{\rm SM}\simeq\frac{3m_h^2}{v}\left[1-\frac{N_c}{3\pi^2}\frac{m_t^4}{v^2m_h^2}\right],
\end{align}
where $N_c$ is the color number of the top quarks.
We determine the deviation of the Higgs trilinear coupling from that in the SM as
\begin{align}
\delta\lambda_{hhh}\equiv\frac{\lambda_{hhh}-\lambda_{hhh}^{\rm SM}}{\lambda_{hhh}^{\rm SM}}.
\end{align}
When $\delta\lambda_{hhh}$ is equal to zero, the coupling has the same value as in the SM.
In the following, we analyze $\delta\lambda_{hhh}$ by the Type-I and -X 2HDMs.

The current limits on the Higgs trilinear coupling from Higgs pair production are $-4.2<\delta\lambda_{hhh}<10.9$ (at $95\%\ {\rm CL}$) from ATLAS \cite{ATLAS-CONF-2019-049}.
At the future measurement, like the HL-LHC, the limit could reach an accuracy of about 50--60\% with 3 ab$^{-1}$ data \cite{Cepeda:2019klc},
while the ILC operating at 500 GeV has the possibility to measure $\delta\lambda_{hhh}$ with 27$\%$ of precision \cite{Fujii:2015jha}.


\begin{figure}[t]
\centering
\begin{tabular}{c}

\begin{minipage}{0.5\hsize}
\centering
\includegraphics[clip, width=7.5cm]{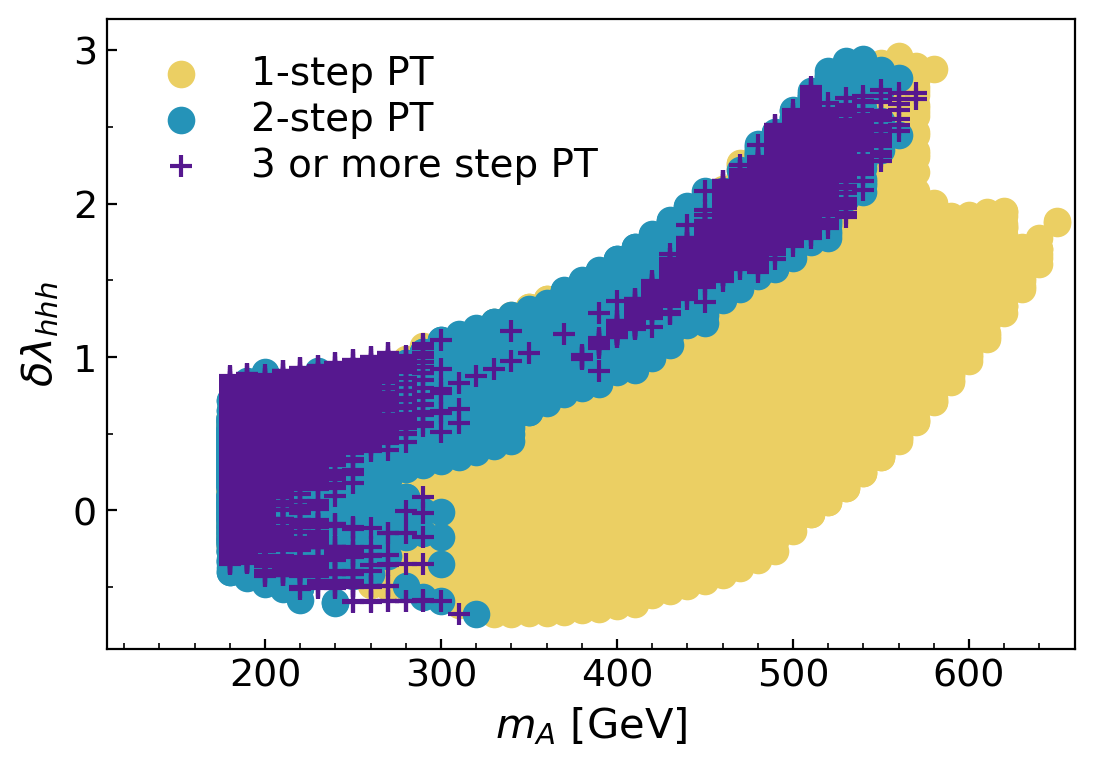}
\end{minipage}

\begin{minipage}{0.5\hsize}
\centering
\includegraphics[clip, width=7.5cm]{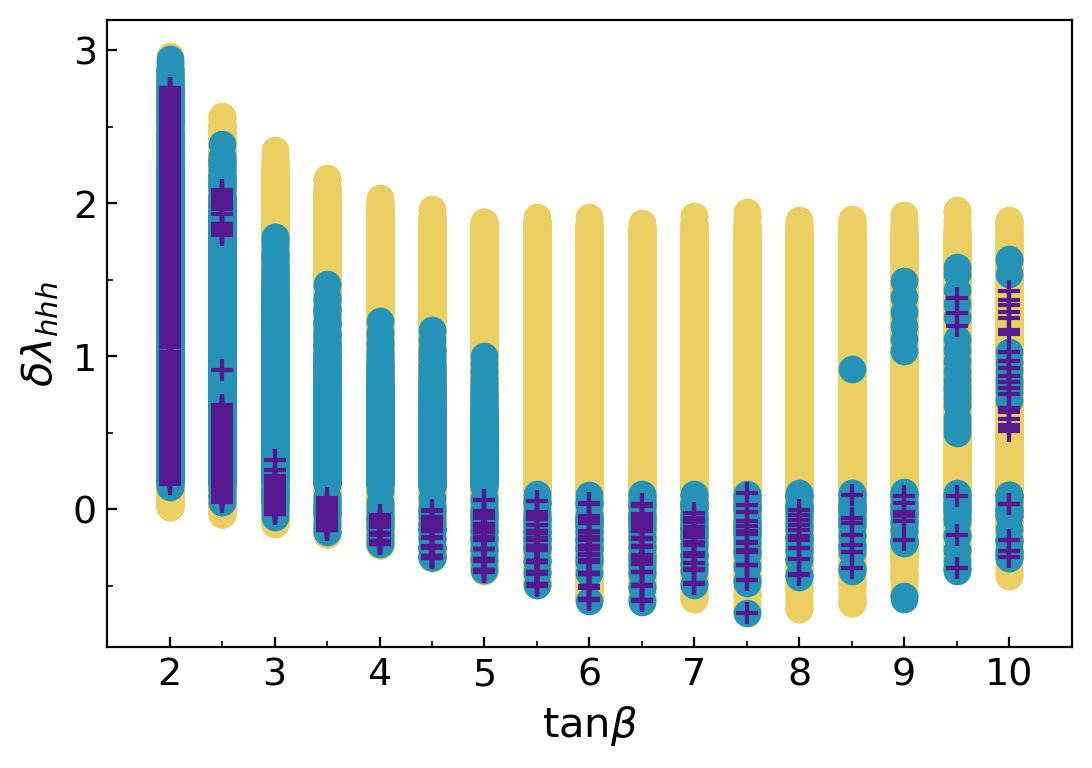}
\end{minipage}\\

\begin{minipage}{0.5\hsize}
\centering
\includegraphics[clip, width=7.5cm]{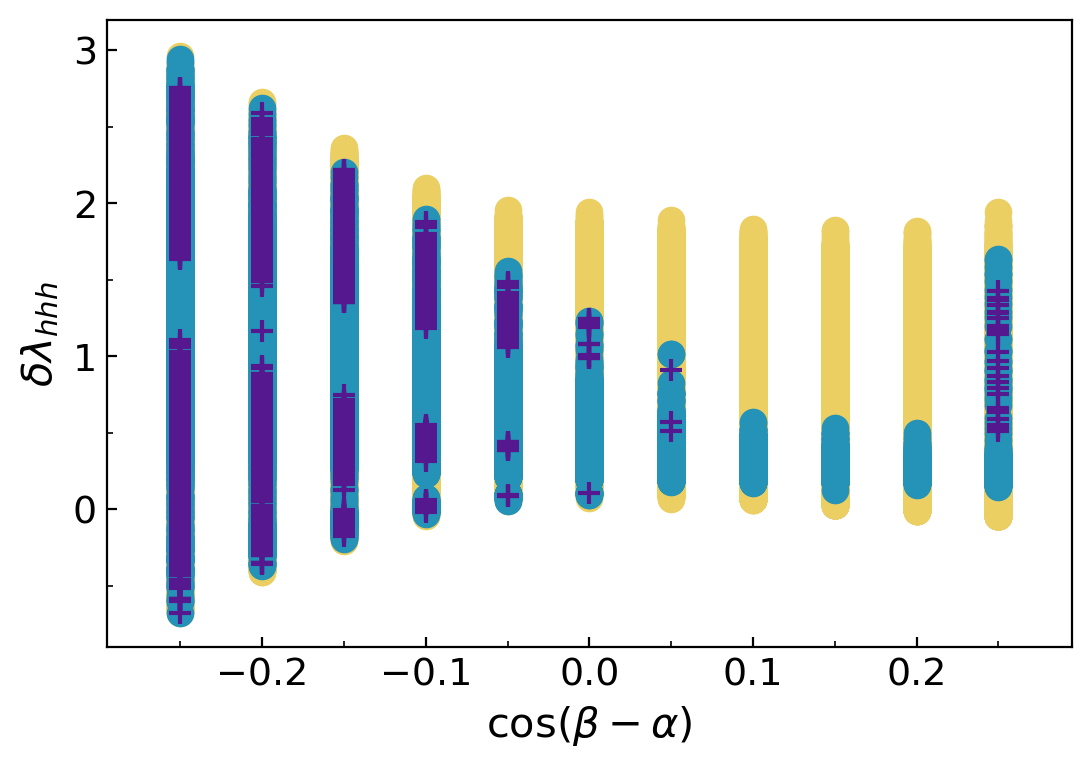}
\end{minipage}

\begin{minipage}{0.5\hsize}
\centering
\includegraphics[clip, width=7.5cm]{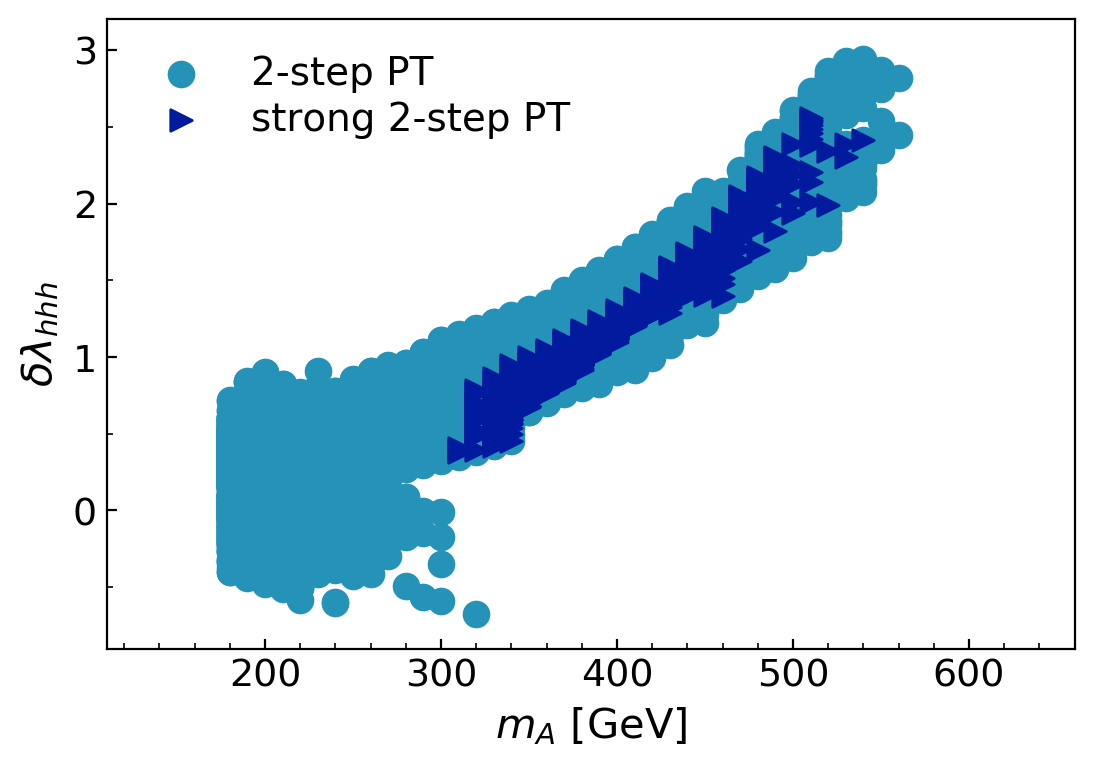}
\end{minipage}

\end{tabular}
\caption{
Predictions for $\delta\lambda_{hhh}$ in the Type-I 2HDM with $m_A=m_{H^\pm}$.
The panels except for the lower right panel show $\delta\lambda_{hhh}$ where the 1-step, 2-step, and 3 or more step PTs occur for $m_A$ (upper left), $\tan\beta$ (upper right), and $\cos(\beta-\alpha)$ (lower left).
The lower right plane shows $\delta\lambda_{hhh}$ where the 2-step and strong 2-step PTs happen for $m_A$.
}
\label{fig:deltalhhhmAI}
\end{figure}

\begin{figure}[t]
\centering
\begin{tabular}{c}

\begin{minipage}{0.5\hsize}
\centering
\includegraphics[clip, width=7.5cm]{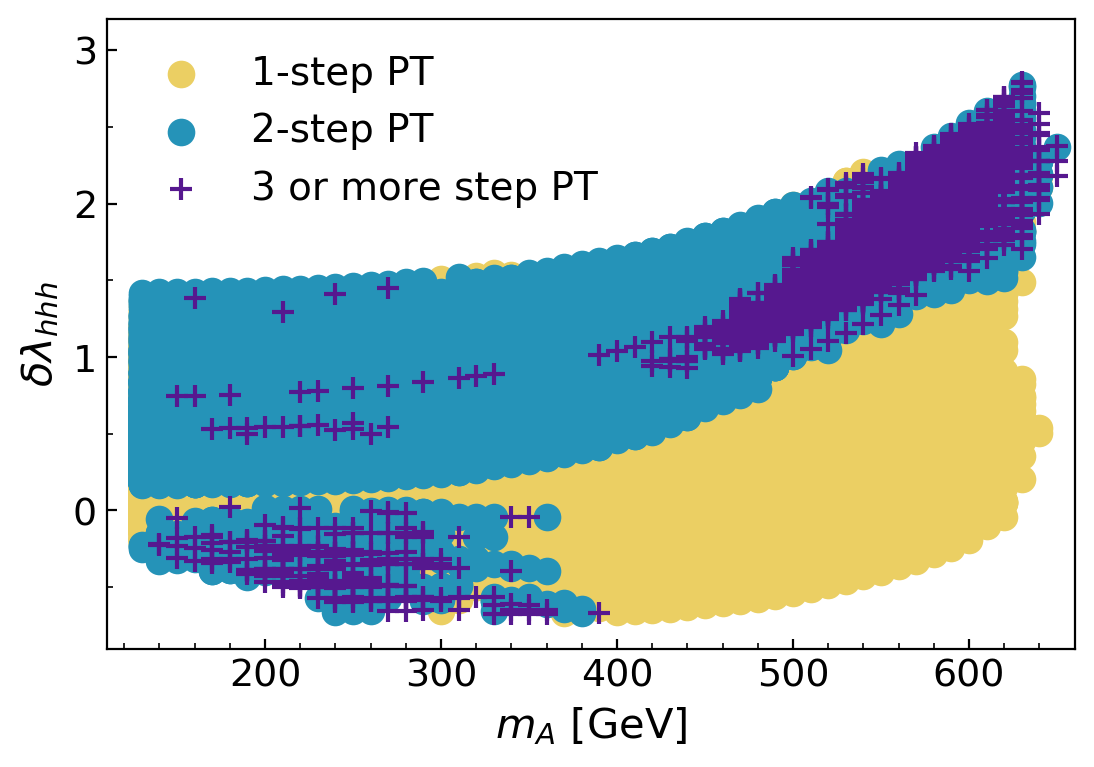}
\end{minipage}

\begin{minipage}{0.5\hsize}
\centering
\includegraphics[clip, width=7.5cm]{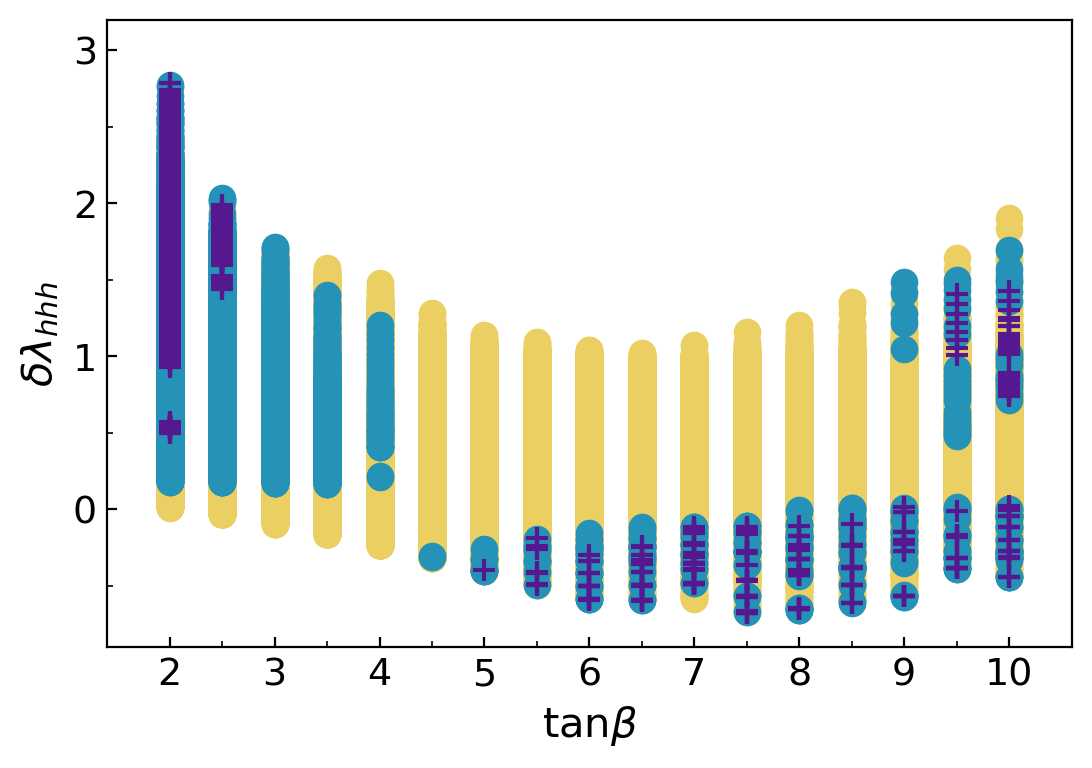}
\end{minipage}\\

\begin{minipage}{0.5\hsize}
\centering
\includegraphics[clip, width=7.5cm]{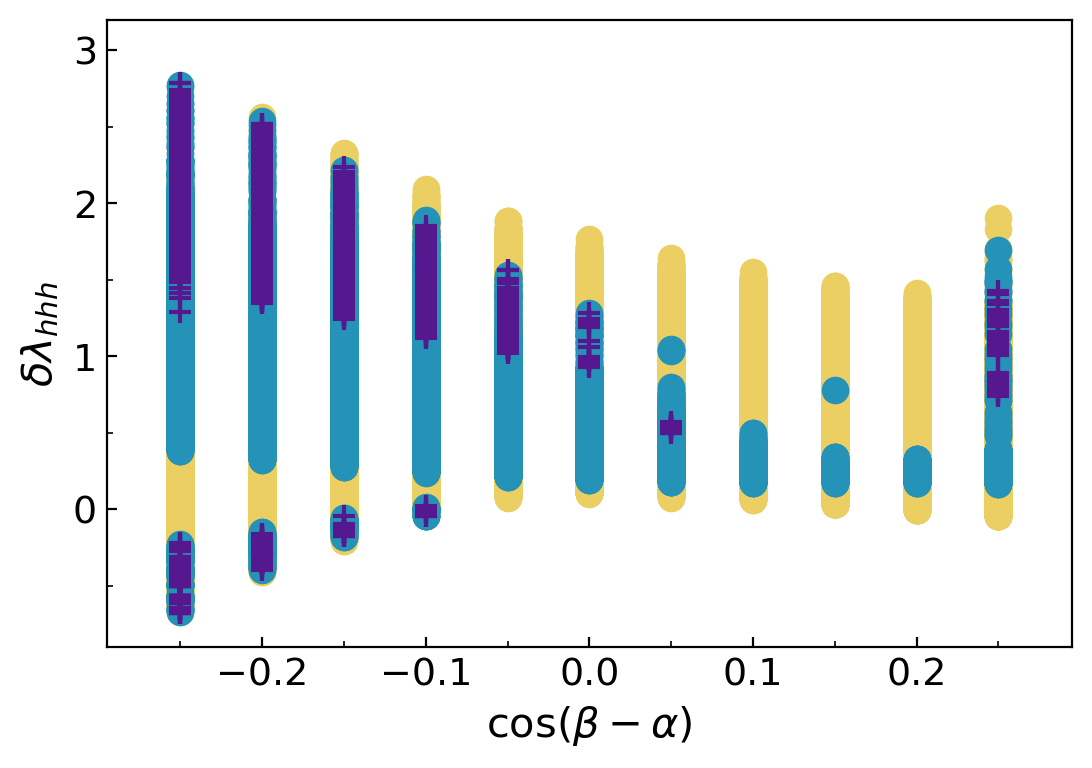}
\end{minipage}

\begin{minipage}{0.5\hsize}
\centering
\includegraphics[clip, width=7.5cm]{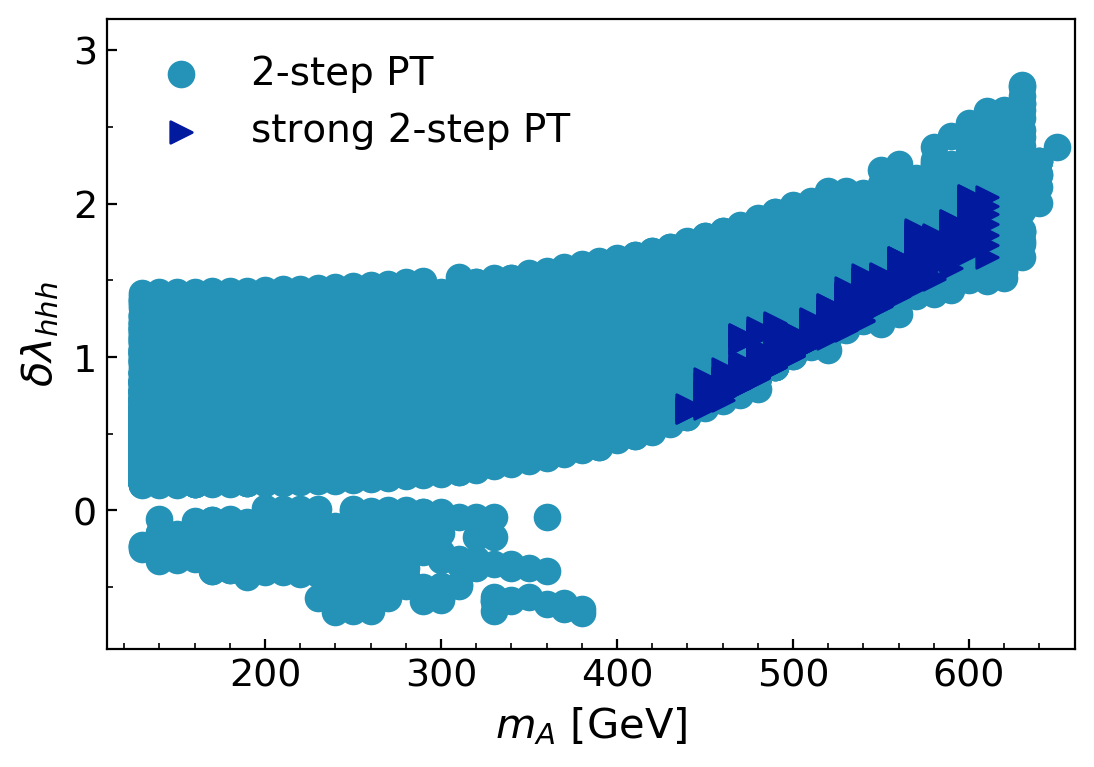}
\end{minipage}

\end{tabular}
\caption{
Same as Fig.~\ref{fig:deltalhhhmAI} but for the Type-I 2HDM with $m_H=m_{H^\pm}$.
}
\label{fig:deltalhhhmHI}
\end{figure}

\begin{figure}[t]
\centering
\begin{tabular}{c}

\begin{minipage}{0.5\hsize}
\centering
\includegraphics[clip, width=7.5cm]{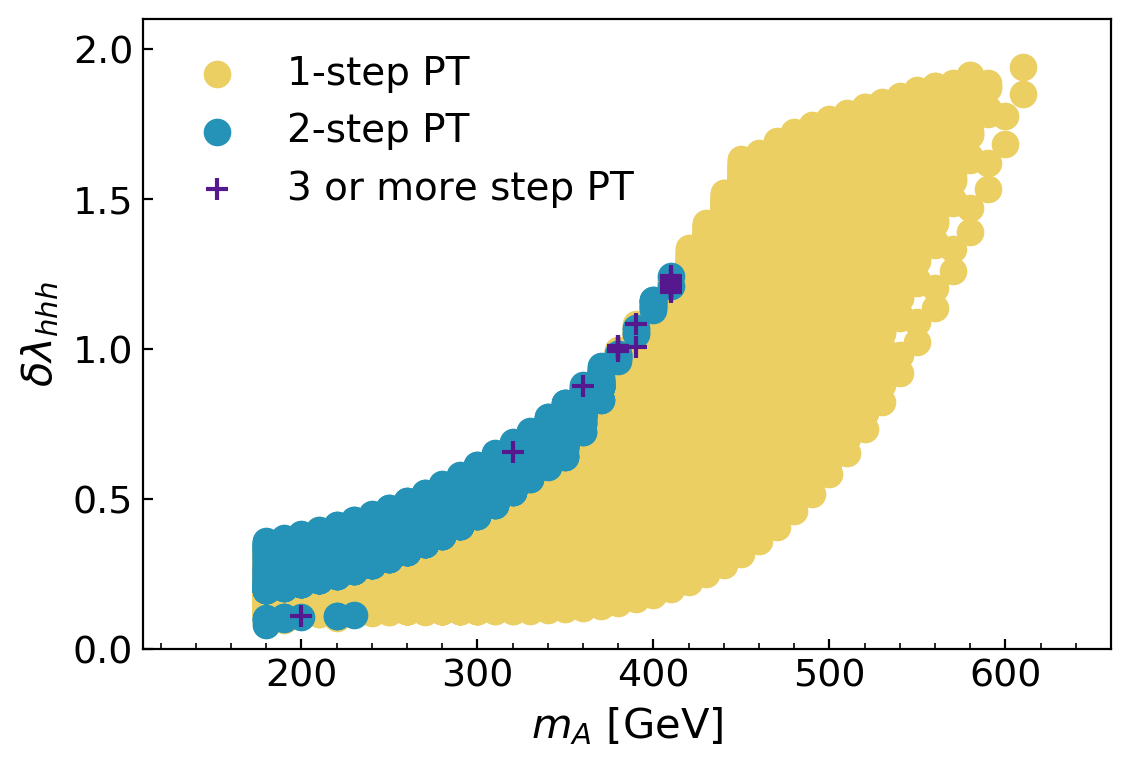}
\end{minipage}

\begin{minipage}{0.5\hsize}
\centering
\includegraphics[clip, width=7.5cm]{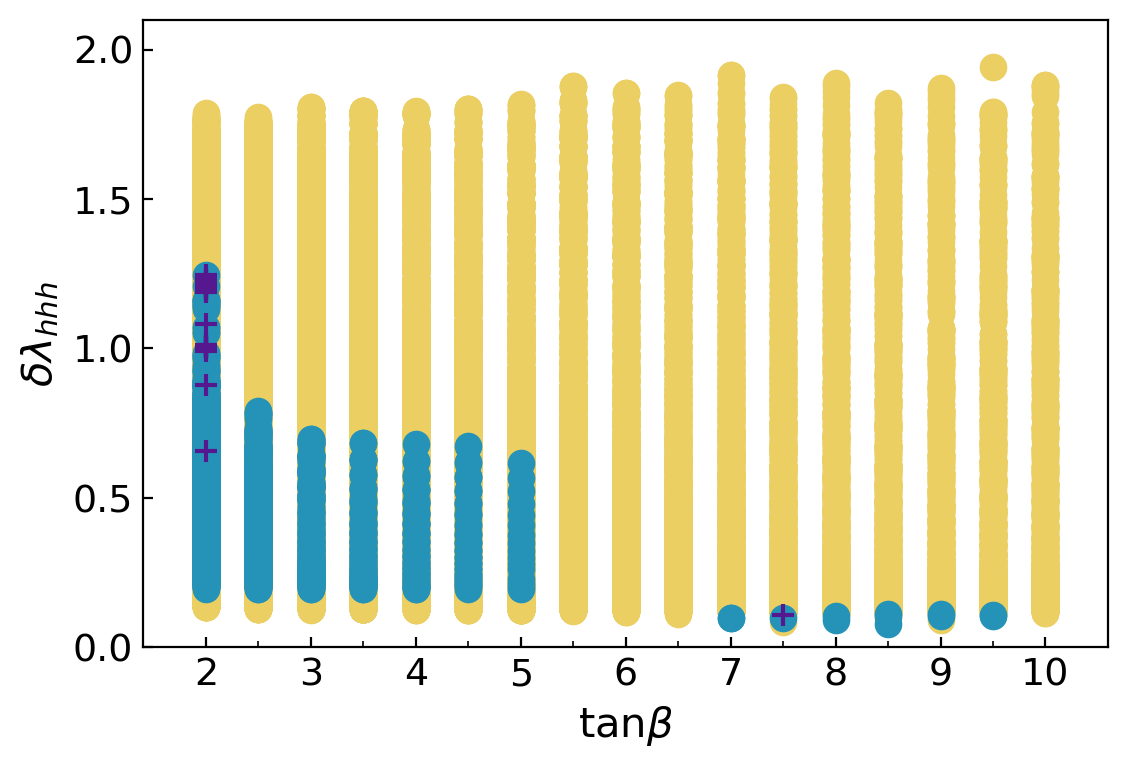}
\end{minipage}\\

\begin{minipage}{0.5\hsize}
\centering
\includegraphics[clip, width=7.5cm]{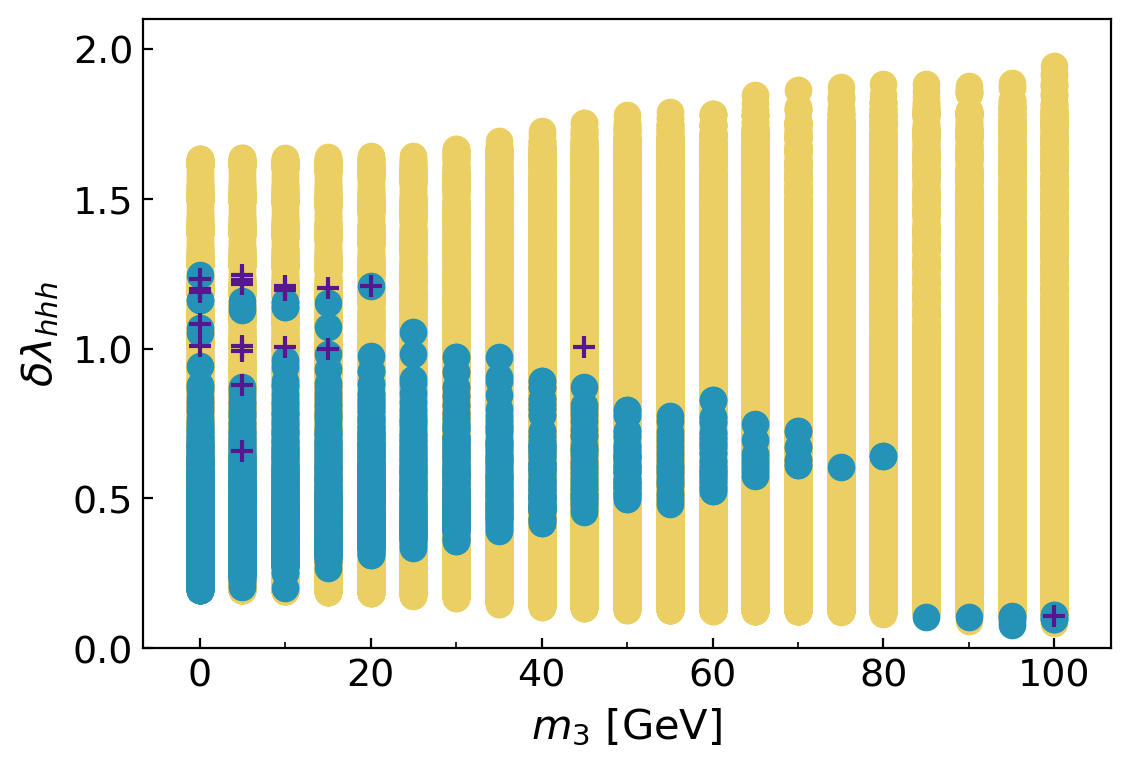}
\end{minipage}

\begin{minipage}{0.5\hsize}
\centering
\includegraphics[clip, width=7.5cm]{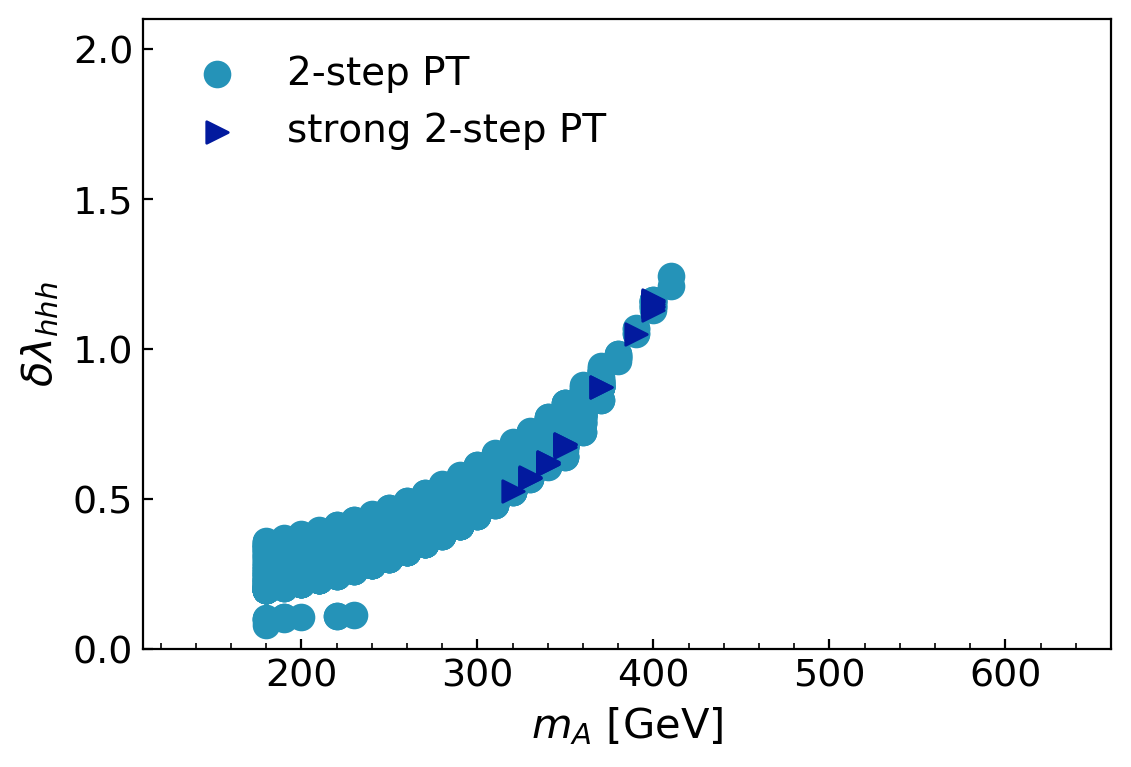}
\end{minipage}

\end{tabular}
\caption{
Predictions for $\delta\lambda_{hhh}$ in the Type-I 2HDM with $m_A=m_{H^\pm}$.
The panels except for the lower right panel show $\delta\lambda_{hhh}$ where the 1-step, 2-step, and 3 or more step PTs occur for $m_A$ (upper left), $\tan\beta$ (upper right), and $m_3$ (lower left).
The lower right plane shows $\delta\lambda_{hhh}$ where the 2-step and strong 2-step PTs happen for $m_A$.
}
\label{fig:deltalhhhmAX}
\end{figure}

\begin{figure}[t]
\centering
\begin{tabular}{c}

\begin{minipage}{0.5\hsize}
\centering
\includegraphics[clip, width=7.5cm]{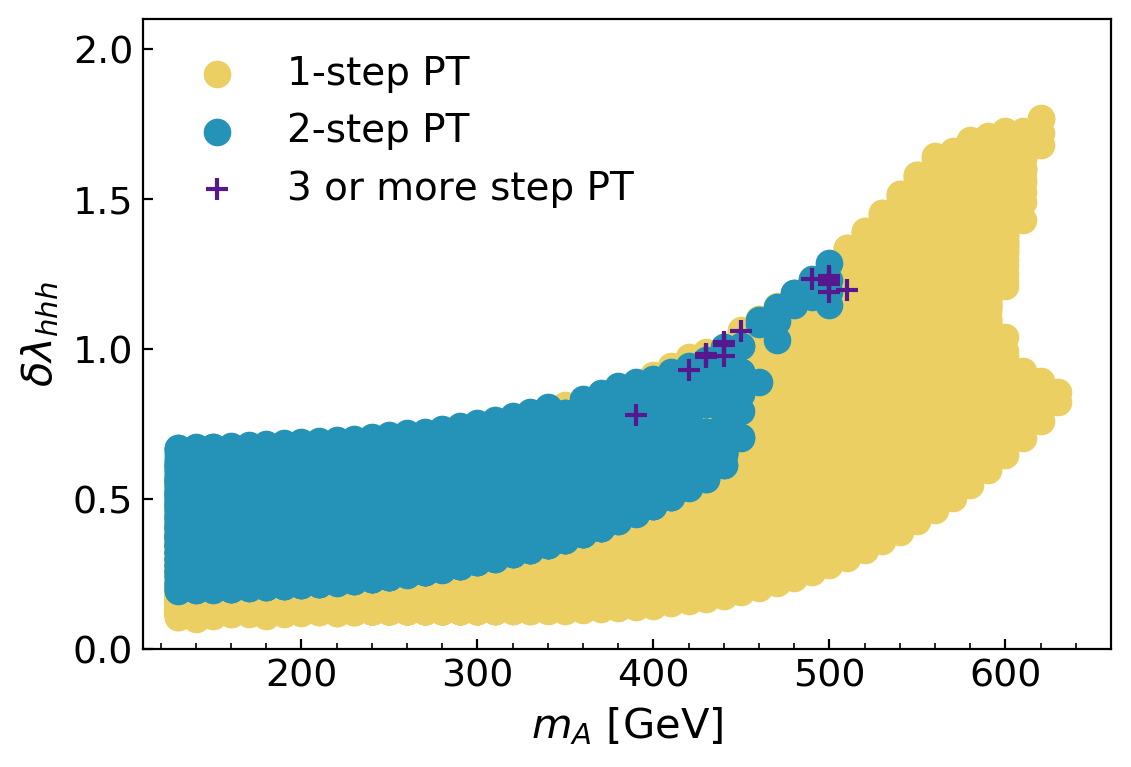}
\end{minipage}

\begin{minipage}{0.5\hsize}
\centering
\includegraphics[clip, width=7.5cm]{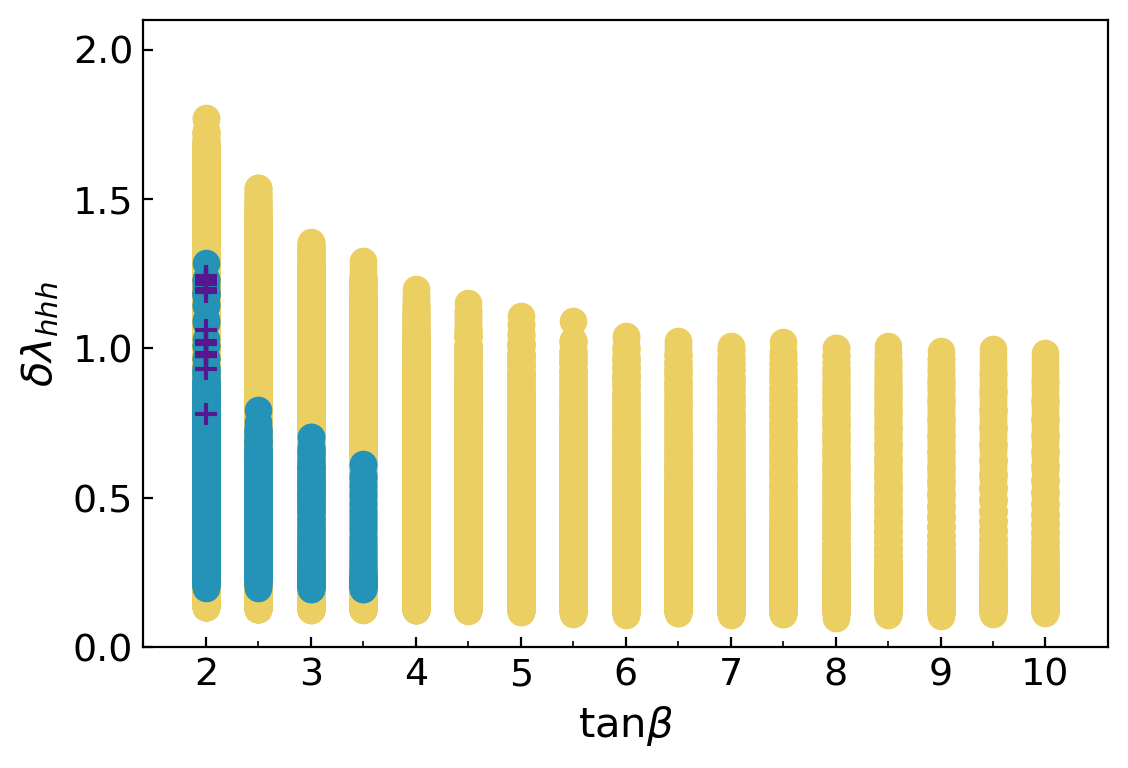}
\end{minipage}\\

\begin{minipage}{0.5\hsize}
\centering
\includegraphics[clip, width=7.5cm]{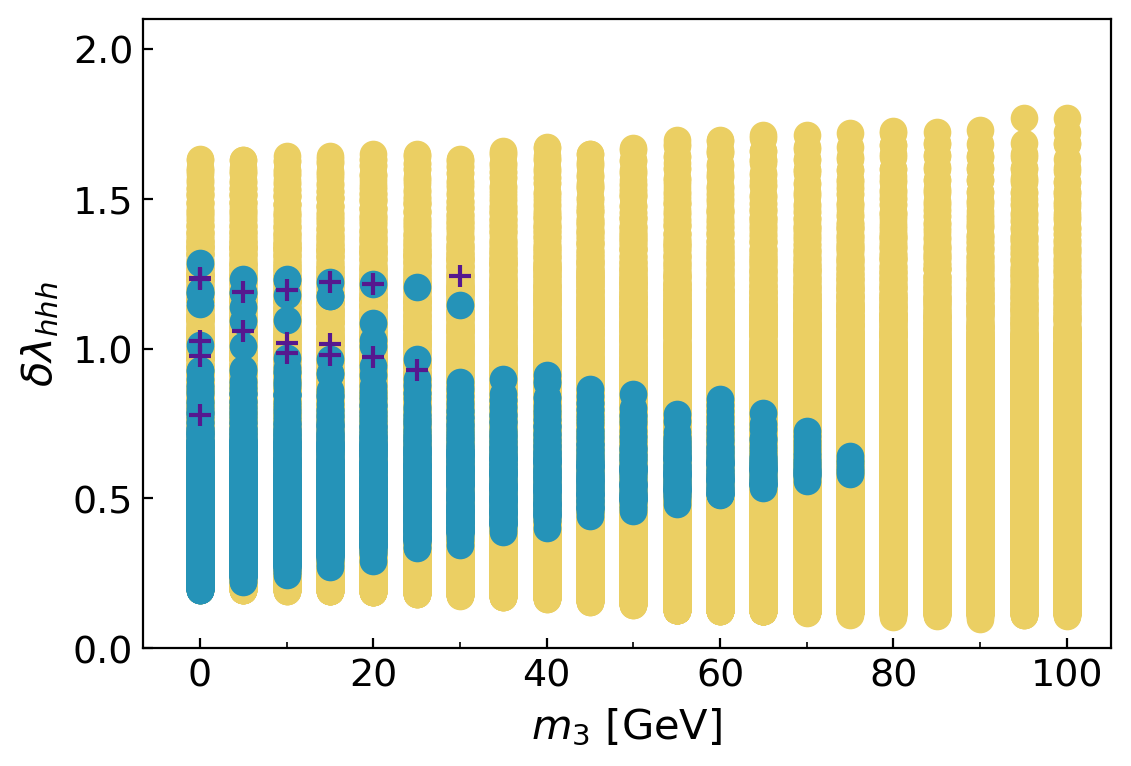}
\end{minipage}

\begin{minipage}{0.5\hsize}
\centering
\includegraphics[clip, width=7.5cm]{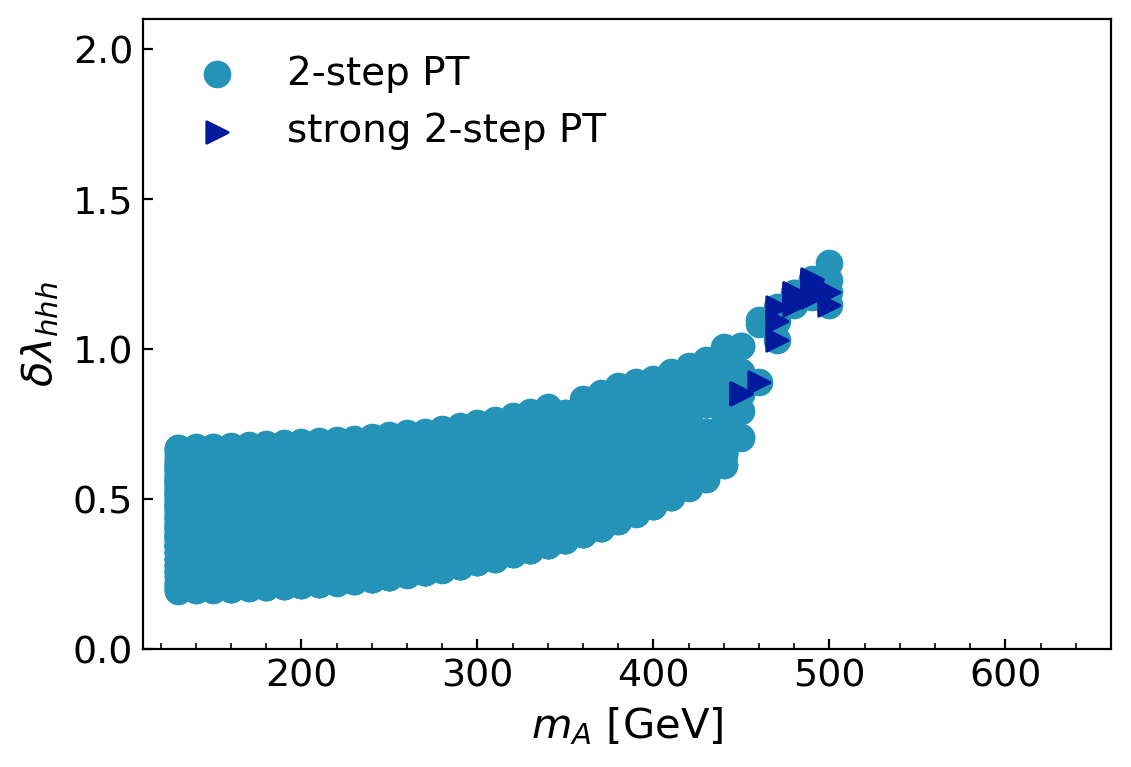}
\end{minipage}

\end{tabular}
\caption{
Same as Fig.~\ref{fig:deltalhhhmAX} but for the Type-X 2HDM with $m_H=m_{H^\pm}$.
}
\label{fig:deltalhhhmHX}
\end{figure}

\begin{figure}[t]
 \centering
 \begin{tabular}{c}

 \begin{minipage}{0.5\hsize}
 \centering
 \includegraphics[clip, width=7.5cm]{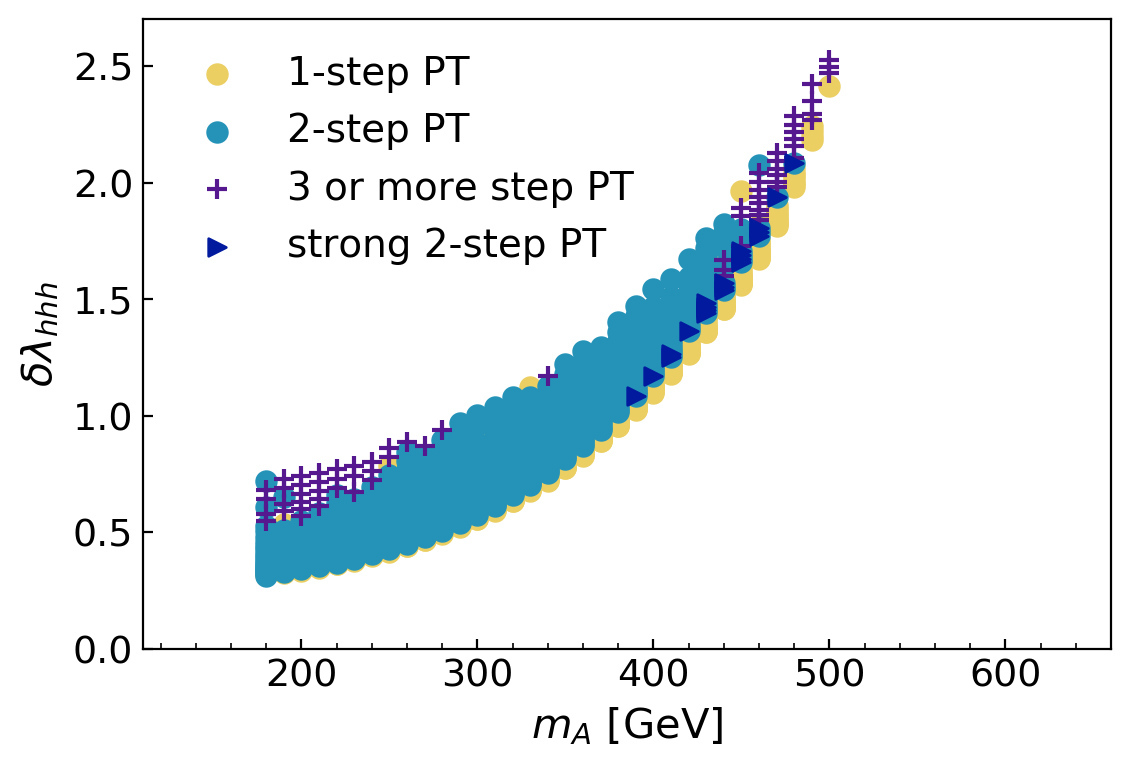}
 \end{minipage}

 \begin{minipage}{0.5\hsize}
 \centering
 \includegraphics[clip, width=7.5cm]{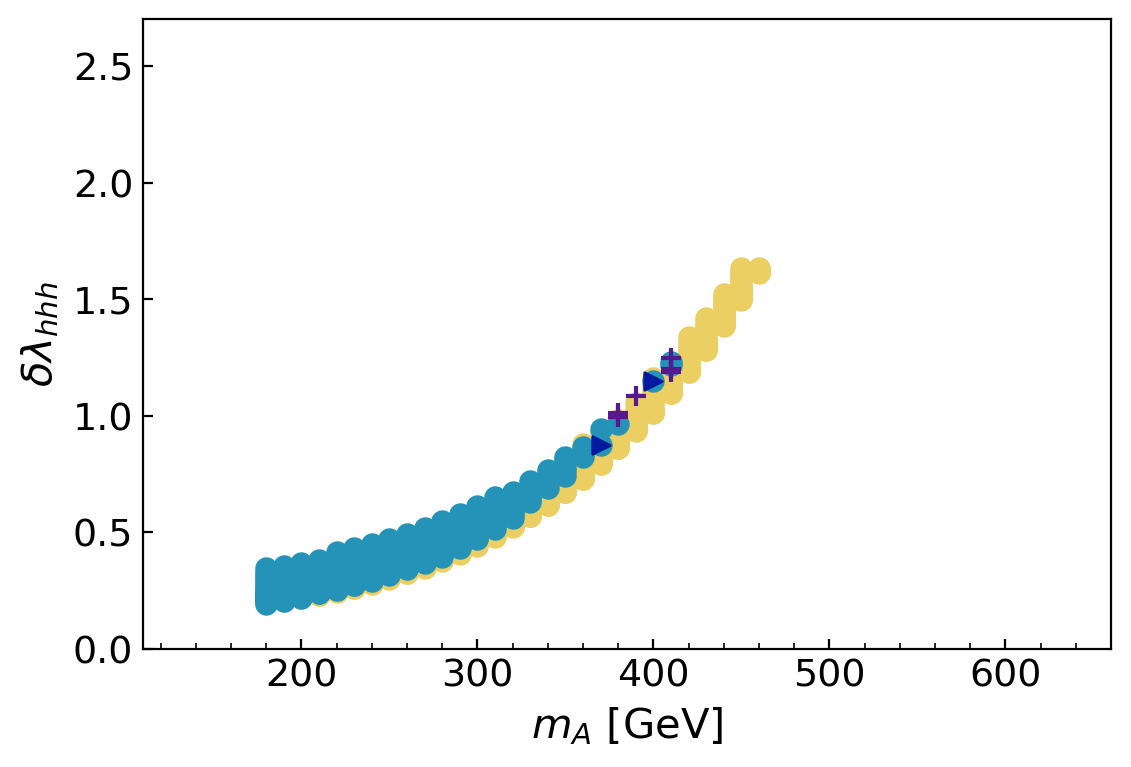}
 \end{minipage}

 \end{tabular}
 \caption{
 Predictions for $\delta\lambda_{hhh}$ where the 1-step, 2-step, strong 2-step, and 3 or more step PTs occur for $m_A$ in the Type-I 2HDM with $m_A=m_{H^\pm}$.
 The left (right) panel shows $\delta\lambda_{hhh}$ at $\tan\beta=2$, $\cos(\beta-\alpha)=-0.2$ ($0$) and $m_3=0$.
 }
 \label{fig:mAdeltalhhhAllPTfixedmAI}
 \end{figure}


\subsubsection{Type-I $(m_A=m_{H^\pm})$}
Fig.~\ref{fig:deltalhhhmAI} shows $\delta\lambda_{hhh}$ in the region where the 1-step, 2-step, and 3 or more step PTs occur in the Type-I 2HDM with $m_A=m_{H^\pm}$ as a function of $m_A$ (upper left), $\tan\beta$ (upper right), and $\cos(\beta-\alpha)$ (lower left).
In the upper left panel, the parameter points where the multi-step PTs happen are located on the upper side of the plots in the region $m_A\gtrsim300$ GeV.
In other words, compared with the results of the 1-step PTs, the values of $\delta\lambda_{hhh}$ for the multi-step PTs have a tendency to be large at the same value of $m_A$.
The upper right panel shows the smaller $\tan\beta$ is, the larger the maximum value of $\delta\lambda_{hhh}$ is for $\tan\beta\lesssim8$.
In the lower left panel, when the multi-step PTs occur,
the maximum value of $\delta\lambda_{hhh}$ becomes larger as $\cos(\beta-\alpha)$ gets smaller except for $\cos(\beta-\alpha)\simeq0.25$, especially in the negative values of $\cos(\beta-\alpha)$.
Note that there are parameter points that have the negative deviations.
They are in $|\cos(\beta-\alpha)|\gtrsim 0.1$ and $m_3\gtrsim50\ {\rm GeV}$ found out by our analysis.
The region where $\delta\lambda_{hhh}\simeq2.5$ for the multi-step PTs is 490 GeV $\lesssim m_A\lesssim$ 560 GeV, $\tan\beta\simeq2$, and $\cos(\beta-\alpha)\lesssim-0.2$ in our parameter space.
In the region,
we find that $m_H$ is in 320 GeV $\lesssim m_H\lesssim$ 410 GeV.

The lower right panel of Fig.~\ref{fig:deltalhhhmAI} represents $\delta\lambda_{hhh}$ for the 2-step and strong 2-step PTs as a function of $m_A$.
It indicates $\delta\lambda_{hhh}$ have the possibility of being large as 0.5--2.5 when the strong 2-step PTs happen.
When $\delta\lambda_{hhh}$ is 2.5 with the strong 2-step PTs, we have found $m_A\simeq510$ GeV, $m_H\simeq320$ GeV, $\tan\beta\simeq2$, and $\cos(\beta-\alpha)\simeq-0.25$.
The deviations $\delta\lambda_{hhh}\simeq$0.5--2.5 would be tested at future colliders such as the HL-LHC and the ILC.

\subsubsection{Type-I $(m_H=m_{H^\pm})$}
The deviations $\delta\lambda_{hhh}$ for the Type-I 2HDM with $m_H=m_{H^\pm}$ are shown in Fig.~\ref{fig:deltalhhhmHI}.
In the upper left panel,
the parameter points where the multi-step PTs occur are located on the upper side in $m_A\gtrsim400$ GeV.
The behavior of the predictions for the multi-step PTs in Fig.~\ref{fig:deltalhhhmHI} are similar to the ones in the Type-I 2HDM with $m_A=m_{H^\pm}$.
From panels except for the lower right panel in Fig.~\ref{fig:deltalhhhmHI}, we find that the region where the multi-step PTs occur with $\delta\lambda_{hhh}\simeq2.5$ is 590 GeV $\lesssim m_A\lesssim$ 640 GeV, $\tan\beta\simeq2$, and $\cos(\beta-\alpha)\lesssim-0.2$.
In the region, the range of $m_H$ is 370 GeV $\lesssim m_H\lesssim$ 420 GeV.
Additionally, in the lower right panel of Fig.~\ref{fig:deltalhhhmHI}, the range of $\delta\lambda_{hhh}$ for the strong 2-step PTs occur is about 0.5--2.0, while the largest value of such $\delta\lambda_{hhh}$ is slightly smaller than that in the Type-I 2HDM with $m_A=m_{H^\pm}$.
When $\delta\lambda_{hhh}$ is 2 with the strong 2-step PTs, we have found 600 GeV $\lesssim m_A\lesssim$ 610 GeV, 300 GeV $\lesssim m_H\lesssim$ 360 GeV, $\tan\beta\simeq2$, and $\cos(\beta-\alpha)\lesssim-0.15$.
Although the constraint from $B\rightarrow\mu^+\mu^-$ excludes the part of the above region (especially the region with $m_H\simeq330$--340 GeV as described in Section~\ref{sec:AnalysismHI}), it still remains and such $\delta\lambda_{hhh}$ would be tested at the future collider experiments.

\subsubsection{Type-X $(m_A=m_{H^\pm})$}
Fig.~\ref{fig:deltalhhhmAX} gives $\delta\lambda_{hhh}$ in the region where the 1-step, 2-step, and 3 or more step PTs occur in the Type-X 2HDM with $m_A=m_{H^\pm}$ as a function of $m_A$ (upper left), $\tan\beta$ (upper right), and $m_3$ (lower left).
We take $\cos(\beta-\alpha)=0$ (alignment limit) in the Type-X 2HDMs.
In the upper left panel, the shape of the region where the multi-step PTs occur is narrow, hence the value of $\delta\lambda_{hhh}$ is predictable when $m_A$ is fixed.
The largest value of $\delta\lambda_{hhh}$ for the multi-step PTs is about 1.2
at $m_A\simeq 400$ GeV, $\tan\beta\simeq2$, and $m_3\lesssim 20$ GeV, and 310 GeV $\lesssim m_H \lesssim$ 340 GeV.
Although the largest value is smaller than that in the Type-I 2HDMs, it can be accessed at the future collider experiments.
The lower left panel shows that the values of $\delta\lambda_{hhh}$ for the multi-step PTs converge to around 0.7 as $m_3$ gets larger.
Such dependence of $\delta\lambda_{hhh}$ on $m_3$ is not seen in the Type-I 2HDMs.
Moreover,
the predicted values of $\delta\lambda_{hhh}$ stay positive in all regions, although the negative $\delta\lambda_{hhh}$ are also predicted
in the Type-I 2HDMs.
These differences between the Type-I and Type-X are mainly due to the range of $\cos(\beta-\alpha)$.
The lower right panel in Fig.~\ref{fig:deltalhhhmAX} shows that the range where the strong 2-step PTs occur is $0.5\lesssim\delta\lambda_{hhh}\lesssim1.2$.
When $\delta\lambda_{hhh}$ is 1.2 with the strong 2-step PTs, we have found $m_A\simeq400$ GeV, $m_H\simeq340$ GeV, and $\tan\beta\simeq2$.
The deviations $\delta\lambda_{hhh}\simeq$0.5--1.2 would be explored at the future colliders like the HL-LHC and the ILC.

\subsubsection{Type-X $(m_H=m_{H^\pm})$}
The deviations $\delta\lambda_{hhh}$ in the Type-X 2HDM with $m_H=m_{H^\pm}$ are shown in Fig.~\ref{fig:deltalhhhmHX}.
We obtain similar features to the ones in the Type-X 2HDM with $m_A=m_{H^\pm}$, except that the region where the multi-step PTs occur gets broad upward.
From Fig.~\ref{fig:deltalhhhmHX} except for the lower right panel (also from the top left panel in Fig.~\ref{fig:AllPTmHX}), we see that $\delta\lambda_{hhh}\simeq 1.2$ for the multi-step PTs
is predicted for 480 GeV $\lesssim m_A\lesssim$ 510 GeV, $\tan\beta\simeq2$, $m_3\lesssim$ 30 GeV, and
310 GeV $\lesssim m_H\lesssim$ 340 GeV.
However, this region is excluded by the constraint from $B\rightarrow\mu^+\mu^-$ as described in Section~\ref{sec:AnalysismHX}.
Nevertheless, when $m_A \simeq 400$ GeV, the maximum value of $\delta\lambda_{hhh}\simeq 0.9$ is allowed by the constraint,
where the other parameters are
$\tan\beta\simeq 2$, $m_3\lesssim$ 40 GeV, and $m_H\simeq$ 350 GeV.
Such a value of $\delta\lambda_{hhh}$ can be tested at the future collider experiments.
In the lower right panel in Fig.~\ref{fig:deltalhhhmHX}, we see that the strong 2-step PTs give $\delta\lambda_{hhh}\simeq1.2$ around $m_A\simeq 490$ GeV.
However, all regions where the strong 2-step PTs occur is excluded by the constraint from $B\rightarrow\mu^+\mu^-$ as mentioned in Section~\ref{sec:AnalysismHX}.

\vspace{0.3in}

Fig.~\ref{fig:mAdeltalhhhAllPTfixedmAI} shows the predictions for $\delta\lambda_{hhh}$ in the two cases in Fig.~\ref{fig:AllPTfixedmAI}.
In the left (right) panel of Fig.~\ref{fig:mAdeltalhhhAllPTfixedmAI}, we take $\tan\beta=2$, $\cos(\beta-\alpha)=-0.2$ $(0)$,
and $m_3=0$ in the Type-I 2HDM with $m_A=m_{H^\pm}$.
Compared with the same value of $m_A$, $\delta\lambda_{hhh}$ for the multi-step PTs have a tendency to be larger than that for the 1-step PTs.
We can also see the largest value of $\delta\lambda_{hhh}$ where the multi-step PTs occur at $\cos(\beta-\alpha)=-0.2$ (left) is greater than that at $\cos(\beta-\alpha)=0$ (right).
Meanwhile, $\delta\lambda_{hhh}$ for the strong 2-step PTs are relatively large as about $\delta\lambda_{hhh}\simeq1$--2 (left)
and $\simeq 1$ (right), respectively.
The regions of the strong 2-step PTs do not receive the constraint from $B\rightarrow \mu\mu$ decays since $m_A(=m_{H^\pm})>340$ GeV.

\subsection{Gravitational waves from multi-step PT}
The first order PT at the EW scale is the source of GW whose typical spectrum has a peak frequency.
Therefore if the first order PT occurs multiple times in a multi-step EWPT, the multi-peaked GW can be observed in the space-based interferometers.
In this subsection, we study such a possibility in the case of the 2-step PT.

The GW spectrum is characterized by two parameters $\alpha_{\rm GW}$ and $\tilde \beta_{\rm GW}$ at the nuclear temperature $T_n$.
Here $T_n$
is determined by the condition that one bubble nucleates per Hubble radius
$S_3/T_n\simeq140$ where $S_3$ is the O(3) symmetric action.
The $\alpha_{\rm GW}$ is given by
\begin{align}
&\alpha_{\rm GW}\equiv\frac{\epsilon(T_n)}{\rho_{\rm rad}(T_n)}\,,
\end{align}
which is the ratio of the latent heat $\epsilon (T_n)$ to the radiation density $\rho_{\rm rad}(T_n)=g_*(\pi^2T_n^4)/30$ where $g_*$ is 110.75 in the 2HDMs.
The latent heat in the first order PT is calculated as
\begin{align}
&\epsilon(T_n)=\left.\left[-\Delta V+T\frac{\partial \Delta V}{\partial T}
\right]\right|_{T=T_n},
\end{align}
where $\Delta V$ is the difference between the effective potential of two phases before and after the PT.
On the other hand, $\tilde \beta_{\rm GW}$ is defined as $\tilde \beta_{\rm GW}\equiv \beta_{\rm GW}/H_n$, where
$H_n$ is the Hubble parameter at $T_n$ and $\beta_{\rm GW}$ is
the inverse time duration of the PT
\begin{align}
\beta_{\rm GW}\equiv
H_nT_n\left.\frac{d}{dT}\left(\frac{S_3(T)}{T}\right)\right|_{T=T_n}\, .
\end{align}

There are three contributions to the GW spectrum at a first order PT:
\al{
h ^2\Omega_\text{GW} (f) =
h ^2\Omega_\varphi(f) +
h ^2\Omega_\text{sw} (f) +h ^2\Omega_\text{turb}(f)\,.
\label{GW-Omega}
}
Here $h$ is the dimensionless Hubble parameter,
$f$ is the frequency of the GW at present,
$\Omega_\varphi$ is the scalar field contribution from collisions of bubble walls
\cite{Kosowsky:1991ua,Kosowsky:1992rz,Kosowsky:1992vn,Kamionkowski:1993fg,Caprini:2007xq,Huber:2008hg},
$\Omega_{\rm sw}$
 is the contribution from sound waves surrounding the bubble walls \cite{Hindmarsh:2013xza,Giblin:2013kea,Giblin:2014qia,Hindmarsh:2015qta}
and $\Omega_{\rm turb}$ is the contribution from
magnetohydrodynamic (MHD) turbulence in plasma
 \cite{Caprini:2006jb,Kahniashvili:2008pf,Kahniashvili:2008pe,Kahniashvili:2009mf,Caprini:2009yp,Binetruy:2012ze}.
Each contribution is given by $\alpha_{\rm GW}$ and $\tilde \beta_{\rm GW}$
with the velocity of bubble wall $v_w$ and the $\kappa_\varphi$, $\kappa_{\mathrm{sw}}$, and $\kappa_{\rm turb}$ which are the fraction of vacuum energy, respectively, converted into gradient energy of scalar field, bulk motion of the fluid, and MHD turbulence.
Numerical simulations and analytic estimates of the individual contributions lead to the following formula:

\begin{itemize}
\setlength{\leftskip}{-5mm}
\item Scalar field contribution $\Omega_{\varphi}$~\cite{Huber:2008hg} :
\begin{align}
\hspace{-10mm}
h^2\,\Omega_{\varphi}(f)=1.67\times 10^{-5}
\tilde\beta_{\rm GW}^{-2}
\left(\frac{\kappa_{\varphi} \alpha_{\rm GW}}{1+\alpha_{\rm GW}}\right)^2
\left(\frac{100}{g_{*}}\right)^{1/3}\left(\frac{0.11v^3_{w}}{0.42+v_w^2 }\right)\frac{3.8(f/f_{\varphi})^{2.8}}{1+2.8(f/f_{\varphi})^{3.8}},
\end{align}
where
the peak frequency is
\begin{align}
f_{\varphi} = 16.5 \times 10^{-6}
\tilde\beta_{\rm GW}
\left(\frac{0.62}{1.8-0.1 v_w+v^2_w }\right)\left( \frac{T_n}{100~\mathrm{GeV}}\right)\left( \frac{g_{*}}{100}\right)^{1/6}~\mathrm{Hz}.
\end{align}
\item Sound-wave contribution $\Omega_{\mathrm{sw}}$~\cite{Hindmarsh:2015qta} :
\begin{align}
\hspace{-10mm}
h^2\,\Omega_{\mathrm{sw}}(f)=
2.65\times 10^{-6}
\tilde\beta_{\rm GW}^{-1}
\left(\frac{\kappa_\text{sw} \alpha_{\rm GW}}{1+\alpha_{\rm GW}}\right)^2\left(\frac{100}{g_{*}}\right)^{1/3} v_{w}(f/f_{\mathrm{sw}})^3\left(\frac{7}{4+3(f/f_{\mathrm{sw}})^2}\right)^{7/2},
\end{align}
where the peak frequency is
\begin{align}
 f_{\mathrm{sw}} = 1.9 \times 10^{-5} v_w ^{-1}
\tilde\beta_{\rm GW}
 \left( \frac{T_n}{100~\mathrm{GeV}}\right)\left( \frac{g_{*}}{100}\right)^{1/6}~\mathrm{Hz}.
\end{align}
\item MHD turbulence contribution $\Omega_{\mathrm{turb}}$~\cite{Caprini:2009yp,Binetruy:2012ze} :
\begin{align}
\hspace{-10mm}
h^2\,\Omega_{\mathrm{turb}}(f)=
3.35\times 10^{-4}
\tilde\beta_{\rm GW}^{-1}
\left(\frac{\kappa_\text{turb}\alpha_{\rm GW}}{1+\alpha_{\rm GW}}\right)^{\frac{3}{2}}\left(\frac{100}{g_{*}}\right)^{1/3}
v_{w}\frac{(f/f_{\mathrm{turb}})^3}{[
1+(f/f_{\mathrm{turb}})]^{\frac{11}{3}}(1+8\pi f/h_n)}
\end{align}
where the peak frequency is
\begin{align}
f_{\mathrm{turb}} = 2.7 \times 10^{-5} v_w ^{-1}
\tilde\beta_{\rm GW}
 \left( \frac{T_n}{100~\mathrm{GeV}}\right)\left( \frac{g_{*}}{100}\right)^{1/6}~\mathrm{Hz}\,,
\end{align}
and
\begin{align}
h_n = 1.65 \times 10^{-5} \left( \frac{T_n}{100~\mathrm{GeV}}\right)\left( \frac{g_{*}}{100}\right)^{1/6}~\mathrm{Hz}\,.
\end{align}
\end{itemize}
We assume the bubble wall velocity as $v_w=1$ for simplicity and set \cite{Kamionkowski:1993fg,Espinosa:2010hh}
\begin{align}
&\kappa_\varphi \simeq \frac{1}{1+0.715\alpha_{\rm GW}}\left(0.715\alpha_{\rm GW}+\frac{4}{27}\sqrt{\frac{3\alpha_{\rm GW}}{2}}\right),\\
&\kappa_{\mathrm{sw}} \simeq \frac{\alpha}{0.73+0.083\sqrt{\alpha_{\rm GW}}+\alpha_{\rm GW}},
\end{align}
and $\kappa_{\rm turb} \approx 0.1 \kappa_{\mathrm{sw}}$~\cite{Hindmarsh:2015qta}.


\begin{figure}[t]
\centering
\begin{tabular}{c}

\begin{minipage}{0.5\hsize}
\centering
\includegraphics[clip, width=7.5cm]{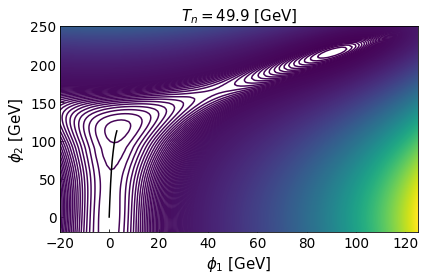}
\end{minipage}

\begin{minipage}{0.5\hsize}
\centering
\includegraphics[clip, width=7.5cm]{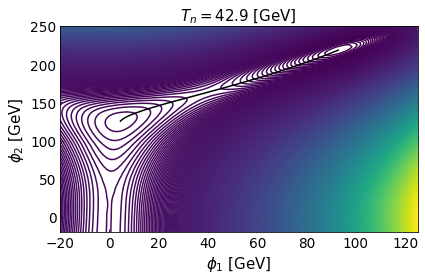}
\end{minipage}

\end{tabular}
\caption{
Left (Right) : Contour plots of the effective potential and the path of the first (second) step PT at $T_n=49.9$ (42.9) GeV for the benchmark point.
}
\label{fig:2-stepPTcausingGW}
\end{figure}

\begin{figure}[t]
\centering
\begin{tabular}{c}

\begin{minipage}{1\hsize}
\centering
\includegraphics[clip, width=9cm]{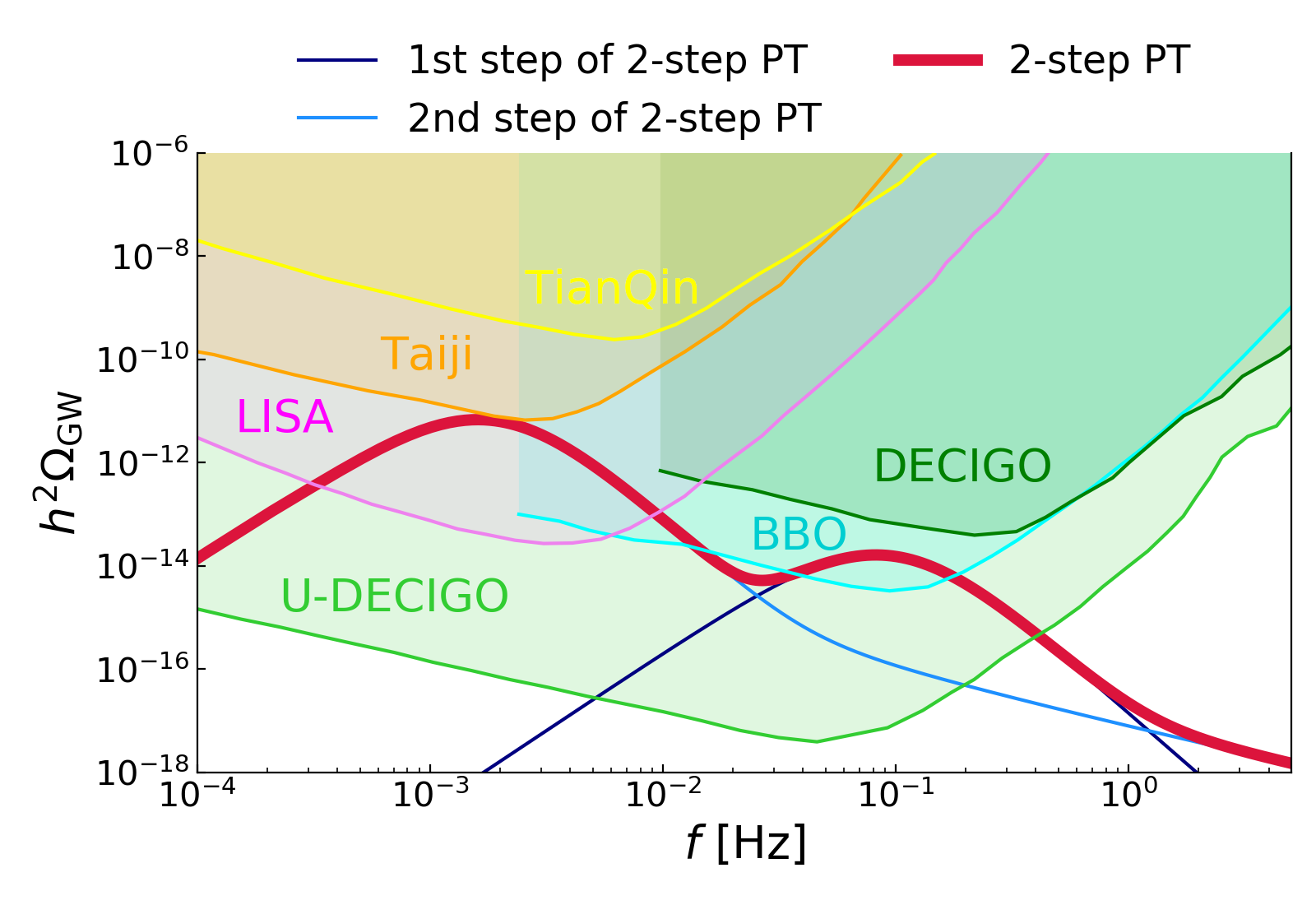}
\end{minipage}

\end{tabular}
\caption{
GW spectrums from the first and second step of the strong 2-step PT for the benchmark point.
The navy and blue lines represent the GW spectrums from the first and second step PTs, respectively.
The red line shows the superposed GW spectrum.
}
\label{fig:2-stepPTGW}
\end{figure}


We compute the GW spectrums from a strong 2-step PT where both the first and second step PTs are first order.
The following parameter set in the Type-I 2HDM is chosen as a benchmark point :
\begin{align}
&m_A=m_{H^\pm}=490\ {\rm GeV},\ m_H=300\ {\rm GeV},\ \tan\beta=2.3,\ \cos(\beta-\alpha)=-0.21,\ m_3=20\ {\rm GeV}.\nonumber
\end{align}
In the left and right panels in Fig.~\ref{fig:2-stepPTcausingGW}, the paths of the first and second PT in the strong 2-step PT
are respectively shown by the black line in the $\phi_2$ vs.~$\phi_1$ plane.
The contour plots of the effective potential at $T_n$ are also given in Fig.~\ref{fig:2-stepPTcausingGW}.
The path of the first step PT runs almost along the $\phi_2$ axis from the origin
to $(\phi_1, \phi_2)\simeq(3\ {\rm GeV},\ 115\ {\rm GeV})$ at $T_n\simeq49.9$ GeV.
The path of the second step PT goes from $(\phi_1, \phi_2)\simeq(5\ {\rm GeV}, 126\ {\rm GeV})$ to $(\phi_1, \phi_2)\simeq(93\ {\rm GeV}, 219\ {\rm GeV})$, which is in the direction of the EW vacuum, at $T_n=42.9$ GeV.
The strengths of the first and second step PT are respectively $\xi=2.1$ and 4.2, then both of them satisfy the criterion $\xi\geq1$.
The values of ($\alpha_{\rm GW}, \tilde\beta_{\rm GW}$) are $(8.1\times 10^{-2}, 8.5\times10^3)$ for the first step and $(0.16, 1.9\times 10^2)$ for the second step.
The GW spectrums $h^2\Omega_{\rm GW}$ from these PTs are shown in Fig.~\ref{fig:2-stepPTGW}.
The observable areas by the future space-based interferometers such as LISA \cite{Caprini:2015zlo, LISA:2017pwj, Caprini:2019egz}, DECIGO \cite{Seto:2001qf, Kawamura:2011zz}, BBO \cite{Corbin:2005ny}, U-DECIGO \cite{Kudoh:2005as}, Taiji \cite{Hu:2017mde, Guo:2018npi}, and TianQin \cite{Luo:2015ght, Hu:2018yqb} are also presented.
The navy and blue lines represent
the GW spectrums from the first and second step PTs which
have the peak frequencies around 0.1 Hz and $2\times 10^{-3}$ Hz, respectively.
The superposed GW spectrum is shown by the red line.
We can see that it has a double peak, which can be observed by BBO or U-DECIGO~\footnote{
Recent studies in Ref.~\cite{Guo:2020grp} suggest the existence of an additional suppression factor for the $\Omega_{\mathrm{sw}}$ due to the finite lifetime of the sound waves.
The factors are about 0.005 and 0.1, respectively, for the first and second step PTs at our benchmark point.
Taking into account the suppressions, the peak of the GW spectrum of the first step PT can be hardly seen.
However, there are still several uncertainties in the calculation of the GW spectrum (see Refs.~\cite{Cutting:2019zws,Wang:2020jrd,Croon:2020cgk,Guo:2021qcq,Gould:2021oba,Giese:2020rtr,Hoeche:2020rsg,Giese:2020znk,Wang:2020nzm,Wang:2020zlf} for recent works).
}.
Additionally, the deviation of the Higgs trilinear coupling $\delta\lambda_{hhh}$ is 2.2 for the benchmark point.
Such $\delta\lambda_{hhh}$ has the possibility to be measured at the HL-LHC and the ILC.
Therefore, the signature of the strong 2-step PT at the benchmark point may be observed in the experiments of both GW and colliders.
With a combination of these signatures, it might be possible to identify whether the strong 2-step PT occurred in the early universe.

\section{Conclusions}\label{sec:Conclusion}
In this paper, we have studied the parameter regions where the multi-step and strong 2-step EWPTs occur by scanning the parameter spaces in the CP-conserving Type-I and Type-X 2HDMs with $m_A$ or $m_H=m_{H^\pm}$.
In the analyses, we have focused on the small $m_3$ as 0 $\leq m_3\leq100$ GeV.
As a result of our scan, areas where the multi-step and strong 2-step PTs occur have been found.
The features of the parameter region where the multi-step PTs likely to occur are: (i) $m_A-m_H$ is negative with large magnitude, (ii) $\tan\beta$ is small, (iii) $\cos(\beta-\alpha)$ is negative and small, (iv) $m_3$ is small.
The features (ii), (iii), and (iv) are preferred for the negative $m_2^2$ with large magnitude, which can yield the minimum point along the $\phi_2$ axis.
By contrast, the strong 2-step PTs occur only when the mass hierarchy $m_A>m_H$ exists in our parameter search, while they have similar features as (ii), (iii), and (iv).
On the other hand, the VEVs after the first step of the multi-step PTs have a tendency to be located along the $\phi_2$ (or $\phi_1$) axis, and the VEVs after the last step PTs likely to lie in the direction of the EW vacuum.

As the possible physical signatures for the multi-step PTs in the collider experiments, we have investigated the deviation of the Higgs trilinear coupling from that in the SM $\delta\lambda_{hhh}$.
The maximum value of $\delta\lambda_{hhh}$ increases as $\tan\beta$ and $\cos(\beta-\alpha)$ (which is zero in the Type-X 2HDMs) becomes smaller respectively in the case where the multi-step PTs occur.
Compared with the results of the 1-step PTs at the same value of $m_A$, the values of $\delta\lambda_{hhh}$ for the multi-step PTs have a tendency to be large.
In particular, when the strong 2-step PTs happen, $\delta\lambda_{hhh}$ are larger than about 0.5 and the largest value of $\delta\lambda_{hhh}$ in the Type-I 2HDMs reach over 2.
Such deviations would be measured at future colliders like the HL-LHC and the ILC.
As the signatures observed by the space-based interferometers, we have computed the GW spectrums from the strong 2-step PT where the first order PT occurs twice.
The superposed GW spectrum has the possibility to have a double peak and be observed by the future observers as BBO and U-DECIGO.
The multi-step EWPT might be confirmed by combining the information obtained from the future collider and GW experiments.


\begin{figure}[t]
\centering
\begin{tabular}{c}

\begin{minipage}{0.5\hsize}
\centering
\includegraphics[clip, width=7.5cm]{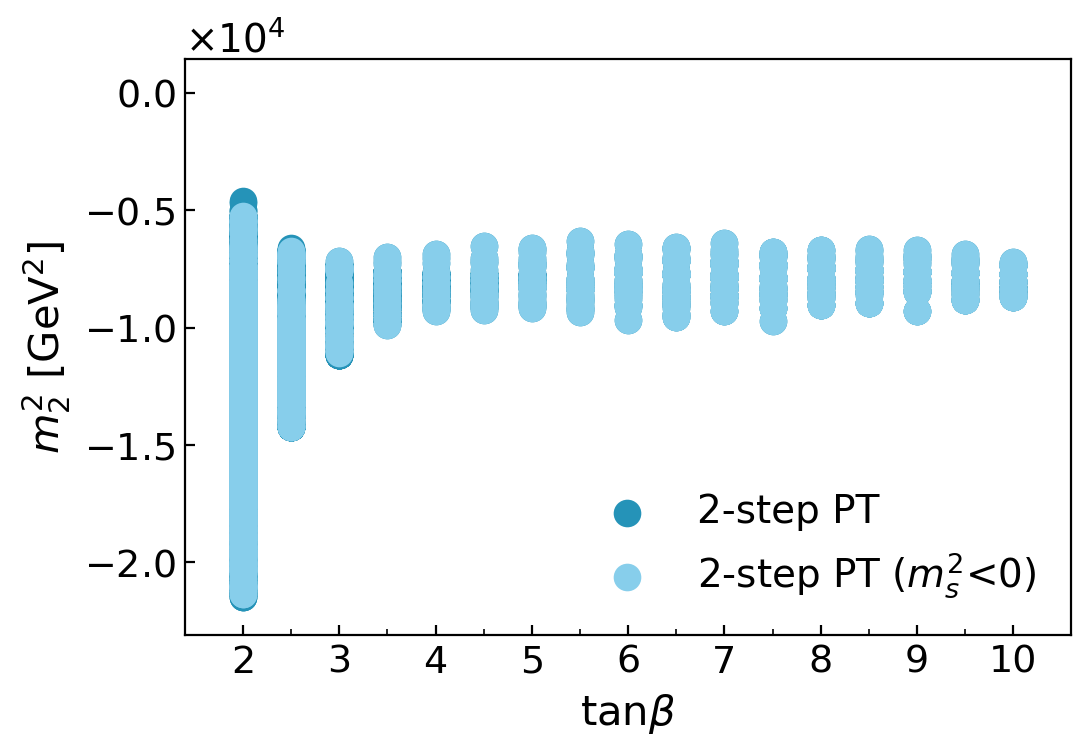}
\end{minipage}

\begin{minipage}{0.5\hsize}
\centering
\includegraphics[clip, width=7.5cm]{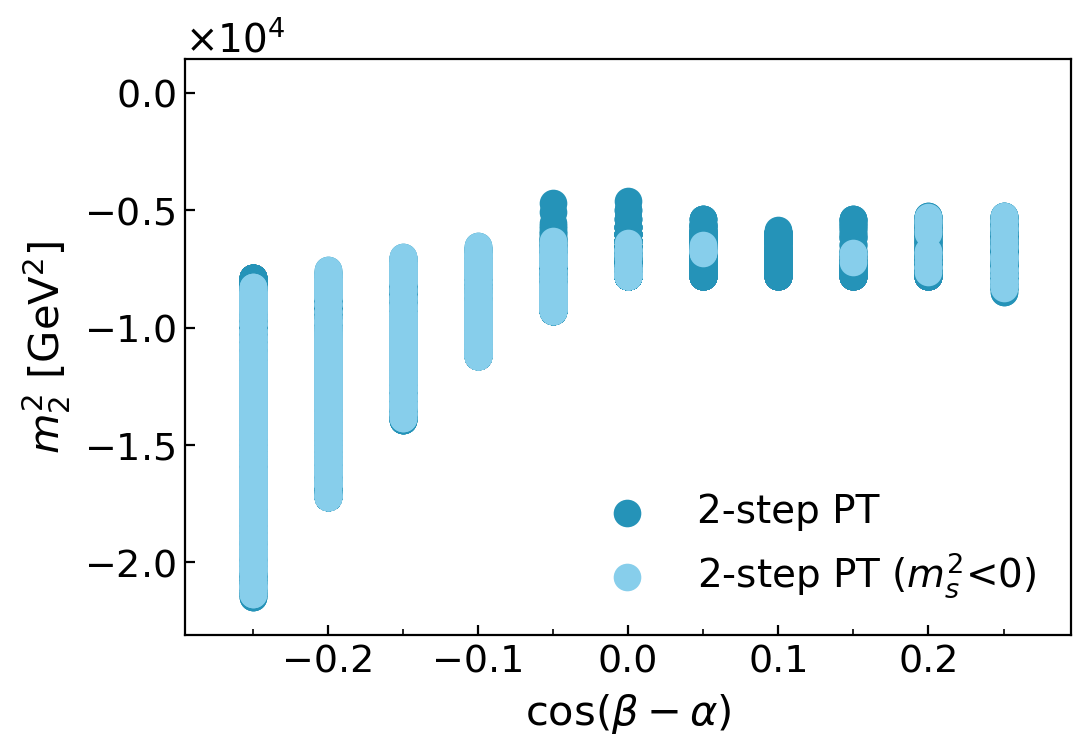}
\end{minipage}

\end{tabular}
\caption{
Parameter points where the 2-step PTs (blue) and those with $m_s^2<0$ at the origin for the first step PTs (light-blue) occur in the $m_2^2$ vs.~$\tan\beta$ (left) and $m_2^2$ vs.~$\cos(\beta-\alpha)$ (right) planes in the Type-I 2HDM with $m_A=m_{H^\pm}$.
}
\label{fig:m2_2stepPTImAI_negative}
\end{figure}

\appendix
\section{Complex effective potential at finite temperature}\label{sec:ComplexPotential}
In this appendix, we comment on the region where the complex effective potential appears at finite temperature.
Fig.~\ref{fig:m2_2stepPTImAI_negative} shows parameter points where the 2-step PTs occur for the negative scalar squared-masses at the origin for the first step PTs colored by light-blue in the Type-I 2HDM with $m_A=m_{H^\pm}$. The scalar squared-mass $m_s^2$ indicates the smallest squared-mass among the scalar fields.
The parameter points for the  2-step PTs (blue) are the same as in Fig.~\ref{fig:m2AllPTImAI}.
The light-blue points almost overlap with the points for the 2-step PTs  except for $\cos(\beta-\alpha)\simeq-0.05$--0.15.
Although the light-blue points are widespread in the parameter space, a ratio of the number of points for the negative squared-masses to that for all points where the 2-step PTs occur is about 15$\%$.
On the other hand, we also find that the parameter points for the 1-step PTs are also widespread, and the number ratio is about 25$\%$.
Ref.~\cite{Delaunay:2007wb} shows that the resummation method can cure the contributions from the negative squared-masses.

\begin{table}[t]
\centering
\begin{tabular}{cc}
\begin{minipage}[c]{1\hsize}
\centering
\begin{tabular}{|l||c|c|c|c|c|c|} \hline
& $m_A$ [GeV] & $m_H$ [GeV] & $m_A-m_H$ [GeV] & $\tan\beta$ & $\cos(\beta-\alpha)$ & $m_3$ [GeV] \\ \hline\hline
Type-I ($m_A=m_{H^\pm}$) & 130--550 ($\sim$7$\%$) & 390 (47$\%$) & $-$250 (100$\%$) & 2 (11$\%$) & $-$0.25 (9$\%$) & 0 (48$\%$) \\ \hline
Type-I ($m_H=m_{H^\pm}$) & 650 (100$\%$) & 360 (35$\%$) & $-$210 (21$\%$) & 2 (13$\%$) & $-$0.25 (9$\%$) & 0 (60$\%$) \\ \hline
Type-X ($m_A=m_{H^\pm}$) & 310 (11$\%$) & 350 (36$\%$) & $-$130 (21$\%$) & 2 (13$\%$) & - & 0 (49$\%$) \\ \hline
Type-X ($m_H=m_{H^\pm}$) & 130--350 ($\sim$14$\%$) & 350 (38$\%$) & $-$210 (50$\%$) & 2 (18$\%$) & - & 0 (60$\%$) \\ \hline
\end{tabular}
\caption{
Values or ranges of input parameters where ${\cal R}_{\rm multi}$ have the maximum values.
The values inside the parentheses represent the maximum values of ${\cal R}_{\rm multi}$.
}
\label{tb:Numbertable1}
\vspace{0.3in}
\end{minipage}\\

\begin{minipage}[c]{1\hsize}
\centering
\begin{tabular}{|l||c|c|c|c|c|c|} \hline
& $m_A$ [GeV] & $m_H$ [GeV] & $m_A-m_H$ [GeV] & $\tan\beta$ & $\cos(\beta-\alpha)$ & $m_3$ [GeV] \\ \hline\hline
Type-I ($m_A=m_{H^\pm}$) & 470 (41$\%$) & 280 (12$\%$) & 210 (100$\%$) & 2.5--4.5 ($\sim10\%$) & $-$0.25 (11$\%$) & 20 (7$\%$) \\ \hline
Type-I ($m_H=m_{H^\pm}$) & 580 (50$\%$) & 280 (8$\%$) & 310 (100$\%$) & 2.5--4 ($\sim11\%$) & $-$0.25 (9$\%$) & 10 (5$\%$) \\ \hline
\end{tabular}
\caption{
Values of input parameters where ${\cal R}_{\rm st2}$ have the maximum values.
The values inside the parentheses represent the maximum values of ${\cal R}_{\rm st2}$.
Note that we omit the results for the Type-X 2HDMs because the number of points for the strong 2-step PTs in these cases are not large enough to consider the dependencies of the ratios on the input parameters.
}
\label{tb:Numbertable2}
\end{minipage}
\end {tabular}
\end{table}

\begin{figure}[t]
 \centering
  \begin{tabular}{c}

   \begin{minipage}{0.5\hsize}
    \centering
     \includegraphics[clip, width=7.5cm]{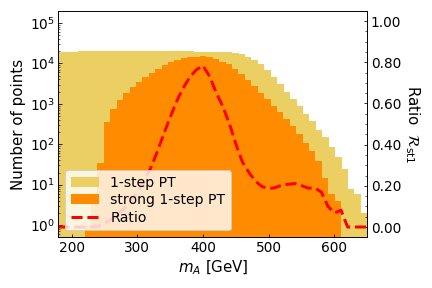}
   \end{minipage}

   \begin{minipage}{0.5\hsize}
    \centering
     \includegraphics[clip, width=7.5cm]{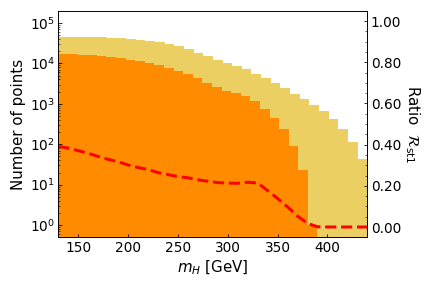}
   \end{minipage}\\

   \begin{minipage}{0.5\hsize}
    \centering
     \includegraphics[clip, width=7.5cm]{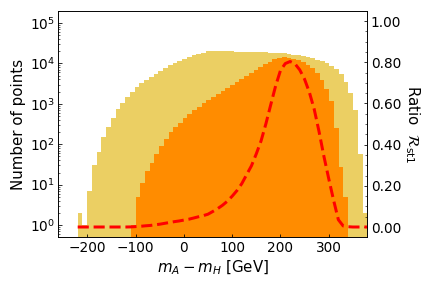}
   \end{minipage}

   \begin{minipage}{0.5\hsize}
    \centering
     \includegraphics[clip, width=7.5cm]{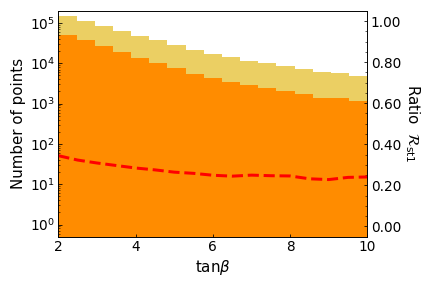}
   \end{minipage}\\

   \begin{minipage}{0.5\hsize}
    \centering
     \includegraphics[clip, width=7.5cm]{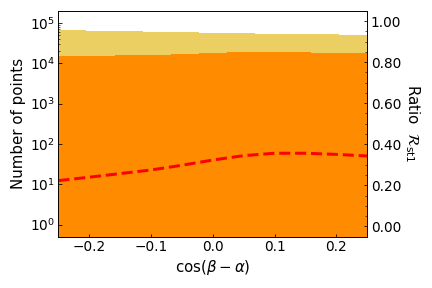}
   \end{minipage}

   \begin{minipage}{0.5\hsize}
    \centering
     \includegraphics[clip, width=7.5cm]{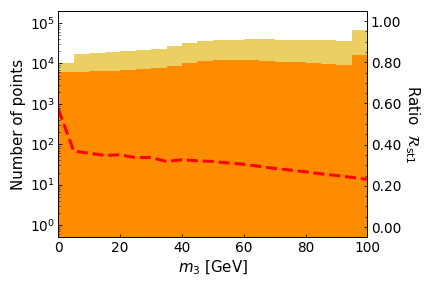}
   \end{minipage}

  \end{tabular}
\caption{Number of points where the 1-step (yellow) and strong 1-step (orange) PTs occur as a function of $m_A$, $m_H$, $m_A- m_H$, $\tan\beta$, $\cos(\beta-\alpha)$, and $m_3$ in the Type-I 2HDM with $m_A=m_{H^\pm}$.
The red dashed line represents $R_{\rm st1}$, which are the ratios of the number of points where the strong 1-step PTs occur to that for the 1-step PTs.}
\label{fig:Numberinput3}
\end{figure}

\begin{table}[t]
\centering
\begin{tabular}{cc}
\begin{minipage}[c]{1\hsize}
\centering
\begin{tabular}{|l||c|c|c|c|c|c|} \hline
& $m_A$ [GeV] & $m_H$ [GeV] & $m_A-m_H$ [GeV] & $\tan\beta$ & $\cos(\beta-\alpha)$ & $m_3$ [GeV] \\ \hline\hline
Type-I ($m_A=m_{H^\pm}$) & 400 (78$\%$) & 130 (39$\%$) & 220 (80$\%$) & 2 (34$\%$) & 0.05--0.25 ($\sim$36$\%$) & 0 (57$\%$) \\ \hline
Type-I ($m_H=m_{H^\pm}$) & 530 (83$\%$) & 130--340 (33$\%$) & 340 (91$\%$) & 2 (35$\%$) & 0.05--0.25 ($\sim$38$\%$) & 0 (66$\%$) \\ \hline
Type-X ($m_A=m_{H^\pm}$) & 400 (91$\%$) & 130 (39$\%$) & 210 (83$\%$) & 2 (40$\%$) & - & 0 (59$\%$) \\ \hline
Type-X ($m_H=m_{H^\pm}$) & 630 (100$\%$) & 360 (43$\%$) & 300 (85$\%$) & 2 (41$\%$) & - & 0 (64$\%$) \\ \hline
\end{tabular}
\caption{
Values or ranges of input parameters where ${\cal R}_{\rm st1}$ have the maximum values.
The values inside the parentheses represent the maximum values of ${\cal R}_{\rm st1}$.
}
\label{tb:Numbertable3}
\vspace{0.1in}
\end{minipage}
\end {tabular}
\end{table}


\section{Tables for number analyses}\label{sec:NumberAnalysis}
We show in Tabs.~\ref{tb:Numbertable1}, and~\ref{tb:Numbertable2},
the values or ranges of input parameters where the ratios ${\cal R}_{\rm multi}$ (\ref{eq:Rmulti}), and ${\cal R}_{\rm st2}$ (\ref{eq:Rst2})
have the maximum values, respectively, for Type-I and -X 2HDMs.
Note that we omit the results for the Type-X 2HDMs in Tab.~\ref{tb:Numbertable2} because the number of points for the strong 2-step PTs in these cases are not large enough to consider the dependencies of the ratios on the input parameters.
We see that {\it e.g.}, in Tab.~\ref{tb:Numbertable1} the multi-step PTs favor $m_A < m_H$ in all four cases.
Similar tendencies in Tabs.~\ref{tb:Numbertable1}, and~\ref{tb:Numbertable2} are seen even if we consider the constraint from $B\rightarrow\mu^+\mu^-$.



Although we do not discuss the strong 1-step PTs in Section~\ref{sec:numericalresults}, we also show the results of the number analyses for them because they are still important in the context of baryogenesis.
Fig.~\ref{fig:Numberinput3} gives the number of points for the 1-step (yellow) and strong 1-step (orange) PTs, as a function of $m_A$ (top), $m_H$ (middle) and $m_A-m_H$ (bottom) in the Type-I 2HDM with $m_A=m_{H^\pm}$.
The red dashed lines represent ${\cal R}_{\rm st1}$, which are the ratios of the number of points for the strong 1-step PTs to that for the 1-step PTs,
\begin{align}
{\cal R}_{\rm st1}=\frac{\# {\rm ~of ~points ~for~ the ~strong~1\mathchar`-step~ PTs}}{\# {\rm ~of ~points ~for~ the~1\mathchar`-step~ PTs}}.
\label{eq:Rst1}
\end{align}
We see in the middle left panel that ${\cal R}_{\rm st1}$ peaks at $m_A-m_H\simeq$ 220 GeV and reaches close to 1.
Hence, the strong 1-step PTs favor the mass hierarchy $m_A>m_H$
(This is consistent with the results in Ref.~\cite{Dorsch:2013wja}).
This is the same feature as the one which the strong 2-step PTs have, shown in the top right panel of Fig.~\ref{fig:strongPTImAI}.

We show in Tab.~\ref{tb:Numbertable3},
the values or ranges of input parameters where the ratios ${\cal R}_{\rm st1}$ (\ref{eq:Rst1})
have the maximum values, respectively, for Type-I and -X 2HDMs.
Similar tendencies in Tabs.~\ref{tb:Numbertable1}, ~\ref{tb:Numbertable2}, and~\ref{tb:Numbertable3} are seen even if we consider the constraint from $B\rightarrow\mu^+\mu^-$.


\begin{acknowledgments}
The work of M.~A. is supported in part by the Japan Society for the
Promotion of Sciences Grant-in-Aid for Scientific Research (Grant
No. 17K05412 and No. 20H00160).
\end{acknowledgments}

\let\doi\relax
\bibliographystyle{JHEP}
\bibliography{2106.03439}

\providecommand{\href}[2]{#2}\begingroup\raggedright\begin{thebibliography}{100}

\bibitem{Aghanim:2018eyx}
{Planck} Collaboration, N.~Aghanim {\em et.~al.,} {\em Astron. Astrophys.} {\bf
  641} (2020) A6 [\href{http://arXiv.org/abs/1807.06209}{{\tt 1807.06209}}].

\bibitem{Sakharov:1967dj}
A.~Sakharov {\em Sov. Phys. Usp.} {\bf 34} (1991), no.~5 392--393.

\bibitem{Kuzmin:1985mm}
V.~Kuzmin, V.~Rubakov and M.~Shaposhnikov,  {\em Phys. Lett. B} {\bf 155}
  (1985) 36.

\bibitem{Kajantie:1995kf}
K.~Kajantie, M.~Laine, K.~Rummukainen and M.~E. Shaposhnikov,  {\em Nucl. Phys.
  B} {\bf 466} (1996) 189--258
  [\href{http://arXiv.org/abs/hep-lat/9510020}{{\tt hep-lat/9510020}}].

\bibitem{Csikor:1998eu}
F.~Csikor, Z.~Fodor and J.~Heitger,  {\em Phys. Rev. Lett.} {\bf 82} (1999)
  21--24 [\href{http://arXiv.org/abs/hep-ph/9809291}{{\tt hep-ph/9809291}}].

\bibitem{Aad:2012tfa}
{ATLAS} Collaboration, G.~Aad {\em et.~al.,} {\em Phys. Lett. B} {\bf 716}
  (2012) 1--29 [\href{http://arXiv.org/abs/1207.7214}{{\tt 1207.7214}}].

\bibitem{ATLAS-CONF-2012-162}
{ATLAS} Collaboration Tech. Rep. ATLAS-CONF-2012-162, CERN, Geneva, Nov, 2012.

\bibitem{Chatrchyan:2012ufa}
{CMS} Collaboration, S.~Chatrchyan {\em et.~al.,} {\em Phys. Lett. B} {\bf 716}
  (2012) 30--61 [\href{http://arXiv.org/abs/1207.7235}{{\tt 1207.7235}}].

\bibitem{CMS-PAS-HIG-12-045}
{CMS} Collaboration Tech. Rep. CMS-PAS-HIG-12-045, CERN, Geneva, 2012.

\bibitem{DOnofrio:2014rug}
M.~D'Onofrio, K.~Rummukainen and A.~Tranberg,  {\em Phys. Rev. Lett.} {\bf 113}
  (2014), no.~14 141602 [\href{http://arXiv.org/abs/1404.3565}{{\tt
  1404.3565}}].

\bibitem{Gavela:1993ts}
M.~B. Gavela, P.~Hernandez, J.~Orloff and O.~Pene,  {\em Mod. Phys. Lett. A}
  {\bf 9} (1994) 795--810 [\href{http://arXiv.org/abs/hep-ph/9312215}{{\tt
  hep-ph/9312215}}].

\bibitem{Huet:1994jb}
P.~Huet and E.~Sather,  {\em Phys. Rev. D} {\bf 51} (1995) 379--394
  [\href{http://arXiv.org/abs/hep-ph/9404302}{{\tt hep-ph/9404302}}].

\bibitem{Gavela:1994dt}
M.~B. Gavela, P.~Hernandez, J.~Orloff, O.~Pene and C.~Quimbay,  {\em Nucl.
  Phys. B} {\bf 430} (1994) 382--426
  [\href{http://arXiv.org/abs/hep-ph/9406289}{{\tt hep-ph/9406289}}].

\bibitem{Dorsch:2013wja}
G.~Dorsch, S.~Huber and J.~No,  {\em JHEP} {\bf 10} (2013) 029
  [\href{http://arXiv.org/abs/1305.6610}{{\tt 1305.6610}}].

\bibitem{Basler:2016obg}
P.~Basler, M.~Krause, M.~Muhlleitner, J.~Wittbrodt and A.~Wlotzka,  {\em JHEP}
  {\bf 02} (2017) 121 [\href{http://arXiv.org/abs/1612.04086}{{\tt
  1612.04086}}].

\bibitem{Bernon:2017jgv}
J.~Bernon, L.~Bian and Y.~Jiang,  {\em JHEP} {\bf 05} (2018) 151
  [\href{http://arXiv.org/abs/1712.08430}{{\tt 1712.08430}}].

\bibitem{Wang:2018hnw}
L.~Wang, J.~M. Yang, M.~Zhang and Y.~Zhang,  {\em Phys. Lett. B} {\bf 788}
  (2019) 519--529 [\href{http://arXiv.org/abs/1809.05857}{{\tt 1809.05857}}].

\bibitem{Su:2020pjw}
W.~Su, A.~G. Williams and M.~Zhang,  {\em JHEP} {\bf 04} (2021) 219
  [\href{http://arXiv.org/abs/2011.04540}{{\tt 2011.04540}}].

\bibitem{Andersen:2017ika}
J.~O. Andersen, T.~Gorda, A.~Helset, L.~Niemi, T.~V.~I. Tenkanen, A.~Tranberg,
  A.~Vuorinen and D.~J. Weir,  {\em Phys. Rev. Lett.} {\bf 121} (2018), no.~19
  191802 [\href{http://arXiv.org/abs/1711.09849}{{\tt 1711.09849}}].

\bibitem{Kainulainen:2019kyp}
K.~Kainulainen, V.~Keus, L.~Niemi, K.~Rummukainen, T.~V.~I. Tenkanen and
  V.~Vaskonen,  {\em JHEP} {\bf 06} (2019) 075
  [\href{http://arXiv.org/abs/1904.01329}{{\tt 1904.01329}}].

\bibitem{Haarr:2016qzq}
A.~Haarr, A.~Kvellestad and T.~C. Petersen,
  \href{http://arXiv.org/abs/1611.05757}{{\tt 1611.05757}}.

\bibitem{Dorsch:2016nrg}
G.~C. Dorsch, S.~J. Huber, T.~Konstandin and J.~M. No,  {\em JCAP} {\bf 05}
  (2017) 052 [\href{http://arXiv.org/abs/1611.05874}{{\tt 1611.05874}}].

\bibitem{Chen:2017com}
C.-Y. Chen, H.-L. Li and M.~Ramsey-Musolf,  {\em Phys. Rev. D} {\bf 97} (2018),
  no.~1 015020 [\href{http://arXiv.org/abs/1708.00435}{{\tt 1708.00435}}].

\bibitem{Kanemura:2020ibp}
S.~Kanemura, M.~Kubota and K.~Yagyu,  {\em JHEP} {\bf 08} (2020) 026
  [\href{http://arXiv.org/abs/2004.03943}{{\tt 2004.03943}}].

\bibitem{Blinov:2015sna}
N.~Blinov, J.~Kozaczuk, D.~E. Morrissey and C.~Tamarit,  {\em Phys. Rev. D}
  {\bf 92} (2015), no.~3 035012 [\href{http://arXiv.org/abs/1504.05195}{{\tt
  1504.05195}}].

\bibitem{Hammerschmitt:1994fn}
A.~Hammerschmitt, J.~Kripfganz and M.~Schmidt,  {\em Z. Phys. C} {\bf 64}
  (1994) 105--110 [\href{http://arXiv.org/abs/hep-ph/9404272}{{\tt
  hep-ph/9404272}}].

\bibitem{Fromme:2006cm}
L.~Fromme, S.~J. Huber and M.~Seniuch,  {\em JHEP} {\bf 11} (2006) 038
  [\href{http://arXiv.org/abs/hep-ph/0605242}{{\tt hep-ph/0605242}}].

\bibitem{Witten:1984rs}
E.~Witten {\em Phys. Rev. D} {\bf 30} (1984) 272--285.

\bibitem{Hogan:1986qda}
C.~Hogan {\em Mon. Not. Roy. Astron. Soc.} {\bf 218} (1986) 629--636.

\bibitem{Caprini:2015zlo}
C.~Caprini {\em et.~al.,} {\em JCAP} {\bf 04} (2016) 001
  [\href{http://arXiv.org/abs/1512.06239}{{\tt 1512.06239}}].

\bibitem{LISA:2017pwj}
{LISA} Collaboration, P.~Amaro-Seoane {\em et.~al.,}
  \href{http://arXiv.org/abs/1702.00786}{{\tt 1702.00786}}.

\bibitem{Caprini:2019egz}
C.~Caprini {\em et.~al.,} {\em JCAP} {\bf 03} (2020) 024
  [\href{http://arXiv.org/abs/1910.13125}{{\tt 1910.13125}}].

\bibitem{Profumo:2007wc}
S.~Profumo, M.~J. Ramsey-Musolf and G.~Shaughnessy,  {\em JHEP} {\bf 08} (2007)
  010 [\href{http://arXiv.org/abs/0705.2425}{{\tt 0705.2425}}].

\bibitem{Espinosa:2011ax}
J.~R. Espinosa, T.~Konstandin and F.~Riva,  {\em Nucl. Phys. B} {\bf 854}
  (2012) 592--630 [\href{http://arXiv.org/abs/1107.5441}{{\tt 1107.5441}}].

\bibitem{Curtin:2014jma}
D.~Curtin, P.~Meade and C.-T. Yu,  {\em JHEP} {\bf 11} (2014) 127
  [\href{http://arXiv.org/abs/1409.0005}{{\tt 1409.0005}}].

\bibitem{Jiang:2015cwa}
M.~Jiang, L.~Bian, W.~Huang and J.~Shu,  {\em Phys. Rev. D} {\bf 93} (2016),
  no.~6 065032 [\href{http://arXiv.org/abs/1502.07574}{{\tt 1502.07574}}].

\bibitem{Huang:2015bta}
F.~P. Huang and C.~S. Li,  {\em Phys. Rev. D} {\bf 92} (2015), no.~7 075014
  [\href{http://arXiv.org/abs/1507.08168}{{\tt 1507.08168}}].

\bibitem{Kurup:2017dzf}
G.~Kurup and M.~Perelstein,  {\em Phys. Rev. D} {\bf 96} (2017), no.~1 015036
  [\href{http://arXiv.org/abs/1704.03381}{{\tt 1704.03381}}].

\bibitem{Kang:2017mkl}
Z.~Kang, P.~Ko and T.~Matsui,  {\em JHEP} {\bf 02} (2018) 115
  [\href{http://arXiv.org/abs/1706.09721}{{\tt 1706.09721}}].

\bibitem{Matsui:2017ggm}
T.~Matsui {\em EPJ Web Conf.} {\bf 168} (2018) 05001
  [\href{http://arXiv.org/abs/1709.05900}{{\tt 1709.05900}}].

\bibitem{Chiang:2017nmu}
C.-W. Chiang, M.~J. Ramsey-Musolf and E.~Senaha,  {\em Phys. Rev. D} {\bf 97}
  (2018), no.~1 015005 [\href{http://arXiv.org/abs/1707.09960}{{\tt
  1707.09960}}].

\bibitem{Hashino:2018zsi}
K.~Hashino, M.~Kakizaki, S.~Kanemura, P.~Ko and T.~Matsui,  {\em JHEP} {\bf 06}
  (2018) 088 [\href{http://arXiv.org/abs/1802.02947}{{\tt 1802.02947}}].

\bibitem{Huang:2018aja}
F.~P. Huang, Z.~Qian and M.~Zhang,  {\em Phys. Rev. D} {\bf 98} (2018), no.~1
  015014 [\href{http://arXiv.org/abs/1804.06813}{{\tt 1804.06813}}].

\bibitem{Chiang:2019oms}
C.-W. Chiang and B.-Q. Lu,  {\em JHEP} {\bf 07} (2020) 082
  [\href{http://arXiv.org/abs/1912.12634}{{\tt 1912.12634}}].

\bibitem{Carena:2019une}
M.~Carena, Z.~Liu and Y.~Wang,  {\em JHEP} {\bf 08} (2020) 107
  [\href{http://arXiv.org/abs/1911.10206}{{\tt 1911.10206}}].

\bibitem{Ghorbani:2020xqv}
P.~Ghorbani \href{http://arXiv.org/abs/2010.15708}{{\tt 2010.15708}}.

\bibitem{Niemi:2021qvp}
L.~Niemi, P.~Schicho and T.~V.~I. Tenkanen,
  \href{http://arXiv.org/abs/2103.07467}{{\tt 2103.07467}}.

\bibitem{Land:1992sm}
D.~Land and E.~D. Carlson,  {\em Phys. Lett. B} {\bf 292} (1992) 107--112
  [\href{http://arXiv.org/abs/hep-ph/9208227}{{\tt hep-ph/9208227}}].

\bibitem{Friedlander:2020tnq}
A.~Friedlander, I.~Banta, J.~M. Cline and D.~Tucker-Smith,  {\em Phys. Rev. D}
  {\bf 103} (2021), no.~5 055020 [\href{http://arXiv.org/abs/2009.14295}{{\tt
  2009.14295}}].

\bibitem{Fabian:2020hny}
S.~Fabian, F.~Goertz and Y.~Jiang,  \href{http://arXiv.org/abs/2012.12847}{{\tt
  2012.12847}}.

\bibitem{Wang:2019pet}
X.~Wang, F.~P. Huang and X.~Zhang,  {\em Phys. Rev. D} {\bf 101} (2020), no.~1
  015015 [\href{http://arXiv.org/abs/1909.02978}{{\tt 1909.02978}}].

\bibitem{Patel:2012pi}
H.~H. Patel and M.~J. Ramsey-Musolf,  {\em Phys. Rev. D} {\bf 88} (2013) 035013
  [\href{http://arXiv.org/abs/1212.5652}{{\tt 1212.5652}}].

\bibitem{Chala:2018opy}
M.~Chala, M.~Ramos and M.~Spannowsky,  {\em Eur. Phys. J. C} {\bf 79} (2019),
  no.~2 156 [\href{http://arXiv.org/abs/1812.01901}{{\tt 1812.01901}}].

\bibitem{Bell:2020gug}
N.~F. Bell, M.~J. Dolan, L.~S. Friedrich, M.~J. Ramsey-Musolf and R.~R. Volkas,
   {\em JHEP} {\bf 05} (2020) 050 [\href{http://arXiv.org/abs/2001.05335}{{\tt
  2001.05335}}].

\bibitem{Niemi:2020hto}
L.~Niemi, M.~Ramsey-Musolf, T.~V. Tenkanen and D.~J. Weir,
  \href{http://arXiv.org/abs/2005.11332}{{\tt 2005.11332}}.

\bibitem{Patel:2013zla}
H.~H. Patel, M.~J. Ramsey-Musolf and M.~B. Wise,  {\em Phys. Rev. D} {\bf 88}
  (2013), no.~1 015003 [\href{http://arXiv.org/abs/1303.1140}{{\tt
  1303.1140}}].

\bibitem{Inoue:2015pza}
S.~Inoue, G.~Ovanesyan and M.~J. Ramsey-Musolf,  {\em Phys. Rev. D} {\bf 93}
  (2016) 015013 [\href{http://arXiv.org/abs/1508.05404}{{\tt 1508.05404}}].

\bibitem{Huang:2017laj}
F.~P. Huang and X.~Zhang,  {\em Phys. Lett. B} {\bf 788} (2019) 288--294
  [\href{http://arXiv.org/abs/1701.04338}{{\tt 1701.04338}}].

\bibitem{Chao:2017vrq}
W.~Chao, H.-K. Guo and J.~Shu,  {\em JCAP} {\bf 09} (2017) 009
  [\href{http://arXiv.org/abs/1702.02698}{{\tt 1702.02698}}].

\bibitem{Ramsey-Musolf:2017tgh}
M.~J. Ramsey-Musolf, P.~Winslow and G.~White,  {\em Phys. Rev. D} {\bf 97}
  (2018), no.~12 123509 [\href{http://arXiv.org/abs/1708.07511}{{\tt
  1708.07511}}].

\bibitem{Vieu:2018nfq}
T.~Vieu, A.~P. Morais and R.~Pasechnik,  {\em JCAP} {\bf 07} (2018) 014
  [\href{http://arXiv.org/abs/1801.02670}{{\tt 1801.02670}}].

\bibitem{Vieu:2018zze}
A.~P. Morais, R.~Pasechnik and T.~Vieu,
  \href{http://arXiv.org/abs/1802.10109}{{\tt 1802.10109}}.

\bibitem{Bian:2018bxr}
L.~Bian and X.~Liu,  {\em Phys. Rev. D} {\bf 99} (2019), no.~5 055003
  [\href{http://arXiv.org/abs/1811.03279}{{\tt 1811.03279}}].

\bibitem{Zhou:2018zli}
R.~Zhou, W.~Cheng, X.~Deng, L.~Bian and Y.~Wu,  {\em JHEP} {\bf 01} (2019) 216
  [\href{http://arXiv.org/abs/1812.06217}{{\tt 1812.06217}}].

\bibitem{Bell:2019mbn}
N.~F. Bell, M.~J. Dolan, L.~S. Friedrich, M.~J. Ramsey-Musolf and R.~R. Volkas,
   {\em JHEP} {\bf 19} (2020) 012 [\href{http://arXiv.org/abs/1903.11255}{{\tt
  1903.11255}}].

\bibitem{Morais:2019fnm}
A.~P. Morais and R.~Pasechnik,  {\em JCAP} {\bf 04} (2020) 036
  [\href{http://arXiv.org/abs/1910.00717}{{\tt 1910.00717}}].

\bibitem{Zhou:2020idp}
L.~Bian, H.-K. Guo, Y.~Wu and R.~Zhou,  {\em Phys. Rev. D} {\bf 101} (2020),
  no.~3 035011 [\href{http://arXiv.org/abs/1906.11664}{{\tt 1906.11664}}].

\bibitem{Baum:2020vfl}
S.~Baum, M.~Carena, N.~R. Shah, C.~E.~M. Wagner and Y.~Wang,
  \href{http://arXiv.org/abs/2009.10743}{{\tt 2009.10743}}.

\bibitem{Ghosh:2020ipy}
T.~Ghosh, H.-K. Guo, T.~Han and H.~Liu,
  \href{http://arXiv.org/abs/2012.09758}{{\tt 2012.09758}}.

\bibitem{Matsui:2021khj}
T.~Matsui, T.~Nomura and K.~Yagyu,  \href{http://arXiv.org/abs/2102.09247}{{\tt
  2102.09247}}.

\bibitem{Kanemura:2002vm}
S.~Kanemura, S.~Kiyoura, Y.~Okada, E.~Senaha and C.~Yuan,  {\em Phys. Lett. B}
  {\bf 558} (2003) 157--164 [\href{http://arXiv.org/abs/hep-ph/0211308}{{\tt
  hep-ph/0211308}}].

\bibitem{Kanemura:2004ch}
S.~Kanemura, Y.~Okada and E.~Senaha,  {\em Phys. Lett. B} {\bf 606} (2005)
  361--366 [\href{http://arXiv.org/abs/hep-ph/0411354}{{\tt hep-ph/0411354}}].

\bibitem{Braathen:2019zoh}
J.~Braathen and S.~Kanemura,  {\em Eur. Phys. J. C} {\bf 80} (2020), no.~3 227
  [\href{http://arXiv.org/abs/1911.11507}{{\tt 1911.11507}}].

\bibitem{Arco:2020ucn}
F.~Arco, S.~Heinemeyer and M.~J. Herrero,  {\em Eur. Phys. J. C} {\bf 80}
  (2020), no.~9 884 [\href{http://arXiv.org/abs/2005.10576}{{\tt 2005.10576}}].

\bibitem{Cepeda:2019klc}
M.~Cepeda {\em et.~al.,} {\em CERN Yellow Rep. Monogr.} {\bf 7} (2019) 221--584
  [\href{http://arXiv.org/abs/1902.00134}{{\tt 1902.00134}}].

\bibitem{Fujii:2015jha}
K.~Fujii {\em et.~al.,} \href{http://arXiv.org/abs/1506.05992}{{\tt
  1506.05992}}.

\bibitem{Corbin:2005ny}
V.~Corbin and N.~J. Cornish,  {\em Class. Quant. Grav.} {\bf 23} (2006)
  2435--2446 [\href{http://arXiv.org/abs/gr-qc/0512039}{{\tt gr-qc/0512039}}].

\bibitem{Kudoh:2005as}
H.~Kudoh, A.~Taruya, T.~Hiramatsu and Y.~Himemoto,  {\em Phys. Rev. D} {\bf 73}
  (2006) 064006 [\href{http://arXiv.org/abs/gr-qc/0511145}{{\tt
  gr-qc/0511145}}].

\bibitem{Barger:1989fj}
V.~D. Barger, J.~L. Hewett and R.~J.~N. Phillips,  {\em Phys. Rev. D} {\bf 41}
  (1990) 3421--3441.

\bibitem{Grossman:1994jb}
Y.~Grossman {\em Nucl. Phys. B} {\bf 426} (1994) 355--384
  [\href{http://arXiv.org/abs/hep-ph/9401311}{{\tt hep-ph/9401311}}].

\bibitem{Aoki:2009ha}
M.~Aoki, S.~Kanemura, K.~Tsumura and K.~Yagyu,  {\em Phys. Rev. D} {\bf 80}
  (2009) 015017 [\href{http://arXiv.org/abs/0902.4665}{{\tt 0902.4665}}].

\bibitem{Quiros:1999jp}
M.~Quiros pp.~187--259, 1, 1999.
\newblock \href{http://arXiv.org/abs/hep-ph/9901312}{{\tt hep-ph/9901312}}.

\bibitem{Cline:2011mm}
J.~M. Cline, K.~Kainulainen and M.~Trott,  {\em JHEP} {\bf 11} (2011) 089
  [\href{http://arXiv.org/abs/1107.3559}{{\tt 1107.3559}}].

\bibitem{Dolan:1973qd}
L.~Dolan and R.~Jackiw,  {\em Phys. Rev. D} {\bf 9} (1974) 3320--3341.

\bibitem{PhysRevD.36.2474}
E.~J. Weinberg and A.~Wu,  {\em Phys. Rev. D} {\bf 36} (1987) 2474--2480.

\bibitem{Parwani:1991gq}
R.~R. Parwani {\em Phys. Rev. D} {\bf 45} (1992) 4695
  [\href{http://arXiv.org/abs/hep-ph/9204216}{{\tt hep-ph/9204216}}]. [Erratum:
  Phys.Rev.D 48, 5965 (1993)].

\bibitem{Arnold:1992rz}
P.~B. Arnold and O.~Espinosa,  {\em Phys. Rev. D} {\bf 47} (1993) 3546
  [\href{http://arXiv.org/abs/hep-ph/9212235}{{\tt hep-ph/9212235}}]. [Erratum:
  Phys.Rev.D 50, 6662 (1994)].

\bibitem{Laine:2017hdk}
M.~Laine, M.~Meyer and G.~Nardini,  {\em Nucl. Phys. B} {\bf 920} (2017)
  565--600 [\href{http://arXiv.org/abs/1702.07479}{{\tt 1702.07479}}].

\bibitem{Carrington:1991hz}
M.~E. Carrington {\em Phys. Rev. D} {\bf 45} (1992) 2933--2944.

\bibitem{Blinov:2015vma}
N.~Blinov, S.~Profumo and T.~Stefaniak,  {\em JCAP} {\bf 07} (2015) 028
  [\href{http://arXiv.org/abs/1504.05949}{{\tt 1504.05949}}].

\bibitem{Deshpande:1977rw}
N.~G. Deshpande and E.~Ma,  {\em Phys. Rev. D} {\bf 18} (1978) 2574.

\bibitem{Sher:1988mj}
M.~Sher {\em Phys. Rept.} {\bf 179} (1989) 273--418.

\bibitem{Nie:1998yn}
S.~Nie and M.~Sher,  {\em Phys. Lett. B} {\bf 449} (1999) 89--92
  [\href{http://arXiv.org/abs/hep-ph/9811234}{{\tt hep-ph/9811234}}].

\bibitem{Kanemura:1999xf}
S.~Kanemura, T.~Kasai and Y.~Okada,  {\em Phys. Lett. B} {\bf 471} (1999)
  182--190 [\href{http://arXiv.org/abs/hep-ph/9903289}{{\tt hep-ph/9903289}}].

\bibitem{Kanemura:1993hm}
S.~Kanemura, T.~Kubota and E.~Takasugi,  {\em Phys. Lett. B} {\bf 313} (1993)
  155--160 [\href{http://arXiv.org/abs/hep-ph/9303263}{{\tt hep-ph/9303263}}].

\bibitem{Akeroyd:2000wc}
A.~G. Akeroyd, A.~Arhrib and E.-M. Naimi,  {\em Phys. Lett. B} {\bf 490} (2000)
  119--124 [\href{http://arXiv.org/abs/hep-ph/0006035}{{\tt hep-ph/0006035}}].

\bibitem{Barroso:2013awa}
A.~Barroso, P.~Ferreira, I.~Ivanov and R.~Santos,  {\em JHEP} {\bf 06} (2013)
  045 [\href{http://arXiv.org/abs/1303.5098}{{\tt 1303.5098}}].

\bibitem{Ivanov:2015nea}
I.~Ivanov and J.~P. Silva,  {\em Phys. Rev. D} {\bf 92} (2015), no.~5 055017
  [\href{http://arXiv.org/abs/1507.05100}{{\tt 1507.05100}}].

\bibitem{Wainwright:2011kj}
C.~L. Wainwright {\em Comput. Phys. Commun.} {\bf 183} (2012) 2006--2013
  [\href{http://arXiv.org/abs/1109.4189}{{\tt 1109.4189}}].

\bibitem{Haber:2010bw}
H.~E. Haber and D.~O'Neil,  {\em Phys. Rev. D} {\bf 83} (2011) 055017
  [\href{http://arXiv.org/abs/1011.6188}{{\tt 1011.6188}}].

\bibitem{Haller:2018nnx}
J.~Haller, A.~Hoecker, R.~Kogler, K.~M\"onig, T.~Peiffer and J.~Stelzer,  {\em
  Eur. Phys. J. C} {\bf 78} (2018), no.~8 675
  [\href{http://arXiv.org/abs/1803.01853}{{\tt 1803.01853}}].

\bibitem{Arhrib:2018ewj}
A.~Arhrib, R.~Benbrik, H.~Harouiz, S.~Moretti and A.~Rouchad,
  \href{http://arXiv.org/abs/1810.09106}{{\tt 1810.09106}}.

\bibitem{Aad:2019mbh}
{ATLAS} Collaboration, G.~Aad {\em et.~al.,} {\em Phys. Rev. D} {\bf 101}
  (2020), no.~1 012002 [\href{http://arXiv.org/abs/1909.02845}{{\tt
  1909.02845}}].

\bibitem{Khachatryan:2016are}
{CMS} Collaboration, V.~Khachatryan {\em et.~al.,} {\em Phys. Lett. B} {\bf
  759} (2016) 369--394 [\href{http://arXiv.org/abs/1603.02991}{{\tt
  1603.02991}}].

\bibitem{Aaboud:2018eoy}
{ATLAS} Collaboration, M.~Aaboud {\em et.~al.,} {\em Phys. Lett. B} {\bf 783}
  (2018) 392--414 [\href{http://arXiv.org/abs/1804.01126}{{\tt 1804.01126}}].

\bibitem{Sirunyan:2019wrn}
{CMS} Collaboration, A.~M. Sirunyan {\em et.~al.,} {\em JHEP} {\bf 03} (2020)
  055 [\href{http://arXiv.org/abs/1911.03781}{{\tt 1911.03781}}].

\bibitem{Kling:2020hmi}
F.~Kling, S.~Su and W.~Su,  {\em JHEP} {\bf 06} (2020) 163
  [\href{http://arXiv.org/abs/2004.04172}{{\tt 2004.04172}}].

\bibitem{Benbrik:2020nys}
S.~Semlali, H.~Day-Hall, S.~Moretti and R.~Benbrik,  {\em Phys. Lett. B} {\bf
  810} (2020) 135819 [\href{http://arXiv.org/abs/2006.05177}{{\tt
  2006.05177}}].

\bibitem{ATLAS-CONF-2019-049}
{ATLAS} Collaboration Tech. Rep. ATLAS-CONF-2019-049, CERN, Geneva, Oct, 2019.

\bibitem{Kosowsky:1991ua}
A.~Kosowsky, M.~S. Turner and R.~Watkins,  {\em Phys. Rev. D} {\bf 45} (1992)
  4514--4535.

\bibitem{Kosowsky:1992rz}
A.~Kosowsky, M.~S. Turner and R.~Watkins,  {\em Phys. Rev. Lett.} {\bf 69}
  (1992) 2026--2029.

\bibitem{Kosowsky:1992vn}
A.~Kosowsky and M.~S. Turner,  {\em Phys. Rev. D} {\bf 47} (1993) 4372--4391
  [\href{http://arXiv.org/abs/astro-ph/9211004}{{\tt astro-ph/9211004}}].

\bibitem{Kamionkowski:1993fg}
M.~Kamionkowski, A.~Kosowsky and M.~S. Turner,  {\em Phys. Rev. D} {\bf 49}
  (1994) 2837--2851 [\href{http://arXiv.org/abs/astro-ph/9310044}{{\tt
  astro-ph/9310044}}].

\bibitem{Caprini:2007xq}
C.~Caprini, R.~Durrer and G.~Servant,  {\em Phys. Rev. D} {\bf 77} (2008)
  124015 [\href{http://arXiv.org/abs/0711.2593}{{\tt 0711.2593}}].

\bibitem{Huber:2008hg}
S.~J. Huber and T.~Konstandin,  {\em JCAP} {\bf 09} (2008) 022
  [\href{http://arXiv.org/abs/0806.1828}{{\tt 0806.1828}}].

\bibitem{Hindmarsh:2013xza}
M.~Hindmarsh, S.~J. Huber, K.~Rummukainen and D.~J. Weir,  {\em Phys. Rev.
  Lett.} {\bf 112} (2014) 041301 [\href{http://arXiv.org/abs/1304.2433}{{\tt
  1304.2433}}].

\bibitem{Giblin:2013kea}
J.~T. Giblin, J and J.~B. Mertens,  {\em JHEP} {\bf 12} (2013) 042
  [\href{http://arXiv.org/abs/1310.2948}{{\tt 1310.2948}}].

\bibitem{Giblin:2014qia}
J.~T. Giblin and J.~B. Mertens,  {\em Phys. Rev. D} {\bf 90} (2014), no.~2
  023532 [\href{http://arXiv.org/abs/1405.4005}{{\tt 1405.4005}}].

\bibitem{Hindmarsh:2015qta}
M.~Hindmarsh, S.~J. Huber, K.~Rummukainen and D.~J. Weir,  {\em Phys. Rev. D}
  {\bf 92} (2015), no.~12 123009 [\href{http://arXiv.org/abs/1504.03291}{{\tt
  1504.03291}}].

\bibitem{Caprini:2006jb}
C.~Caprini and R.~Durrer,  {\em Phys. Rev. D} {\bf 74} (2006) 063521
  [\href{http://arXiv.org/abs/astro-ph/0603476}{{\tt astro-ph/0603476}}].

\bibitem{Kahniashvili:2008pf}
T.~Kahniashvili, A.~Kosowsky, G.~Gogoberidze and Y.~Maravin,  {\em Phys. Rev.
  D} {\bf 78} (2008) 043003 [\href{http://arXiv.org/abs/0806.0293}{{\tt
  0806.0293}}].

\bibitem{Kahniashvili:2008pe}
T.~Kahniashvili, L.~Campanelli, G.~Gogoberidze, Y.~Maravin and B.~Ratra,  {\em
  Phys. Rev. D} {\bf 78} (2008) 123006
  [\href{http://arXiv.org/abs/0809.1899}{{\tt 0809.1899}}]. [Erratum:
  Phys.Rev.D 79, 109901 (2009)].

\bibitem{Kahniashvili:2009mf}
T.~Kahniashvili, L.~Kisslinger and T.~Stevens,  {\em Phys. Rev. D} {\bf 81}
  (2010) 023004 [\href{http://arXiv.org/abs/0905.0643}{{\tt 0905.0643}}].

\bibitem{Caprini:2009yp}
C.~Caprini, R.~Durrer and G.~Servant,  {\em JCAP} {\bf 12} (2009) 024
  [\href{http://arXiv.org/abs/0909.0622}{{\tt 0909.0622}}].

\bibitem{Binetruy:2012ze}
P.~Binetruy, A.~Bohe, C.~Caprini and J.-F. Dufaux,  {\em JCAP} {\bf 06} (2012)
  027 [\href{http://arXiv.org/abs/1201.0983}{{\tt 1201.0983}}].

\bibitem{Espinosa:2010hh}
J.~R. Espinosa, T.~Konstandin, J.~M. No and G.~Servant,  {\em JCAP} {\bf 06}
  (2010) 028 [\href{http://arXiv.org/abs/1004.4187}{{\tt 1004.4187}}].

\bibitem{Seto:2001qf}
N.~Seto, S.~Kawamura and T.~Nakamura,  {\em Phys. Rev. Lett.} {\bf 87} (2001)
  221103 [\href{http://arXiv.org/abs/astro-ph/0108011}{{\tt
  astro-ph/0108011}}].

\bibitem{Kawamura:2011zz}
S.~Kawamura {\em et.~al.,} {\em Class. Quant. Grav.} {\bf 28} (2011) 094011.

\bibitem{Hu:2017mde}
W.-R. Hu and Y.-L. Wu,  {\em Natl. Sci. Rev.} {\bf 4} (2017), no.~5 685--686.

\bibitem{Guo:2018npi}
W.-H. Ruan, Z.-K. Guo, R.-G. Cai and Y.-Z. Zhang,  {\em Int. J. Mod. Phys. A}
  {\bf 35} (2020), no.~17 2050075 [\href{http://arXiv.org/abs/1807.09495}{{\tt
  1807.09495}}].

\bibitem{Luo:2015ght}
{TianQin} Collaboration, J.~Luo {\em et.~al.,} {\em Class. Quant. Grav.} {\bf
  33} (2016), no.~3 035010 [\href{http://arXiv.org/abs/1512.02076}{{\tt
  1512.02076}}].

\bibitem{Hu:2018yqb}
X.-C. Hu, X.-H. Li, Y.~Wang, W.-F. Feng, M.-Y. Zhou, Y.-M. Hu, S.-C. Hu, J.-W.
  Mei and C.-G. Shao,  {\em Class. Quant. Grav.} {\bf 35} (2018), no.~9 095008
  [\href{http://arXiv.org/abs/1803.03368}{{\tt 1803.03368}}].

\bibitem{Guo:2020grp}
H.-K. Guo, K.~Sinha, D.~Vagie and G.~White,  {\em JCAP} {\bf 01} (2021) 001
  [\href{http://arXiv.org/abs/2007.08537}{{\tt 2007.08537}}].

\bibitem{Cutting:2019zws}
D.~Cutting, M.~Hindmarsh and D.~J. Weir,  {\em Phys. Rev. Lett.} {\bf 125}
  (2020), no.~2 021302 [\href{http://arXiv.org/abs/1906.00480}{{\tt
  1906.00480}}].

\bibitem{Wang:2020jrd}
X.~Wang, F.~P. Huang and X.~Zhang,  {\em JCAP} {\bf 05} (2020) 045
  [\href{http://arXiv.org/abs/2003.08892}{{\tt 2003.08892}}].

\bibitem{Croon:2020cgk}
D.~Croon, O.~Gould, P.~Schicho, T.~V.~I. Tenkanen and G.~White,  {\em JHEP}
  {\bf 04} (2021) 055 [\href{http://arXiv.org/abs/2009.10080}{{\tt
  2009.10080}}].

\bibitem{Guo:2021qcq}
H.-K. Guo, K.~Sinha, D.~Vagie and G.~White,
  \href{http://arXiv.org/abs/2103.06933}{{\tt 2103.06933}}.

\bibitem{Gould:2021oba}
O.~Gould and T.~V.~I. Tenkanen,  {\em JHEP} {\bf 06} (2021) 069
  [\href{http://arXiv.org/abs/2104.04399}{{\tt 2104.04399}}].

\bibitem{Giese:2020rtr}
F.~Giese, T.~Konstandin and J.~Van De~Vis,  {\em JCAP} {\bf 07} (2020), no.~07
  057 [\href{http://arXiv.org/abs/2004.06995}{{\tt 2004.06995}}].

\bibitem{Hoeche:2020rsg}
S.~H\"oche, J.~Kozaczuk, A.~J. Long, J.~Turner and Y.~Wang,  {\em JCAP} {\bf
  03} (2021) 009 [\href{http://arXiv.org/abs/2007.10343}{{\tt 2007.10343}}].

\bibitem{Giese:2020znk}
F.~Giese, T.~Konstandin, K.~Schmitz and J.~Van De~Vis,  {\em JCAP} {\bf 01}
  (2021) 072 [\href{http://arXiv.org/abs/2010.09744}{{\tt 2010.09744}}].

\bibitem{Wang:2020nzm}
X.~Wang, F.~P. Huang and X.~Zhang,  {\em Phys. Rev. D} {\bf 103} (2021), no.~10
  103520 [\href{http://arXiv.org/abs/2010.13770}{{\tt 2010.13770}}].

\bibitem{Wang:2020zlf}
X.~Wang, F.~P. Huang and X.~Zhang,  \href{http://arXiv.org/abs/2011.12903}{{\tt
  2011.12903}}.

\bibitem{Delaunay:2007wb}
C.~Delaunay, C.~Grojean and J.~D. Wells,  {\em JHEP} {\bf 04} (2008) 029
  [\href{http://arXiv.org/abs/0711.2511}{{\tt 0711.2511}}].

\end{thebibliography}\endgroup

\end{document}